%% file: MUNThesis.tex
\documentclass[12pt]{MUNThesis}
\usepackage[left=3.8cm, right=2.5cm, top=3cm, bottom=3cm]{geometry}
\usepackage{hyperref}
\usepackage{paralist}
\usepackage{bbding}
\usepackage{listings}
\usepackage{wasysym}
\usepackage{appendix}
\usepackage{grffile}
\usepackage{graphicx}
\usepackage{url}
\usepackage{subfigure}
\usepackage{fancyhdr}
\usepackage{amsthm}
 \usepackage{color}
\usepackage{footmisc}
\usepackage{algorithmic}
\usepackage{listings}
\usepackage{fancyvrb}
\usepackage{cite}
\usepackage{epigraph}
\usepackage{mathtools}

\title{Classical gravitational scattering in the relativistic Kepler problem}

\author{Michael Y. Grudich}

\deg{Bachelor of Science (Honours)}

\fac{Department of Physics and Physical Oceanography \\
Department of Mathematics and Statistics}

\date{March 2014}

\copyrightyear{2014}

\AtBeginDocument{%
      \setlength\abovedisplayskip{0pt}
      \setlength\belowdisplayskip{0pt}}
\begin{document}
\muntitlepage
\setcounter{secnumdepth}{3} \setcounter{tocdepth}{3}

\pagenumbering{roman} \setcounter{page}{1}
\pagebreak

\input{abstract}


\renewcommand{\contentsname}{Table of Contents}
\tableofcontents{}
\addcontentsline{toc}{chapter}{Table of Contents}
\listoffigures{}
\addcontentsline{toc}{chapter}{List of Figures}

\doublespacing
\clearpage
\pagenumbering{arabic} 

\input{intro}
\input{theory}
\input{schwarzschild}

\input{PNResults}
\input{conclusion}
\begin{appendix}
\input{AppendixB}
\end{appendix}
\input{bibliography}

\end{document}

%% file: abstract.tex
\doublespacing
\setlength{\topmargin}{-.5in}
\chapter*{Abstract}

\addcontentsline{toc}{chapter}{Abstract}


Black holes are an ubiquitous end state of stellar evolution and successfully explain some of the most extreme physics encountered in astronomical observations. The Kerr geometry is the known exact solution to Einstein's equations for a static, eternal black hole within the framework of general relativity, and hence is of great importance in relativistic astrophysics. An understanding of the orbital dynamics of test bodies and light rays in the Kerr spacetime is therefore fundamental to the physics of a black hole. In this work, the scattering and capturing properties of unbound, ``hyperbolic'' orbits in the spacetime are studied. In particular, the differential scattering cross section and capture cross section are derived over the parameter space of energies, impact parameters and black hole spin orientation and magnitude. The problem is then generalized to the motion of two massive objects on a hyperbolic encounter, and the added effects of gravitational radiation and finite mass ratio studied within the post-Newtonian formalism.

%% file: intro.tex
\chapter{Introduction}
\epigraph{``We are to admit no more causes of natural things than such as are both true and sufficient to explain their appearances.''}{Isaac Newton}

The understanding of the force driving the motion of celestial objects has been marked by a series of refinements, each requiring greater mathematical sophistication than the last, but also reducing the number of assumptions from which the motion is derived. The crystal spheres of Aristotle, while technically flawed, posited celestial motion by purely mechanical means, certainly an improvement over the stories of deities racing across the sky common to various ancient cultures. Ptolemy's model of epicycles, rooted in the geometry of Hipparchus, provided an explanation for the retrograde motion of so-called {\it asteres planetai}, literally ``wandering stars'', now known to be the planets. Kepler, making use of the precise astrometric data of Tycho Brahe and a more sophisticated understanding of geometry, showed that the planets orbited not along circles but along ellipses, focused at the Sun. Newton, in his seminal 1687 work, showed that Kepler's laws could be explained by an attractive force between all bodies in direct proportion to the product of their masses and inverse proportion to the square of their separation.

Following in the trend of its predecessor theories, the theory of general relativity has been tested and vindicated largely through its application in the problem of orbital motion. The relativistic theory of gravity is in a sense simpler than even Newton's, dispensing with the concept of a gravitational ``force'' and instead positing that celestial objects are carried by their own inertia along paths through spacetime which, while appearing spatially curved, are in a sense the ``straightest'' possible within the geometry of spacetime. This geometry in turn is coupled to mass (or equivalently, energy) and momentum, and this coupling is expressed through famous field equation relating the geometric quantities $R_{\mu\nu}$ and $g_{\mu\nu}$ to the physical stress-energy $T_{\mu\nu}$ \cite{einstein1915}:
\begin{equation}
R_{\mu\nu} - \frac{1}{2}Rg_{\mu\nu} = \frac{8 \pi G}{c^4} T_{\mu\nu} \,.
\end{equation}
Einstein showed that according to this theory, the orbits about a gravitating body deviated from those obtained from Newtonian gravity, and was able to explain successfully the then-anomalous precession of the orbit of Mercury, deriving a formula for the shift in argument of perihelion each orbit \cite{mercury}:
\begin{equation}
\delta \phi = \frac{6 \pi GM}{c^2 A (1-e^2)} + \mathcal{O}\left(\frac{1}{c^4}\right)\,.
\label{Mercury}
\end{equation}
The deviation from the Newtonian result was detected for Mercury in particular because it is the deepest in the Sun's gravity well of all planets, and correspondingly is moving the fastest; it is a general result that in the limit of small velocity there is a correspondence between the results of general relativistic and Newtonian gravity. The above expression is in fact of order $\frac{v^2}{c^2}$ where $v$ is the orbital velocity.

In addition to providing small corrections to otherwise overwhelmingly Newtonian physics, general relativity made predictions pertaining to the motion of light itself. As was verified by Eddington in 1919, a massive object such as the Sun deflects passing light rays (or relativistic particles such as neutrinos) toward it, and Einstein found the angle of deflection to be, to lowest order in the approach distance $R$:
\begin{equation}
\hat{\theta} = \frac{4 G M}{R c^2} + \mathcal{O}\left(\frac{1}{c^4}\right)\,.
\end{equation}
This is in fact twice the result obtained by na\"{i}vely applying Newton's laws to a test body with an initial approach velocity of $c$. Therefore, unlike the subtle corrections to the relatively slow motion of the planets, the predictions of GR for relativistic orbital motion contrast sharply to those of Newtonian gravity.

Shortly after the publication of equation \ref{Mercury}, Karl Schwarzschild found an exact solution to the field equations which reproduced Einstein's result, now known as the Schwarzschild metric:
\begin{equation}
ds^2 = -\left(1-\frac{2 G M}{c^2 r}\right)c^2 dt^2 + \left(1 -\frac{2 G M}{c^2 r}\right)^{-1} dr^2 + r^2\left(d\theta^2 + \sin^2 \theta d\phi^2 \right)\,.
\end{equation}
This is the exterior geometry of any non-rotating, spherically-symmetric body of mass $M$ and hence was recognized as the basic relativistic model of the gravity of a star.

It was not immediately realized that regime of orbital dynamics well beyond any approximation of precessing conic sections or subtly deflected light rays was in fact physical, as the theoretical existence of objects compact enough to have such a strong field was not yet established in the theory of stellar evolution. It was found by Chandrasekhar that there is a maximum mass for a white dwarf (approximately $1.44M_\odot$) beyond which it is unstable against collapse. It was subsequently shown by Oppenheimer and Snyder \cite{Oppenheimer1939} that, should a compact object of mass $M$ have a diameter on the order of its gravitational radius $R_G=GM/c^2$, it would inevitably collapse to a singularity. The existence of objects possessing such singularities, dubbed ``black holes'' by John Wheeler in 1967, has since become widely accepted as one of the possible end states of stellar evolution. A large number of likely black hole candidates have been identified \cite{BHCandidates}. In particular, it is common to find evidence for the presence of a supermassive black hole at the center of a galaxy.

The Schwarzschild metric probably does not describe the gravitational field close to an astrophysical black hole very well; nature is not so kind as to provide situations of such high symmetry. The metric lacks an important property of a black hole, which is its spin angular momentum $\vec{J}$. In the case of stellar black holes, generally the progenitor star has non-vanishing angular momentum, some of which may be shed during gravitational collapse but the rest of which remains in the resulting black hole. Furthermore, if the black hole has an accretion disk, matter will gradually spiral in, losing some angular momentum in the process, but the angular momentum remaining as it reaches the innermost stable circular orbit is essentially fed into the black hole, spinning it up \cite{frolov}. The spins of various black hole candidates have been measured, and in most cases found to be quite significant \cite{cygnus}\cite{NGC1365}\cite{SgrSpin}\cite{BHSpins}.

The exact geometry of a spinning black hole was derived by Roy Kerr in 1963 \cite{Kerr1963} and today bears his name. The black hole, while a rather bizarre object, is perhaps the simplest macroscopic object in existence. While stars and planets are composed of matter which may be heterogeneous and dynamical, and their gravitational fields determined by their many freely specifiable mass multipole moments, the Kerr geometries are parametrized in only two quantities: the black hole's mass and its intrinsic angular momentum. Therefore, the problem of orbital motion around a black hole is arguably the simplest and the most fundamental in general relativity. As such, the detailed solution of this problem is the first focus of this work. 

The equations of motion for Kerr geodesics are of course well-studied, in particular since they were revealed to be separable in Carter's seminal work \cite{Carter1968} which revealed the necessary fourth constant of motion. Chandrasekhar's opus on black hole physics \cite{Chandra1983} contains possibly the most thorough treatment and overview of the solutions of the Kerr geodesic problem, and this work makes no attempt to achieve the same scope. Rather, we direct our attention to one particular sub-case of the problem, namely the scattering and capturing of test particles approaching from infinity. By solving exactly for the deflection angle of these orbits, a black hole can be studied from the perspective of scattering physics, wherein the physical, gauge-invariant observables are the capture cross section and the differential scattering cross section.

The situation of a test particle orbiting a fixed black hole is the simplest dynamical gravitational system in that it demands no information about the nature of the test particle and only two pieces of information about the nature of the black hole: its mass and its spin. In the approach to the problem the particle is formally considered to have a mass, however the dynamics only depend on specific energies and momenta, and not explicitly on the mass itself. This is an entirely valid assumption when talking about a neutrino or a photon around a stellar mass black hole, however once the mass of the orbiting object becomes comparable to that of the black hole it evidently is not: the black hole should move under the influence of the other body. The next logical step in this exploration of orbital motion is therefore to address the problem of two bodies scattering under a mutual interaction, and so this constitutes the second part of this work.

The simplest version of this problem does not involve black holes; while a \emph{static} astrophysical black hole is characterized by two quantities, two black holes moving under mutual gravity constitutes a \emph{dynamical} situation wherein the spacetime is not static and admits a wide parameter space of initial conditions. For example, the influence of one black hole will perturb the horizon geometry of the other, causing what is effectively a tidal interaction. The simplest general relativistic 2-body problem, actually, is that of two \emph{point masses}, free of any internal structure and characterized only by their masses and spin angular momenta. 

This problem, however, is unphysical; general relativity does not permit the existence of point masses, as any sufficiently dense collection of matter should collapse to a black hole. Nevertheless, the post-Newtonian (PN) formalism \emph{assumes} such structureless point masses as its starting point for solving the field equations, and is able to obtain results which must agree with those of physical black holes up to the order at which tidal effects become important, which turns out to be higher order than the known equations of motion in any case \cite{Blanchet2006}. The details of the black hole's near-zone geometry are effectively effaced in a large region of parameter space, allowing the problem of motion to be studied in the post-Newtonian approximation.

Investigations in numerical relativity have also found that the orbital dynamics of neutron stars closely match those of black holes when tidal effects are small, as one would expect \cite{me1}. Because the size of a neutron star is only a few gravitational radii, tidal effects only become important when the orbit is on the order of this length scale. This is where the PN approximation breaks down, and as such, the post-Newtonian results presented which are actually physically reasonable should be equally applicable to double black hole, black hole-neutron star and double neutron star binaries.

Unlike the Newtonian 2-body problem, which has effectively the same dynamics as the problem of a test body moving in fixed gravitational potential, the relativistic version will have some dependence on the binary mass ratio $q$, smoothly recovering the test particle dynamics in the limit $q\rightarrow 0$. There is also an additional complication: relativistic binaries lose energy and angular momentum by emitting gravitational radiation in a way very much analogous to the process of bremmstrahlung in electrodynamics \cite{Kovacs1978}. This work therefore takes particular interest in the effects of mass ratio and radiation reaction on the observables in the 2-body scattering problem.

%% file: theory.tex
\chapter{Theory}

\section{Scattering Physics}
\epigraph{``What comes around is all around.''}{Ricky, \it{Trailer Park Boys}}
In a scattering event, two objects approach each other, are deflected by some mutual interaction, and proceed away from each other. If the interaction vanishes at large separations, it is possible to define initial and final velocities at infinity in a meaningful way, as the trajectories become asymptotically straight as the interaction vanishes.
\begin{figure}
\centering
\includegraphics[width=0.8\textwidth]{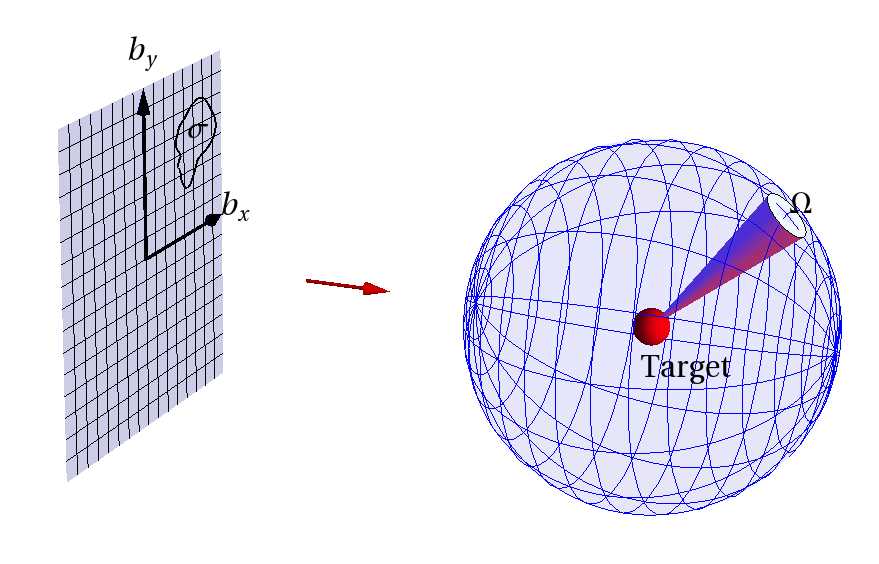}
\caption{A scattering trajectory can be abstracted to a map from a plane at infinity to a sphere at infinity.}
\label{planesphere}
\end{figure}
For a free body scattering off a fixed target, the possible trajectories can be parametrized by the initial speed (or equivalently energy) and an initial position on a ``plane at infinity'' whose normal is parallel to the initial velocity. In classical scattering it is possible to calculate the unique final trajectory after the body has been deflected. Since this trajectory can be specified by two angles in 3 dimensional space, the scattering trajectory maps a position on the plane to a point on the unit sphere $S^2$ (Figure \ref{planesphere}). Therefore, the result of a given scattering event is encoded in a map $\psi:\mathbb{R}^2 \rightarrow S^2$ which takes the initial position on the plane and gives the angles specifying  the body's direction after being deflected.
\begin{figure}
\centering
\includegraphics[width=\textwidth]{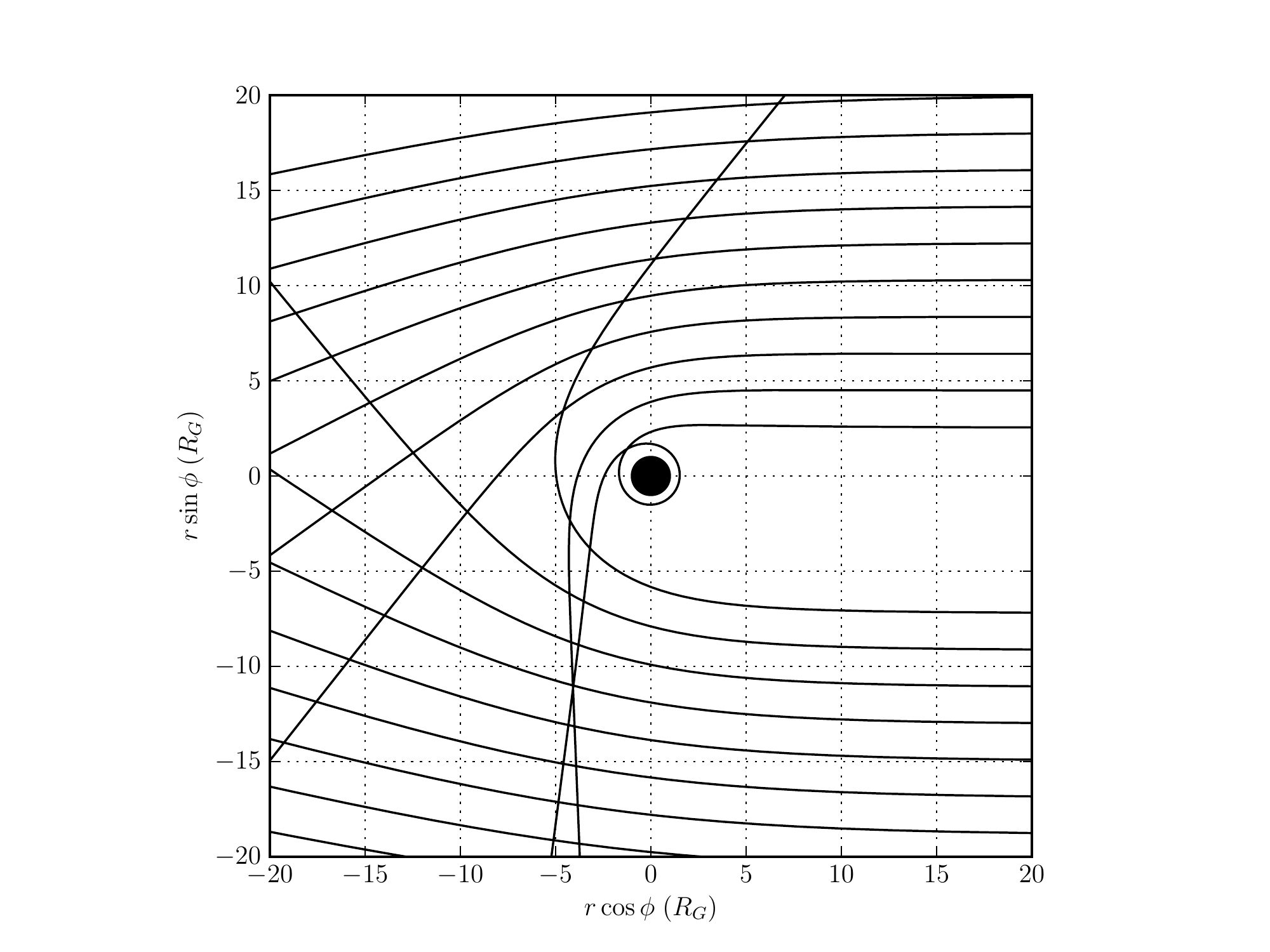}
\caption{Trajectories of scattered light in the equatorial plane of a near-maximally spinning black hole ($\alpha = 0.998$).}
\end{figure}
$S^2$ can be parametrized with the standard spherical coordinates $\hat{\theta}$ and $\hat{\phi}$; in all scattering calculations to follow, $\hat{\theta} = \pi$ will correspond to the direction of approach. The most natural set of coordinates on the plane are of course Cartesian ones, and so we define the coordinates (``impact parameters'') $b_x$ and $b_y$ as horizontal and vertical Cartesian coordinates on the plane with the origin $O$ at the point of intersection with the axis of the target. In many scattering problems the target is spherically symmetric; this effectively makes the parameter space of trajectories one-dimensional modulo rotations, so it is standard to instead specify the impact parameter $b=\sqrt{b_x^2+b_y^2}$.


A {\it cross section} is simply an area in the $b_x$-$b_y$ plane. For example, one could ask what is the area of the subset of the plane whose trajectories end up in a certain solid angular cone: this is the {\it scattering cross section} for that cone. One could also ask what the area of the region of the plane whose trajectories end up captured by the target is: this is the {\it capture cross section}.

To integrate over $\mathbb{R}^2$ in the coordinates of $S^2$, we can simply take the pullback of the area element $d\sigma = db_x \wedge db_y$ under the inverse map $\psi^{-1}:S^2 \rightarrow \mathbb{R}^2$ and integrate it over the sphere:
\begin{equation}
\int_O d\sigma = \int_O d b_x \wedge d b_y = \int_{\psi(O)} \left(\frac{\partial{b_x}}{\partial{\hat{\theta}}} \frac{\partial{b_y}}{\partial{\hat{\phi}}} - \frac{\partial{b_x}}{\partial{\hat{\phi}}} \frac{\partial{b_y}}{\partial{\hat{\theta}}}\right) d\hat{\theta} \wedge d\hat{\phi} \equiv \int_{\psi(O)} \det J\, d\hat{\theta} \wedge d\hat{\phi}\,.
\label{pullback}
\end{equation}

Here J is the Jacobian of $\psi^{-1}$. In defining the inverse scattering function $\psi^{-1}(\hat{\theta},\hat{\phi})$ it was assumed implicitly that $\psi(b_x,b_y)$ was in fact invertible. This is usually locally possible because $\psi$ is usually differentiable, but not necessarily globally possible. For example, it may be the case, and in fact is the case for black hole orbits, that the angle of deflection may be greater than $\pi$. Indeed, unbound black hole orbits exist which orbit the hole arbitrarily many times. As such, there are infinitely many values of $b$ that send the particle in any given direction. $\psi$ is in this case not invertible because it is not injective. It is then necessary to partition $\psi^{-1}$ into branch cuts $\psi^{-1}_n:S^2 \rightarrow O_n$ where $\lbrace O_n \rbrace$ is a partition of $\mathbb{R}^2$. The cross section is then the sum over branches.
\begin{figure}
\centering
\includegraphics[width=0.5\textwidth]{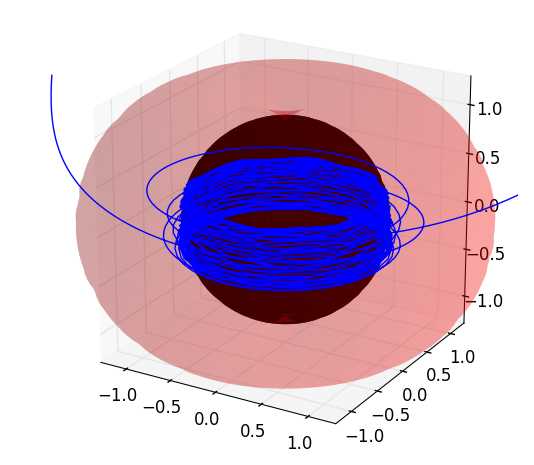}
\caption[Kerr scattering trajectory with multiple orbits]{An example of a light trajectory which travels inside a near-extremal black hole's ergosphere (red) and orbits many times before escaping.}
\end{figure}
Comparing equation \ref{pullback} with the angular area element $d\Omega = \sin \hat{\theta}\, d\hat{\theta} \wedge d\hat{\phi}$ leads to a general definition of the differential scattering cross section:
\begin{align}
\frac{d\sigma}{d\Omega} := \frac{\det J}{\sin \hat{\theta}}\,.
\end{align}

There is a connection between the differential cross section and the result of a scattering experiment: if a beam of particles of uniform intensity is fired at the target, the cross section is proportional to the probability distribution function of the particles' scattering angles. If one were to shine a light source at a target which deflects light (such as a black hole), $\displaystyle\frac{d\sigma}{d\Omega}$ would be proportional to the luminous intensity of the scattered light. Analogously with subluminal particles, if a dust cloud of uniform velocity and negligible self-gravity were to encounter a black hole, $\displaystyle \frac{d\sigma}{d\Omega}$ would be proportional to the post-encounter directional velocity distribution.

\subsection{Newtonian Solution}
\begin{figure}
\centering
\includegraphics[width=0.6\textwidth]{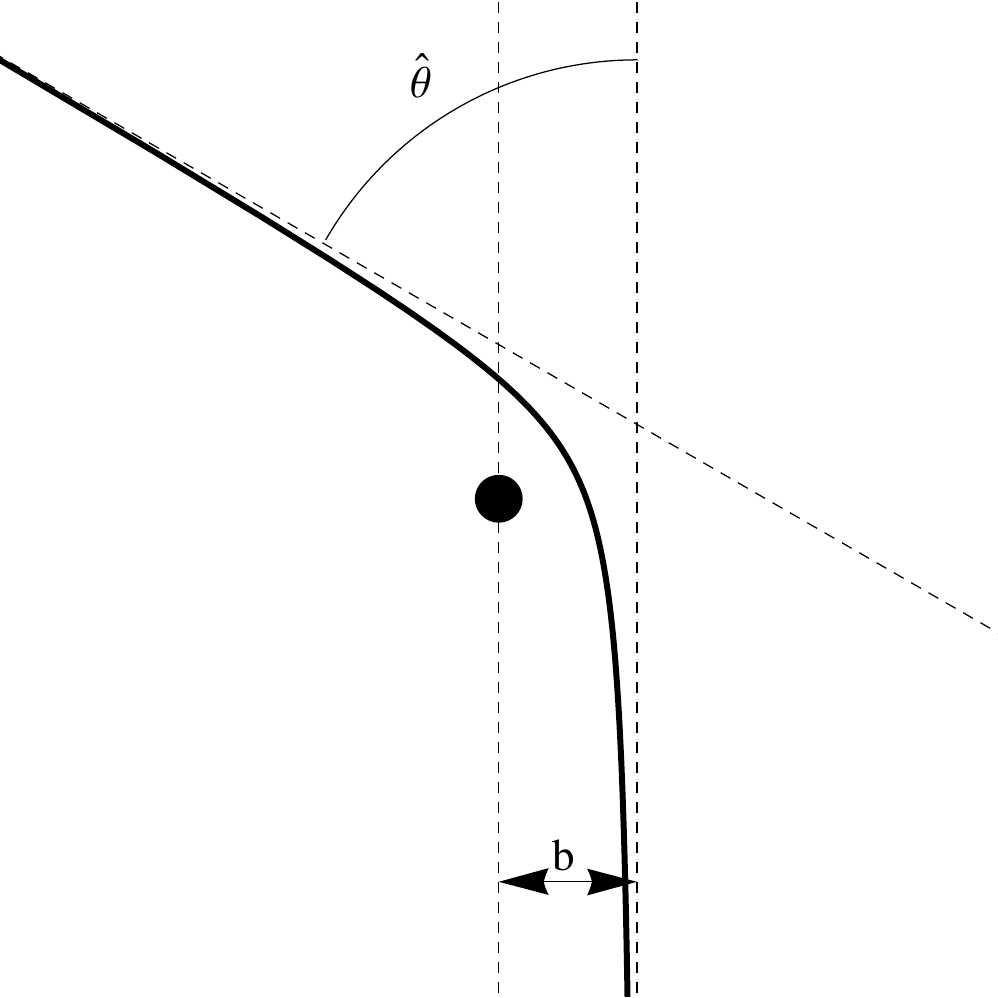}
\caption{Unbound scattering orbit in Newtonian gravity.}
\label{NewtonianOrbit}
\end{figure}
It is illustrative to derive the scattering angle solution for the Newtonian Kepler problem (Figure \ref{NewtonianOrbit}), and the system of dimensionless variables used will be equally applicable to the relativistic problem. Starting with the centre of mass frame Hamiltonian of two moving point masses $m_1$ and $m_2$ with gravitational interaction:
\begin{equation}
\mathcal{H} = \frac{L^2}{2\mu R^2} + \frac{p_R^2}{2\mu} - \frac{G M \mu}{R^2}\,.
\end{equation}
Here $R$ is the distance between the masses, $M=m_1 + m_2$ is the total mass, $\mu = \displaystyle{m_1 m_2}{m_1 + m_2}$ is the reduced mass, $L = p_\phi$ is the total (orbital) angular momentum and $p_R = \mu \dot{R}$. The motion of the system is equivalent to that of particle of mass $\mu$ orbiting about a fixed particle of mass $M$. For an unbound orbit, the Hamiltonian is equal to the kinetic energy at infinity:
\begin{equation}
\mathcal{H} = \frac{L^2}{2\mu R^2} + \frac{p_R^2}{2\mu} - \frac{G M \mu}{R^2} = \frac{1}{2}\mu V_\infty^2\,.
\end{equation}
Introducing the dimensionless variables $r = R/R_G = R/\left(\frac{GM}{c^2}\right)$, $h = \displaystyle\frac{L}{\mu c R_G}$, $p_r = \displaystyle\frac{p_R}{\mu c}$, and $v_\infty = V_\infty/c$:
\begin{equation}
\frac{\mathcal{H}}{\mu c^2} = \frac{p_r^2}{2} + \frac{h^2}{2 r^2} - \frac{1}{r} = \frac{1}{2} v_\infty^2\,.
\end{equation}
The angular displacement is then found by integrating:
\begin{equation}
\phi - \phi_0 = \int d\phi = \int \frac{\dot{\phi}}{\dot{r}} dr = \int \frac{h}{r^2}\frac{1}{\pm\sqrt{v_\infty^2 + \displaystyle{\frac{2}{r} - \frac{h^2}{r^2}}}} dr \,.
\end{equation}
$p_r$ is negative on the approach of the orbit, reaches 0 at the periastron radius $r_p$ and is positive on the escape, hence:
\begin{equation}
\phi - \phi_0 = 2 \int_{r_p}^\infty \frac{h}{\sqrt{v_\infty^2 r^4 + 2 r^3 - h^2 r^2}} dr = \pi + 2 \cot^{-1}\left(b v_\infty^2\right) = \pi + 2 \cot^{-1}\left(b v_\infty^2\right)\,.
\end{equation}
where $b = h/v_\infty$ is the dimensionless impact parameter. The deflection angle $\hat{\theta}$ is then the displacement from the undeflected trajectory $\phi = \pi$:
\begin{equation}
\hat{\theta} =  2 \cot^{-1}\left(b v_\infty^2\right) \rightarrow b = \frac{\cot(\frac{\hat{\theta}}{2})}{v_\infty^2}\,.
\label{NewtonDeflection}
\end{equation}
Notice that an object on a Newtonian unbound orbit can be deflected by at most $\pi$; the orbit is constrained to be a conic section whose asymptotes cannot intersect. This allows the expression for $\hat{\theta}$ to be inverted for $b$ without requiring any partitioning into branch cuts.

The area element in the $b$-plane in polar coordinates is $d\sigma =b\, db \wedge d\varphi$, with $\varphi$ the polar angle. Hence its pullback to the angular sphere is $b\left(\displaystyle \frac{db}{d\hat{\theta}} \frac{d\varphi}{d\hat{\phi}} - \frac{db}{d\hat{\phi}} \frac{d\varphi}{d\hat{\theta}}\right) d\hat{\theta} \wedge d\hat{\phi} = b \displaystyle\frac{db}{d\hat{\theta}} d\hat{\theta} \wedge d\hat{\phi}$, so the differential scattering cross section is:
\begin{equation}
\frac{d\sigma}{d\Omega} = \left|\frac{b}{\sin\hat{\theta}}\frac{db}{d\hat{\theta}}\right| = \frac{\csc^4(\frac{\hat{\theta}}{2})}{4\sin\hat{\theta} v_\infty^4}\,.
\end{equation}
$\displaystyle\frac{d\sigma}{d\Omega}$ diverges in the limit $\hat{\theta}\rightarrow 0$ because the annuli in the $b$-plane whose trajectories are scattering through a given angle $\hat{\theta}$ become larger and larger in area (proportional to $b$) in this limit. On the other end, the cross section vanishes at $\hat{\theta} = \pi$ because the annuli about the origin whose trajectories reach angles close to $\pi$ become vanishingly small in area.
\begin{figure}
\centering
\includegraphics[width=\textwidth]{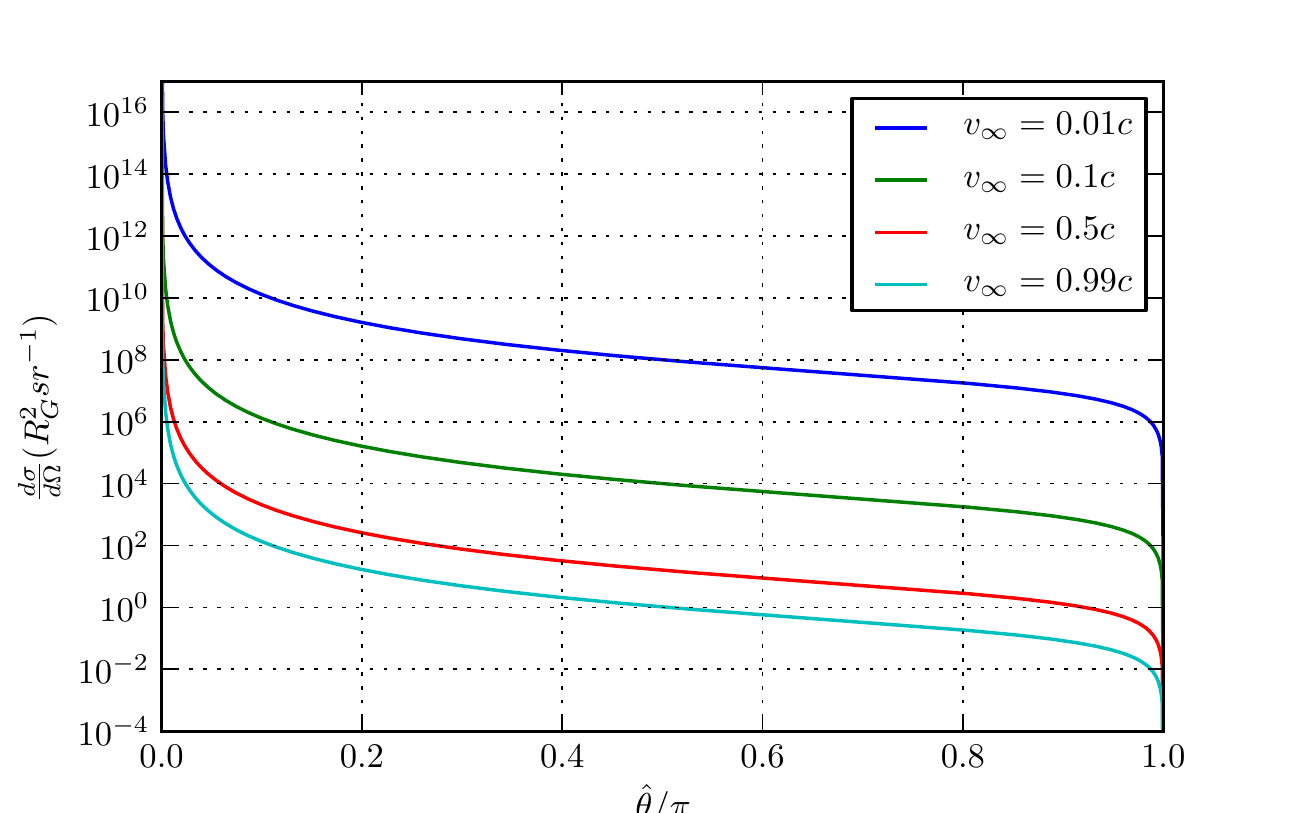}
\caption[Newtonian differential scattering cross section]{Newtonian differential scattering cross section.}
\end{figure}

\section{Geodesics in General Relativity}
\epigraph{``Spacetime tells matter how to move; matter tells spacetime how to curve.''}{John Wheeler}

The theory of general relativity presents the universe as a 4-dimensional manifold $\mathcal{M}$, a space for which around every point there exists an open region which can be mapped smoothly into an open subset of $\mathbb{R}^4$ and back. Such a map defines a local coordinate system $(x^0, x^1, x^2, x^3)$. $\mathcal{M}$ is equipped with a Lorentzian metric $\mathbf{g}$ which is a smooth, symmetric, bilinear map from the tangent vector space of $\mathcal{M}$ to $\mathbb{R}$. $\bold{g}$ may be expressed as a line element in local coordinates $x^i$ in terms of the basis 1-forms $dx^i$ of the cotangent space:

\begin{equation}
\bold{g} = g_{\mu\nu} dx^\mu \otimes dx^\nu\,.
\end{equation}

The path a particle follows through $\mathcal{M}$ may be viewed as a map $f:\mathbb{R} \rightarrow \mathcal{M}$ known as its {\it worldline}, and is the locus of all events at which the particle is present. If the worldline is parametrized in a set of local coordinates $f(\lambda) = x^i(\lambda)$ then its tangent vector $\bold{T}$ is:
\begin{equation}
\bold{T} = T^i \frac{\partial}{\partial x^i} = \frac{d x^i (\lambda)}{d\lambda} \frac{\partial}{\partial x^i}\,.
\end{equation}

$\bold{T}$ is known as {\it timelike} if its length $g_{\mu\nu} T^\mu T^\nu < 0$, {\it null} if $g_{\mu\nu}T^\mu T^\nu = 0$, and {\it spacelike} otherwise. Similarly, these terms apply to the curve if the curve's tangent vector satisfies one of the conditions everywhere on the curve.

The geometric length of the curve between two points $f(a)$ and $f(b)$ is naturally the integral of absolute length of the tangent vector, and for timelike and null curves is proportional to the proper time $\tau$ elapsed on a clock traveling along the curve between these two events. The geometric length and proper time can therefore be calculated by integrating the differential line element:
\begin{equation}
ds^2 = -c^2 d\tau^2 = g_{\mu \nu} dx^{\mu} dx^{\nu} = g_{\mu \nu} \frac{dx^{\mu}}{d\lambda} \frac{dx^{\nu}}{d\lambda} d\lambda^2\,.
\label{lineelement}
\end{equation}
Here $\lambda$ may be any parametrization of the curve, but in the study of geodesic motion it is useful to define $\lambda$ as an affine parameter proportional to the proper time via the equation $m\lambda = \tau$ for a test particle of mass $m$, as the parameter behaves well in the limit $m\rightarrow 0$, allowing the trajectory of a light ray to be recovered from the general solution for a test mass by taking this limit. The tangent vector obtained by differentiating the curve with respect to $\lambda$ is the 4-momentum $p$:
\begin{equation}
\bold{p} \equiv m \frac{dx^i}{d\tau} \frac{\partial}{\partial x^i} = \frac{dx^i}{d\lambda} \frac{\partial}{\partial x^i}\,.
\end{equation}
Hence a normalization condition on $\bold{p}$ may be obtained as a constant of motion from equation \ref{lineelement}:
\begin{equation}
g_{\mu\nu}p^\mu p^\nu = g_{\mu\nu} \frac{dx^{\mu}}{d\lambda} \frac{dx^{\nu}}{d\lambda} = -c^2\left(\frac{d\tau}{d\lambda}\right)^2 = -m^2 c^2\,.
\end{equation}

The orbits of particles are the geodesics of the spacetime, curves of locally extremal length, and hence may be obtained from an action principle with a suitable Lagrangian $\mathcal{L}(x^i, \dot{x}^i)$: 
\begin{equation}
\mathcal{S} = \int \mathcal{L} d\lambda  \text{ where }\mathcal{L} = g_{\mu \nu} \frac{dx^{\mu}}{d\lambda} \frac{dx^{\nu}}{d\lambda}\,.
\end{equation}
The variational equation $\delta S = 0$ is then solved by the Euler-Lagrange equation:
\begin{equation}
\frac{\partial \mathcal{L}}{\partial x^\alpha} = \frac{d}{d\lambda} \frac{\partial \mathcal{L}}{\partial \dot{x}^\alpha} \text{ where } \dot{x}^a = \frac{dx^\alpha}{d\lambda}\,.
\end{equation}
Analogously to classical mechanics, it is possible to express the equations of motion in canonical form by constructing the ``super-Hamiltonian'' \cite{Misner1973}:
\begin{equation}
\mathcal{H} = \frac{1}{2} g^{\mu\nu}p_\mu p_\nu\,.
\end{equation}
The geodesic equations of motion then follow from Hamilton's equations with the affine parameter $\lambda$ taking the place of the classical absolute time $t$:
\begin{align}
\dot{x}^\alpha &= \frac{\partial \mathcal{H}}{\partial p_\alpha} \,,\\
\dot{p}_\alpha &= -\frac{\partial \mathcal{H}}{\partial x^\alpha}\,.
\end{align}

\include{kerr}

\include{PN}

%% file: kerr.tex
\epigraph{``I look at the world and I notice it's turning...''}{The Beatles}


The Kerr line element encodes the geometry of a black hole with mass $M$ and spin angular momentum $J$, and may be expressed in Boyer-Lindquist coordinates $(t,r,\phi,\theta)$ as \cite{Misner1973}:
\begin{align}
ds^2 = - \frac{\Delta}{\Sigma}\left(c\,dt - a \sin^2 \theta\, d\phi \right)^2 + \frac{\sin ^2 \theta}{\Sigma} \left((r^2 + a^2)\,d\phi -a\,dt\right)^2 + \frac{\Sigma}{\Delta}dr^2 + \Sigma d\theta^2\,,
\end{align}
where $\Sigma = r^2 + a^2 \cos ^2 \theta$, $\Delta = r^2 - \displaystyle\frac{2GMr}{c^2} + a^2$, and $a = \displaystyle\frac{J}{Mc}$.
A natural feature to expect of a gravitational field is that it vanishes infinitely far away from the source. This is indeed the case with the Kerr metric, whose limit at infinity can be recognized as Minkowski spacetime. This will be important for scattering physics, as energy and momenta can be defined unambiguously in the asymptotic region in a "preferred" frame in which the black hole is at rest.

By computing the inverse metric and contracting it with a body's 4-momentum, the Hamiltonian of geodesic motion is obtained:
\begin{equation}
\mathcal{H} = \frac{1}{2}g^{\mu\nu}p_\mu p_\nu = -\frac{\left((r^2+a^2)p_t/c + a p_\phi \right)^2}{2\Delta \Sigma} + \frac{(p_\phi + a \sin^2\,\theta p_t/c)^2}{2\Sigma \sin^2\,\theta} + \frac{\Delta p_r^2}{2\Sigma} + \frac{p_\theta^2}{2\Sigma}\,.
\end{equation}
Geodesic motion in the Kerr spacetime is separable: the number of constants of motion is equal to the dimension of the configuration space. The first constant is $\mathcal{H} = -\frac{1}{2}m^2 c^2$, simply as a consequence of the normalization of $\bold{p}$.  The constants $E$ and $L_z$ are found immediately by applying Hamilton's equations, as a result of the axial and time symmetries of the spacetime:
\begin{eqnarray}
\dot{p_t} = -\frac{\partial \mathcal{H} }{\partial{t}} = 0 \implies p_t = g_{tt}\dot{t} + g_{t\phi} \dot{\phi} = -E\,, \\
\dot{p_\phi} = -\frac{\partial \mathcal{H}}{\partial \phi} = 0 \implies p_\phi = g_{\phi\phi} \dot{\phi} + g_{t\phi} \dot{t} = L_z\,.
\end{eqnarray}
$E$ is in fact the total energy (kinetic plus rest) of the particle at infinity, where it is well-defined within the framework of special relativity due to the aforementioned asymptotic flatness of the metric:
\begin{equation}
E = -g_{tt}\dot{t} - g_{t\phi} \dot{\phi} \rightarrow c^2 \frac{dt}{d\lambda} = \frac{m c^2}{\sqrt{1-v^2/c^2}}\,.
\end{equation}
Similarly, $L_z$ is the component of the angular momentum about the $z$-axis, assuming the familiar form at infinity:
\begin{equation}
L_z = g_{\phi\phi} \dot{\phi} + g_{t\phi} \dot{t} \rightarrow r^2 \sin^2\theta \dot{\phi}\,.
\end{equation}
The energy of a particle that is at rest at infinity and falls toward the black hole is therefore its rest energy $E_0=mc^2$. Hence one may reason that any particle with $E < mc^2$ is gravitationally bound, and if not, it is on an escape trajectory. The study of scattering orbits of particles originating at infinity therefore restricts itself to orbits with $E > mc^2$.

The fourth constant of motion is less obvious, and emerges from the Hamilton-Jacobi equation for $\mathcal{H}$ \cite{Carter1968}. Defining $S(t,r,\phi,\theta,\lambda)$ to be Hamilton's principal function, where $\frac{\partial S}{\partial x^i} = p_i$, the equation is \cite{Goldstein}:
\begin{equation}
\frac{\partial S}{\partial{\lambda}} = -\mathcal{H}\,.
\label{HJE}
\end{equation}
As $H$ has no explicit dependence on $\lambda$, $t$ or $\phi$, the separable solution requires that $S$ assume the following form:

\begin{equation}
S = \frac{1}{2}m^2c^2 \lambda - Et + L_z \phi + S_r(r) + S_\theta(\theta)\,.
\end{equation}

\noindent Substituting this into equation \ref{HJE} gives:
\begin{align}
\left(\frac{dS_\theta}{d\theta}\right)^2 + a^2 m^2 c^2 \cos^2 \theta + \left(aE\sin \theta/c - \frac{L_z}{\sin \theta}\right)^2 \nonumber\\
 = -\Delta \left(\frac{dS_r}{dr}\right)^2  + 2 \left((r^2+a^2)E/c-aL_z \right) \frac{dS_r}{dr} -m^2 c^2 r^2\,.
\label{HJE2}
\end{align}
Each side of this equation depends on a different variable, and hence both must be equal to a constant, known as Carter's constant $K$. Substituting the momenta into \ref{HJE2}:
\begin{align}
K &= p_\theta^2 + (aE \sin \theta/c - \frac{L_z}{\sin \theta})^2 + a^2 m^2 c^2 \cos^2\theta \\
&= -\Delta p_r^2 + 2 \left((r^2+a^2)E/c-aL_z \right) p_r -m^2 c^2 r^2 \,.
\end{align}
In the non-relativistic limit, $K$ is equal to the square of the total orbital angular momentum. Another useful quantity is $Q \coloneqq K - (L - aE/c)^2$, which is 0 if and only if the orbit lies in the equatorial plane. It corresponds to the square of the component of $\vec{L}$ projected onto the orbital plane. By differentiating $S$ with respect the the four constants of motion, the integral equations describing test particle orbits are obtained: 
\begin{align}
\int^\theta \frac{d\theta}{\sqrt{\Theta(\theta)}} &= \int^r \frac{dr}{\sqrt{R(r)}}\, , \\
\lambda &= \int^\theta \frac{a^2 \cos^2\theta}{\sqrt{\Theta(\theta)}}\,d\theta + \int^r \frac{r^2}{\sqrt{R(r)}} \,dr\, , \\
t &= \int^\theta \frac{a(L_z-aE \sin^2\theta/c)}{\sqrt{\Theta(\theta)}} \,d\theta + \int^r \frac{(r^2 + a^2)P(r)}{\Delta \sqrt{R(r)}}\, dr\,, \\
\phi &= \int^\theta \frac{L_z - a E \sin^2 \theta/c}{\sin^2 \theta \sqrt{\Theta(\theta)}}\,d\theta + \int^r \frac{a P(r)}{\Delta \sqrt{R(r)}} \,dr\,.
\end{align}
where the functions $\Theta(\theta)$, $P(r)$ and $R(r)$ are thus defined:
\begin{align}
\Theta(\theta) &= Q - \cos^2 \theta\left[a^2(m^2c^2 - E^2/c^2) + L_z^2/\sin^2\theta \right]\,, \\
P(r) &= E(r^2 + a^2)/c - L_z a\, ,\\
R(r) &= P(r)^2 - \Delta(m^2 c^2 r^2 + K)\,.
\end{align}
By differentiating and taking linear combinations of these equations, the geodesic equations of motion may be expressed via four first order differential equations:
\begin{align}
	\Sigma \dot{r} &= \pm \sqrt{R(r)} \label{drdt} \, ,\\
	\Sigma \dot{\theta} &= \pm \sqrt{\Theta(\theta)} \, ,\\
	\Sigma \dot{t} &= a\left(L_z - a E \sin^2\theta\right/c) + \frac{r^2+a^2}{\Delta}\left(\sqrt{R(r)}-P(r)\right) \, ,\\
	\Sigma \dot{\phi} &= \left(L_z \sin^2 \theta - a E/c \right) + \frac{a}{\Delta}\left(\sqrt{R(r)}-P(r)\right)\, .
\end{align}
The dimensional quantities of these equations merely establish the relevant energy, length and timescales. It is advantageous to solve the problem in dimensionless form by expressing the relevant quantities in terms of the gravitational length scale $R_G = \frac{G M}{c^2}$, the speed of light $c$, and the mass of the particle $m$:
\begin{equation}
\mathcal{E} \coloneqq \frac{E}{mc^2} \, ,
\end{equation}
\begin{equation}
h \coloneqq \frac{L}{m c R_G} \, ,
\end{equation}
\begin{equation}
\mathcal{K} \coloneqq \frac{K}{m^2 c^2 R_G^2} \, ,
\end{equation}
\begin{equation}
\mathcal{Q} \coloneqq \frac{Q}{m^2 c^2 R_G^2} \, ,
\end{equation}
\begin{equation}
\alpha \coloneqq \frac{a}{R_G} \, .
\end{equation}
Additionally, the coordinate $r$ shall be rescaled, related to the Boyer-Lindquist coordinate radius $r_{BL}$ by $r_{BL} = R_G r$, and the coordinate $\mu = \cos\theta$ will be used to eliminate all trigonometric functions.  The integral equations of motion which are relevant to the scattering problem then assume a dimensionless form:
\begin{equation}
\int^r \frac{dr}{\sqrt{\mathcal{R}(r)}} = \int ^\mu \frac{d\mu}{\sqrt{\mathcal{M(\mu)}}} \, ,
\label{ThetaDeflection}
\end{equation}
\begin{equation}
\phi -\phi_0= \int ^\mu \frac{h/(\mu^2 - 1) - \alpha \mathcal{E}}{\sqrt{\mathcal{M}(\mu)}} d\mu + \int^r \frac{\alpha}{r^2 - 2r + \alpha^2} \frac{(\alpha^2 + r^2) \mathcal{E} - \alpha h}{\sqrt{\mathcal{R}(r)}} dr \, .
\label{PhiDeflection}
\end{equation}
With the new dimensionless polynomials $\mathcal{R}(r)$ and $\mathcal{M}(\mu)$ defined as:
\begin{align}
\mathcal{R}(r) &= (\mathcal{E}^2 - 1)r^4 +2r^3 + \left(\alpha^2(\mathcal{E}^2-1) - h^2 - \mathcal{Q}\right)r^2 + 2 \mathcal{K} r - \alpha^{2} \mathcal{Q} \, , \\
\mathcal{M}(\mu) &= \alpha^2(1-\mathcal{E}^2) \mu^4 + \left(\alpha^2(\mathcal{E}^2 - 1) - h^2 -\mathcal{Q}\right) \mu^2 + \mathcal{Q} \, .
\end{align}


%% file: PN.tex
\section{Post-Newtonian Theory}
\epigraph{``Essentially, all models are wrong, but some are useful.''}{George E. P. Box}
Beyond the test particle limit, the 2-body problem of general relativity does not admit a closed form solution. One cannot obtain the solution through simple substitutions into the solution for a test particle in a potential, as is possible in Newtonian physics \cite{taylor2005classical}. This is because the picture of gravity of general relativity is fundamentally different from the Newtonian picture: in addition to the two bodies there is also a fully dynamical gravitational field coupled to them through the field equations. It is therefore necessary to resort to approximate methods to solve for the motion of two bodies moving under the influence of gravity. The most general approach is to solve Einstein's equations numerically \cite{Pretorius2005}, which is necessary to study the highly dynamical, strong-field physics of merging compact objects. This approach is a relatively recent development, and is at this time relatively computationally expensive, with simulations of a few orbits requiring tens of thousands of CPU hours \cite{PfeifferNR}. Because of this, simpler, approximate methods have been developed which apply to more restricted regions of the 2-body problem's parameter space.

The flagship of such methods is the post-Newtonian approximation, in which the field equations are solved perturbatively in powers of $\displaystyle\frac{1}{c}$ \cite{Blanchet2006}. By convention, an approximation of order $\displaystyle\frac{1}{c^n}$ is said to be at ``$\displaystyle\frac{n}{2}$PN'' order.  Given suitable gauge conditions, usually harmonic coordinates, it is possible to describe the motion of two point particles in terms of two coordinate 3-vectors $\vec{x_1}(t)$ and $\vec{x_2}(t)$. The equations of motion for these coordinates are then formulated as the Newtonian equations of motion for two gravitating bodies, with conservative corrections entering at $\mathcal{O}\left(\displaystyle\frac{1}{c^2}\right)$ \cite{EIH1938}, including spin-spin and spin-orbit interactions \cite{Kidder1995}, and dissipative radiation reaction components entering at order $\displaystyle\frac{1}{c^5}$. The expansion is generally accurate at all binary mass ratios in the limit $r >> \displaystyle\frac{GM}{c^2}$, and reproduces relativistic effects familiar from black hole geodesics, such as periastron precession and the presence of an innermost stable circular orbit.

The theory of general relativity, viewed from a dynamical perspective, is at a glance incompatible with the concept of a Dirac $\delta$-function mass distribution, as any such singularity should be censored by an event horizon. Indeed, when performing the perturbative analysis to derive the PN equations, certain divergent integrals are encountered. Nevertheless, somewhat amazingly, when these divergences are regularized away, a set of equations which faithfully recovers the weak-field orbital dynamics of a compact object binary is obtained.

In this work the harmonic coordinates formulation is used, however as we are only concerned with gauge-invariant observables in the asymptotic region any other formulation should obtain the same result. We also use the same system of dimensionless quantities as the Newtonian solution in Chapter 2, with all velocities expressed as fractions of $c$ and distances expressed in terms of $R_G=\displaystyle\frac{G(m_1+m_2)}{c^2}$. To 2.5PN order, and neglecting spin-spin and spin-orbit coupling, the dimensionless equation of motion for the separation vector $\vec{x}=\vec{x}_2-\vec{x}_1$ is \cite{Blanchet2006}:
\begin{equation}
\ddot{\vec{x}} = -\frac{1}{r^2}\left(\left(1+\mathcal{A}\right)\vec{n} + \mathcal{B} \vec{v}\right)
\label{PNEqn}
\end{equation}
where the PN coefficients $\mathcal{A}$ and $\mathcal{B}$ are the sum of terms from the various PN orders:
\begin{align}
\mathcal{A} &= \mathcal{A}_{PN}+\mathcal{A}_{2PN}+\mathcal{A}_{2.5PN}+\mathcal{O}(v^6) \,,\\
\mathcal{B} &= \mathcal{B}_{PN}+\mathcal{B}_{2PN}+\mathcal{B}_{2.5PN}+\mathcal{O}(v^6)\,.
\end{align}
where:
\begin{align}
\mathcal{A}_{PN} &= -\frac{\dot{r}^2 \eta}{2} + \left(1 + 3\eta\right) v^2 - \left(4+2\eta\right)/r  \,,\\
\mathcal{A}_{2PN}&= \frac{15\dot{r}^4 \eta}{8}\left(1-3\eta\right) +\frac{3 \dot{r}^2 v^2 \eta}{2} \left(4\eta - 3\right) + \eta v^4 \left(3 - 4\eta\right)\nonumber \,,\\
&+  \frac{1}{r}\left(-\dot{r}^2\left(2+25\eta+2\eta^2\right) + \frac{\eta v^2}{2}\left(4\eta-13\right)\right) + \frac{1}{r^2}\left(9 + \frac{87\eta}{4}\right)\,, \\
\mathcal{A}_{2.5PN} &= -\frac{8 \eta\dot{r}}{15}\left(\frac{9 v^2}{r} + \frac{17}{r^2} \right) \\
\mathcal{B}_{PN} &= 2\dot{r}\left(\eta-2\right)\,, \\
\mathcal{B}_{2PN} &= \frac{3 \eta \dot{r}^3}{2}\left(3 + 2\eta\right) - \frac{\eta\dot{r}v^2}{2}\left(15 + 4\eta\right) + \frac{\dot{r}}{r}\left(2 + \frac{41\eta}{2} + 4 \eta^2 \right)\,, \\
\mathcal{B}_{2.5PN} &= \frac{8\eta}{5}\left(\frac{v^2}{r} + \frac{3}{r^2}\right)\,.
\end{align}
$\eta$ is known as the symmetric mass ratio and is related to the true mass ratio $q$ via $\eta = \displaystyle\frac{q}{1+q^2}$. It is useful because its value for $q$ is equal to its value for $1/q$, and indeed these two situations should have the same physics by symmetry. The value of $\eta$ is at least $0$ in the test particle limit and at most $1/4$ in the equal mass limit.

Clearly, the equation of motion is not manifestly covariant; one may derive entirely different looking ``forces'' given different coordinate conditions. In particular, an alternate formulation exists which is based on iteratively constructing a Hamiltonian \cite{3PNHamiltonian} within the formalism of Arnowitt, Deser and Misner (ADM) \cite{ADM}. The coordinates themselves are not physically meaningful except in the asymptotic region, in which they coincide with a global Lorentz frame. Once again, this is not an issue if we wish to do scattering physics: the inner structure of the spacetime might as well be a black box whose input is the energy and impact parameter and whose output is the scattering angle, both being observables in the asymptotic region.

It was realized that the PN and 2PN terms in the equation of motion can be obtained from a certain Hamiltonian (as opposed to the ADM formulation, wherein the Hamiltonian was obtained first by construction). The conserved energy to 2PN order is \cite{Kidder1995}:
\begin{align}
E &= \left(\mathcal{E}_N + \mathcal{E}_{PN}+\mathcal{E}_{2PN}\right)\mu c^2\\
\mathcal{E}_N &= \frac{1}{2}v^2 -\frac{1}{r} \,,\\
\mathcal{E}_{PN} &= \frac{3}{8}\left(1-3\eta\right) v^4 + \frac{1}{2r}\left(\left(3 + \eta\right) v^2 + \eta \dot{r}^2\right) + \frac{1}{2r^2}\,,\\
\mathcal{E}_{2PN} &= \frac{5}{16}\left(1 - 7\eta + 13 \eta^2\right) v^6 + \frac{3\eta\dot{r}^4}{8r} \left(1-3\eta\right)+\frac{v^4}{8r}\left(21-23\eta-27\eta^2\right) \nonumber \,,\\
&+ \frac{v^2}{8r^2}\left(14-55\eta+4\eta^2\right)+\frac{\eta v^2 \dot{r}^2}{4r}\left(1-15\eta\right) - \frac{1}{4r^3}\left(2+15\eta\right) + \frac{\dot{r}^2}{8r^2}\left(4+69\eta+12\eta^2\right) \,.
\end{align}
\begin{figure}
\centering
\includegraphics[width=0.9\textwidth]{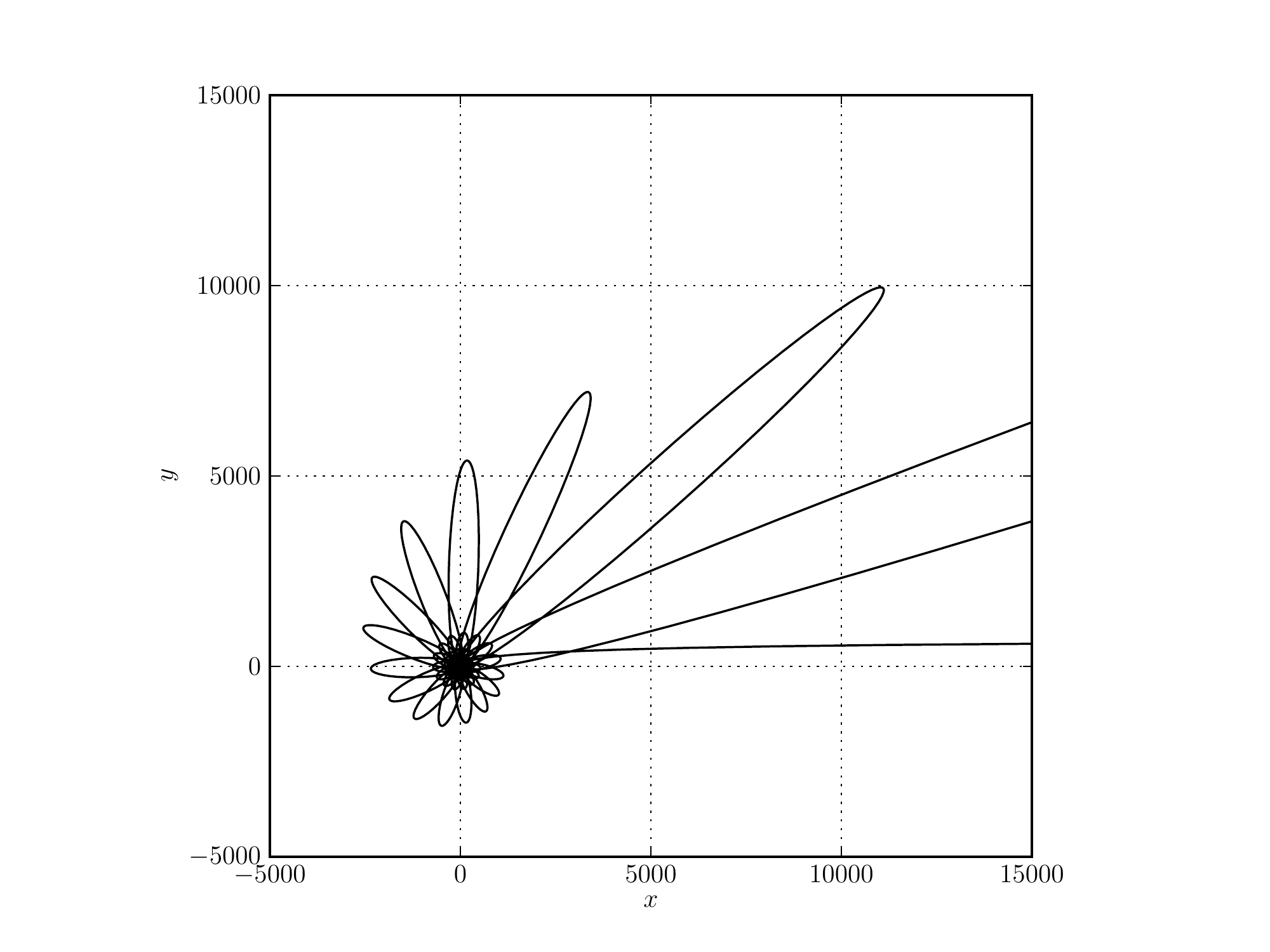}
\caption[Radiative gravitational capture of an equal-mass binary]{Example of an instance where an equal-mass binary on a hyperbolic encounter (with each mass initially traveling at $0.005c$) radiates enough energy to end up on a highly elliptical bound orbit which precesses and decays with each periastron passage.}
\label{PNCapture}
\end{figure}
The 2.5PN and 3.5PN terms of the PN acceleration have a dissipative effect on this energy; they encode the effect of radiation reaction on the masses, and the energy loss associated with it can be reconciled with the gravitational wave flux in the wave zone. This non-conservation of energy gives rise to a fundamental difference between the binary problem and the test particle problem: the mechanism of capture in the test particle problem is the passing of the event horizon, but the mechanism of capture in the binary problem is the emission of gravitational waves. When the masses pass near each other, it is possible for them to lose enough orbital energy to turn their unbound orbit into a bound one. Once in a bound orbit, the masses continue to lose energy and fall into a lower and lower orbit until they eventually merge. The PN approximation cannot be expected to provide accurate physics all the way into the merger, however the initial capture event may occur entirely within the PN regime (see figure \ref{PNCapture}).

Any PN calculations in which the bodies approach to within a few gravitational radii of each other must be taken with a grain of salt; only a full solution of the field equations can yield the highly nonlinear dynamics that occur on the gravitational length scale. To estimate the regime of validity of the PN expansion for a given problem, it is useful to compute the solution to several different orders and require that the series converges.

%% file: schwarzschild.tex

\chapter{Scattering and Capturing of Test Particles}
\epigraph{``Abandon all hope, ye who enter here.''}{Dante's \emph{Inferno}, Canto III}
\section{Capture Cross Section}
The essential property of a black hole is that an object that plunges past its event horizon can never escape. In particular for a Kerr black hole, this event horizon is a coordinate sphere in Boyer-Lindquist coordinates located at $r = \left(1 + \sqrt{1 -\alpha^2} \right)GM/c^2$ \cite{1972}. Therefore, the crossing of the event horizon is a type of capture event, and for a given $\mathcal{E}$, $\theta_0$ and $a$ there exists a certain region $\mathcal{C}$ of the $b$-plane containing all capture orbits, and the area of $\mathcal{C}$ is the capture cross section .

Several properties of $\mathcal{C}$ can be deduced intuitively. Clearly it is easier for a slow-moving particle to fall into the hole than a fast-moving one: the size of $\mathcal{C}$ should therefore increase as $\mathcal{E}$ decreases. As $\mathcal{E} \rightarrow \infty$, $\mathcal{C}$ should converge to a certain limit corresponding to the capture cross section for a photon. In the case of a Schwarzschild hole, $\mathcal{C}$ should evidently be rotationally symmetric about the origin. It corresponds to the region in which the angular momentum corresponding to $b$ is less than or equal to some critical value. Therefore, $\mathcal{C}$ is a disk with a certain radius. It will become apparent that the region retains its disk topology for all spin values, however it becomes geometrically deformed as the symmetry is broken.

A scattering orbit must at some point reach some radius of periastron $r_p$ at which the radial motion has a turning point. Thus, one can interpret capture orbits to be those orbits which lack this lower bound in the radial coordinate. From equation $\ref{drdt}$:
\begin{equation}
\frac{1}{2}\dot{r}^2 - \frac{R(r)}{\Sigma^2} \equiv T + V = 0 \,.
\label{Veff}
\end{equation}

\begin{figure}[ht!]
\centering
\includegraphics[width=0.9\textwidth]{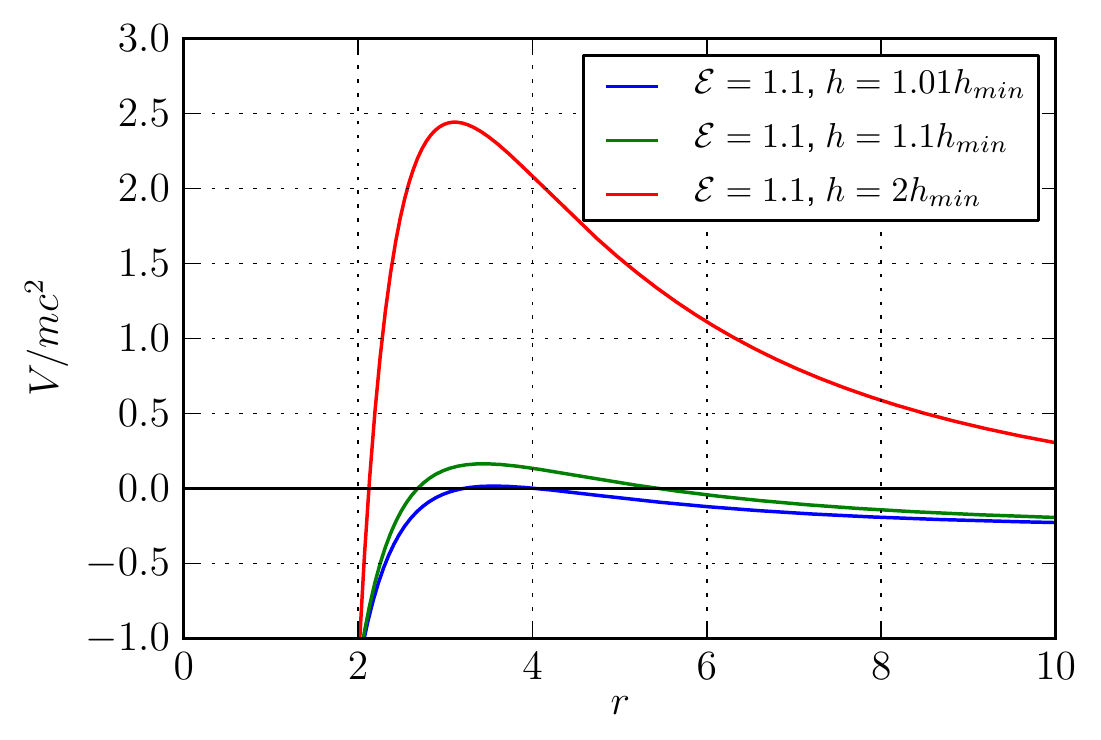}
\caption[Effective potential for radial motion of Kerr geodesics]{The effective potential $V$ from equation \ref{Veff} for a Schwarzschild black hole with $\mathcal{E}$ fixed at 1.1. As the angular momentum approaches the critical value, the potential peak approaches 0 and hence the particle is able to overcome the barrier and fall into the hole.}
\end{figure}
This is analogous to the energy balance equation for a particle moving in a potential defined by the second term, which is defined by the orbital parameters $\mathcal{E}$, $\mathcal{Q}$ and $h$. $r_p$ is the coordinate at which the particle ``bounces off'' the potential barrier, where $R(r)=0$. If a set of orbital parameters lies on the boundary between bouncing off the barrier and going over it, it must be that $r_p$ is located at the peak of the barrier, where $R'(r)=0$. Hence, the boundary of the capture region is defined by the simultaneous equations $R(r_p)=0$ and $R'(r_p)=0$.

If $\alpha = 0$, an equatorial orbit can be assumed without loss of generality, and hence $\mathcal{Q}=0$. The equations can then be solved simultaneously for $h$ and $r_p$:
\begin{align}
r_p &= \frac{8}{\mathcal{E} \left(\sqrt{9 \mathcal{E}^2-8}-3 \mathcal{E}\right)+4}\,, \\
h_{min}^2 &= \frac{8 - 36\mathcal{E}^2 + 27\mathcal{E}^4 + \mathcal{E}(9\mathcal{E}^2 - 8)^{3/2}}{2(\mathcal{E}^2-1)}\,.
\label{Schwhmin}
\end{align}

\begin{figure}[ht!]
\centering
\includegraphics[width=0.9\textwidth]{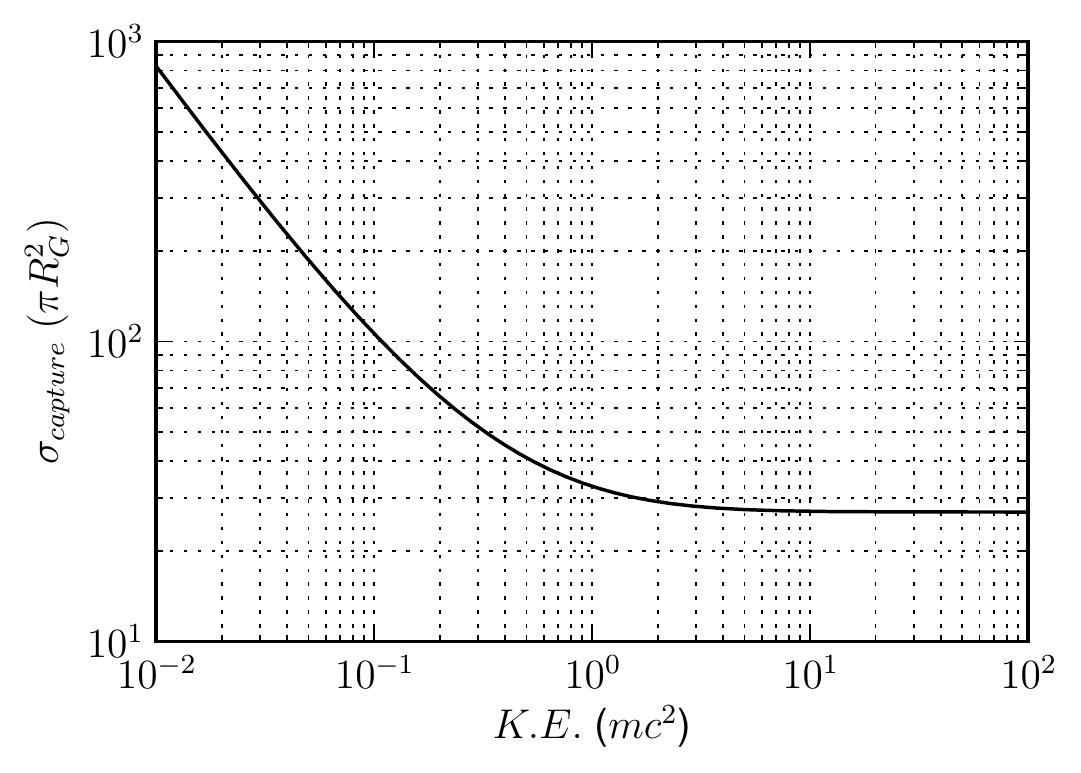}
\caption[Capture cross section of Schwarzschild black hole as a function of kinetic energy]{Capture cross section of a Schwarzschild black hole as a function of the particle kinetic energy (equal to $(\mathcal{E}-1)mc^2$). In the limit of small energy, the cross section is inversely proportional to the kinetic energy, while in the ultrarelativistic regime it approaches the constant value of $27\pi R_G^2$.}
\label{SigmaVsE}
\end{figure}

Hence the capture region in the $b$-plane of a Schwarzschild black hole is a disk bounded by the circle of radius $b_{min} =\displaystyle \frac{h_{min}}{\sqrt{\mathcal{E}^2 - 1}}$. Symmetry also demands that the capture region be circular for orbits approaching along the axis of a Kerr black hole, that is, where the initial $\theta$ coordinate is either $0$ or $\pi$. In this case $h = 0$, and it is possible to solve for $\mathcal{Q}$ and $r_p$ similarly, however not in entirely closed form: $r_p$ is the largest real solution of a quintic polynomial equation:
\begin{align}
&\left(\mathcal{E}^2-1\right)r_p^5+\left(4-3 \mathcal{E}^2\right) r_p^4+\left(2 \alpha^2 \mathcal{E}^2-2 \alpha^2-4\right) r_p^3 +\left(4 a^2-2 \alpha^2 \mathcal{E}^2\right) r_p^2  \nonumber \\
&+ \alpha^4\left(\mathcal{E}^2-1\right) r_p +\alpha^4 \mathcal{E}^2= 0\,.
\label{quintic}
\end{align}
\noindent$\mathcal{Q}$ can then be expressed in terms of $r_p$:
\begin{align}
\mathcal{Q} =& \frac{\left(\mathcal{E}^2 - 1 \right) r_p^4 + 2 r_p^3 + \alpha^2 \left(\mathcal{E}^2 - 1 \right)r_p^2 + 2 \alpha^2 \mathcal{E}^2 r_p}{r_p^2 - 2r_p + \alpha^2}\,.
\end{align}

When $\theta$ is $0$ or $\pi$, $\mathcal{Q} = \left(\mathcal{E}^2 - 1 \right) (b^2 - \alpha^2)$, and hence for polar orbits:
\begin{equation}
b_{min}^2 = \frac{\left(\alpha ^2+r_p^2\right) \left(\alpha ^2 \left(\mathcal{E}^2-1\right)+r_p \left(\left(\mathcal{E}^2-1\right) r_p+2\right)\right)}{\left(\mathcal{E}^2-1\right) \left(\alpha^2+\left(r_p-2\right) r_p\right)} \,.
\end{equation}

In the general case, neither $h$ nor $\mathcal{Q}$ is necessarily 0, so the equations admit a continuum of solutions for $h$ and $\mathcal{Q}$ parametrized in $r_p$.
These can then be related to $b_x$ and $b_y$ via the relations $b_x =\displaystyle \frac{h}{\sin \theta_0 \sqrt{\mathcal{E}^2 - 1}}$ and $b_y^2 = \displaystyle\frac{\mathcal{Q}}{\mathcal{E}^2 - 1} - \cos^2\theta_0 \left(b_x^2 - \alpha^2 \right)$. Physically reasonable solutions are obtained on the interval $\left[r_{p-},r_{p+}\right]$ where $r_{p\pm}$ are the solutions to $b_y(r_p) = 0$, found on the intervals $\left[1, r_{p,polar}\right]$ and $\left[r_{p,polar}, 6\right]$, with $r_{p,polar}$ the solution to equation \ref{quintic}. These roots are most easily found using a root finding algorithm that takes advantage of these bounds, such as bisection or Brent's method. The solutions are farthest apart when $\theta_0 = \pi/2$, and both converge to $r_{p,polar}$ as $\theta_0$ approaches $0$ or $\pi$.
\begin{figure}
\centering
\includegraphics[width=0.9\textwidth]{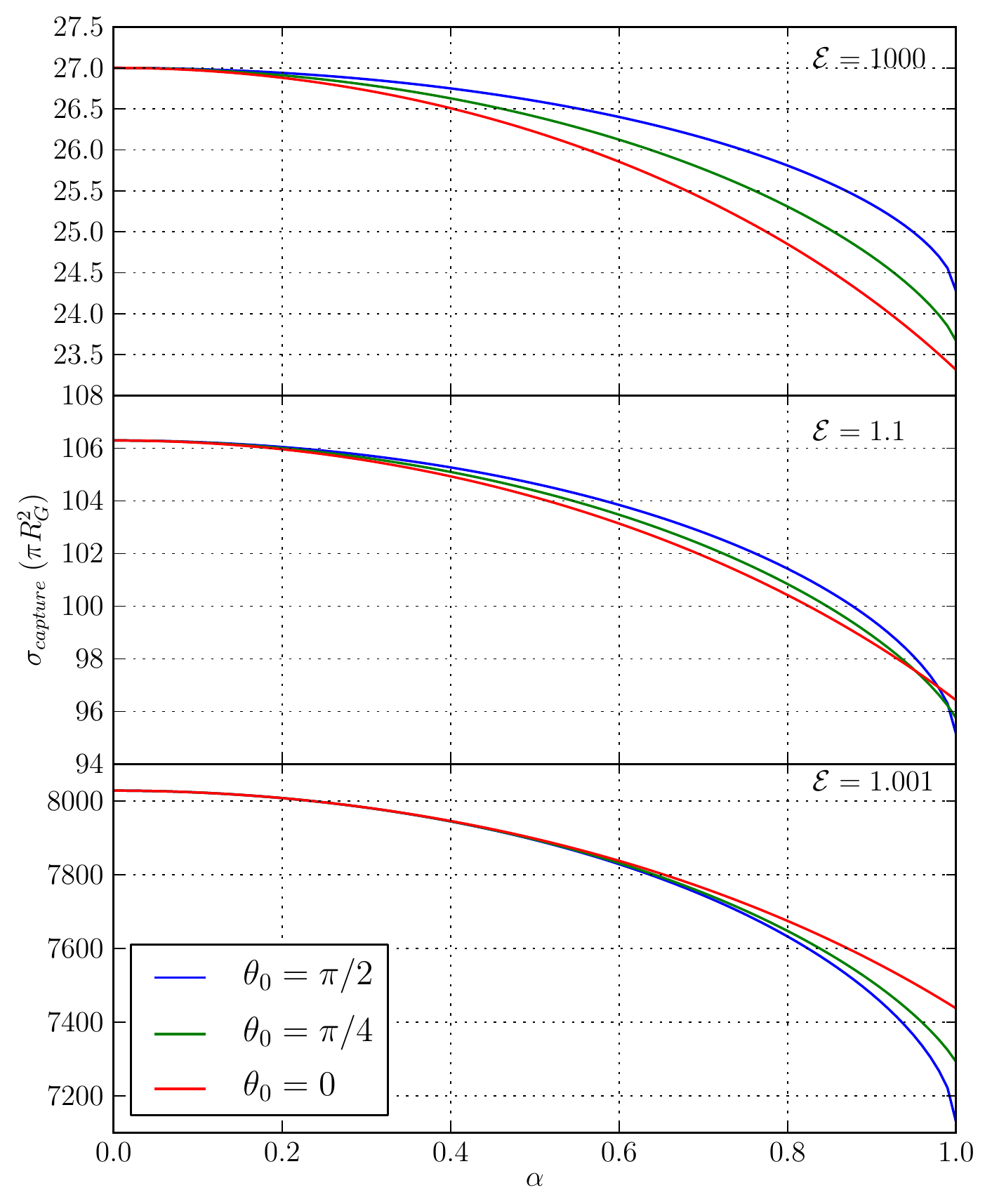}
\caption[Capture cross section of a Kerr black hole]{The total capturing area of a black hole always decreases with increasing spin magnitude and decreasing specific energy. It is also depends on the angle of approach $\theta_0$ to a certain extent.}
\label{CaptureArea}
\end{figure}
With these bounds for $r_p$, the capture region may be visualized by plotting $\left(b_x(r_p), b_y(r_p)\right)$ parametrically. The capture cross section $\sigma_{capture}$ is then the area of the resulting closed curve, which can be computed via numerical integration. 

\begin{figure}
\centering
\hspace{-70pt}\includegraphics[width=1.1\textwidth]{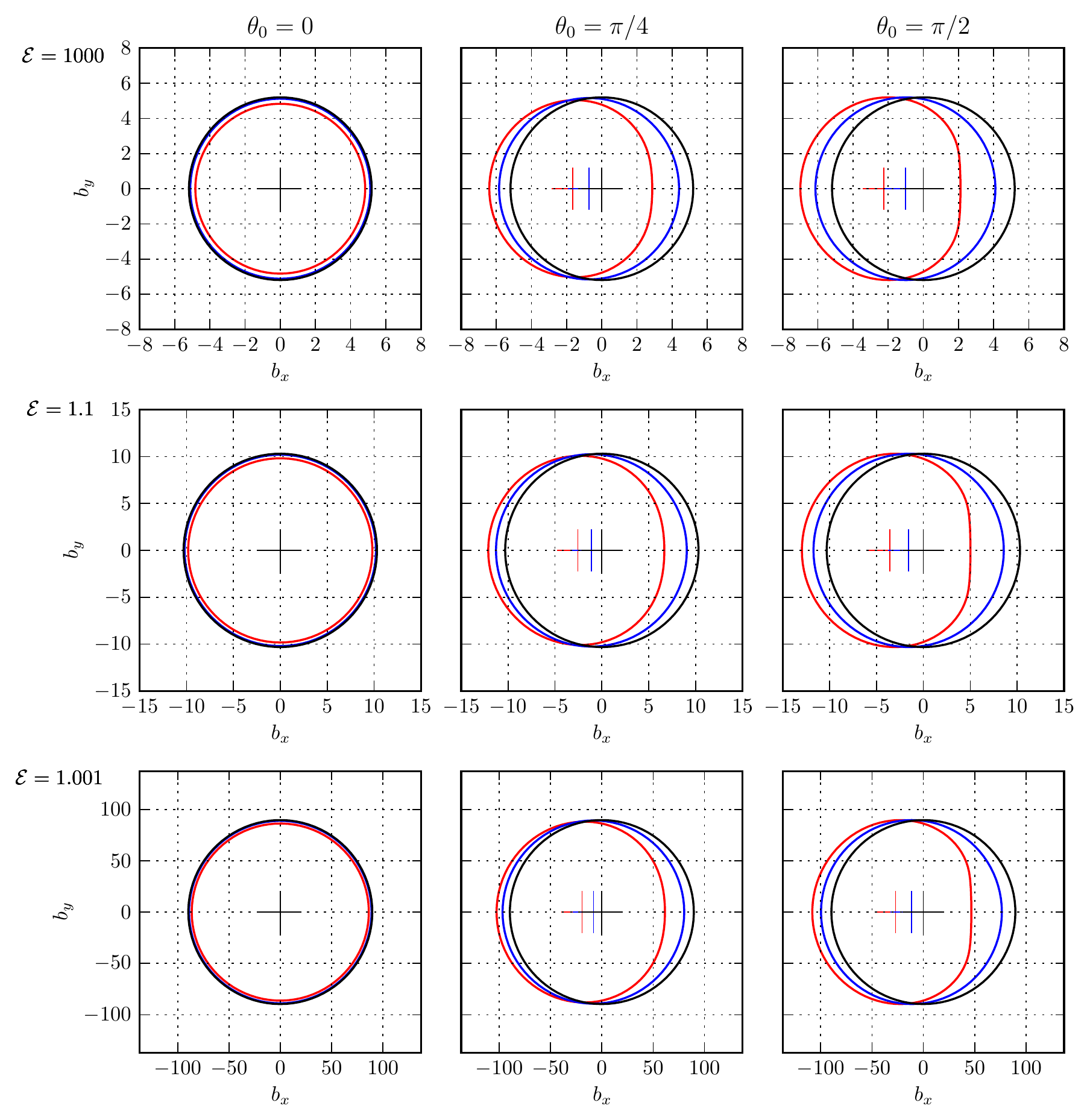}
\caption[Shape of the capturing region of a Kerr black hole]{The shape of the capture region in the $b$-plane for slow ($\mathcal{E}=1.001$), mildly relativistic ($\mathcal{E} = 1.1$) and ultrarelativistic ($\mathcal{E} = 1000$) initial velocities, and $\alpha = 0$ (black), $0.5$ (blue) and $0.998$ (red). For nonzero spin, the shape is skewed in the retrograde direction, with the displaced geometric centroids indicated. The shape of the region is independent of the approach velocity.}
\label{CaptureShape}
\end{figure}

As can be seen from figures \ref{CaptureArea} and \ref{CaptureShape}, the qualitative features of the capture regions predicted at the beginning of this section are correct: the capture cross section scales upward at lower specific energies, and is a topological disk in the plane. As $\alpha$ increases, the cross section always decreases in area, and unless one looks along the axis, it becomes skewed toward the retrograde side while the prograde side flattens out. Because the deflection of prograde orbits is reduced, the cross section for the capture of these orbits drops with increasing spin, and vice versa for retrograde orbits. Hence the net effect of spin on $\sigma_{capture}$ is not particularly drastic, with the increase in retrograde capture area largely making up for the decrease in the prograde.

\section{Calculation of Scattering Angles}

Once the parameter space of scattering orbits (those orbits not lying in the capture region) is known, the scattering angles may be calculated. All scattering orbits in black hole spacetimes share certain properties which are intuitive. Particles on these orbits all come in from an asymptotically straight trajectory, are deflected as they approach the gravitating body, reach the minimal periastron coordinate $r_p$, and proceed to escape to infinity on another asymptotically straight trajectory. These boundary conditions are used to determine the appropriate limits of integration.

\subsection{The Schwarzschild Metric}

The simplest subcase of the problem is evidently when $\alpha=0$. Due to the spherical symmetry of the spacetime, all geodesic motion is confined to a plane, so only motion in the plane $\theta = \pi/2$ ($\mu = 0$) need be considered. The motion of an inclined orbit simply follows from a coordinate transformation.

First, the radial coordinate of periastron $r_3$ is found by solving $\mathcal{R}(r)=0$ from the equation of motion. The solution set $\left\{0, r_1,r_2,r_3\right\}$ of this quartic equation, by Descartes' rule of signs, always has one negative and two positive roots when $h^2>h_{min}^2$; the periastron coordinate is the largest of the two positive roots.

The total deflection angle of the orbit is then obtained by substituting $\alpha=0$ in equation \ref{PhiDeflection}:
\begin{equation}
\phi - \phi_0 = \int^r \frac{h dr}{\sqrt{\mathcal{R}(r)}} \,.
\end{equation}
As the particle approaches, it gets closer to the black hole, so the sign of $\dot{r}$ is negative on the approach, and similarly positive on the escape. Thus, this expression is integrated in two parts:
\begin{equation}
\phi - \phi_0 = \int_\infty^{r_3} \frac{-hdr}{\sqrt{\mathcal{R}(r)}} +\int_{r_3}^{\infty} \frac{hdr}{\sqrt{\mathcal{R}(r)}} = 2 \int_{r_3}^{\infty} \frac{hdr}{\sqrt{\mathcal{R}(r)}} \,.
\end{equation}

The result is an elliptic integral of the first kind:
\begin{align}
\delta\phi &= \frac{4h}{\sqrt{(\mathcal{E}^2-1)(r_2(r_3-r_1))}}  F\left(\alpha,m\right) \, ,\\
\alpha &= \arcsin\left(\sqrt{\frac{r_2}{r_3}}\right) \nonumber \,,\\
m&=\frac{r_3(r_2-r_1)}{r_2(r_3-r_1)} \,. \nonumber
\end{align}

where the incomplete elliptic integral F is defined in the typical way:

\begin{equation}
F(\alpha, m) = \int_0^{\alpha} \frac{d\theta}{\sqrt{1-m \sin^2 \theta}} \,.
\label{EllipticF}
\end{equation}

When computing scattering angles numerically, it is better to express any elliptic integrals in terms of the symmetric Carlson integrals $R_F$ and $R_J$,\cite{NIST}, as efficient and robust methods are available to evaluate these functions \cite{CarlsonDuplication} \cite{NR}. Furthermore, these functions both have a homogeneity property allowing constants to be brought outside, allowing for expressions which are numerically better conditioned. Evaluating deflection angles analytically using these functions can be as much as $\mathcal{O}(10^2)$ times faster than integrating the geodesic equations with an ODE solver\cite{DexterAgol2009}. In terms of Carlson's elliptic integral of the first kind $R_F(x, y, z)$, equation \ref{EllipticF} takes a compact form:
\begin{align}
\phi - \phi_0 &= \frac{4h}{\sqrt{\mathcal{E}^2 -1}} I_r \label{schwdef} \,,\\
I_r &= R_F \left((r_3-r_1)(r_3-r_2), r_3(r_3-r_2), r_3(r_3-r_1)\right) \,.\nonumber
\end{align}
To determine the asymptotic coordinates of orbits not lying in the equatorial plane, one can use the above formulas with $b = \sqrt{b_x^2 + b_y^2}$ to determine the azimuthal angle $\phi'$ within the ``tilted'' coordinates of the orbital plane, and then ``untilt'' these coordinates to obtain the coordinates in the preferred coordinate system by composing two rotations.

\subsection{Equatorial Orbits}
The next important sub-case of the geodesic scattering problem is the deflection of orbits lying in the equatorial plane around a black hole of arbitrary spin. Indeed, to capture the essence of the effect of spin on the scattering angle, it suffices to consider just this case. Because $\theta$ (and hence $\mu$) is constant, the first integral term in equation \ref{PhiDeflection} might appear to vanish. However, recalling that for equatorial orbits $\mathcal{Q} = 0$, the denominator of the integral can be seen to be 0, hence the integral is undefined in this form. It is therefore necessary to substitute the identity of equation \ref{ThetaDeflection} and insert $\mu = 0$ to obtain an integral entirely in $r$:
\begin{equation}
\phi - \phi_0 = 2\int_{r_p}^\infty \frac{h - \alpha \mathcal{E}}{\sqrt{\mathcal{R}(r)}}dr -2\alpha \int_{r_p}^\infty \frac{\mathcal{E}(r^2+\alpha^2) - \alpha h}{(r^2 - 2r + \alpha^2)\sqrt{\mathcal{R}(r)}}dr \,.
\label{Equatorial}
\end{equation}

The constant term of $\mathcal{R}(r)$ vanishes when $\mathcal{Q} = 0$, so it can be factored like so, with $r_1$, $r_2$, and $r_3$ ordered from least to greatest:
\begin{equation}
\mathcal{R}(r) = (\mathcal{E}^2 - 1)r(r-r_1)(r-r_2)(r-r_3) \,.
\end{equation}
The first integral is then of the same kind evaluated in the non-spinning case:
\begin{align}
I_1 &:= 2\int_{r_3}^\infty \frac{h - \alpha \mathcal{E}}{\sqrt{\mathcal{R}(r)}}dr = \frac{4(h-\alpha \mathcal{E})}{\sqrt{\mathcal{E}^2 -1}} I_r\\
I_r &= R_F(X, Y, Z) \nonumber \,\\
X &= r_3 (r_3 - r_1) \nonumber\\
Y &= (r_3 - r_1)(r_3-r_2) \nonumber \,\\
Z &= r_3(r_3 - r_2) \,.\nonumber
\end{align}

The second integral can be reduced to standard elliptic form through partial fraction decomposition:
\begin{align}
I_2 &:= 2\alpha \int_{r_3}^\infty \frac{\mathcal{E}(r^2+\alpha^2) - \alpha h}{(r^2 - 2r + \alpha^2)\sqrt{\mathcal{R}(r)}}dr \nonumber\\
&= 2\alpha \mathcal{E}\int_{r_3}^\infty dr \left(\frac{1}{\sqrt{\mathcal{R}(r)}} + \frac{\alpha^2 - \alpha h/\mathcal{E} + r_{+}^2}{r_{+}-r_{-}} \frac{1}{(r-r_{+})\sqrt{\mathcal{R}(r)}} \right.\nonumber \\
&\left. -\frac{\alpha^2 - \alpha h/\mathcal{E} + r_{-}^2}{r_{+}-r_{-}}  \frac{1}{(r-r_{-})\sqrt{\mathcal{R}(r)}}\right) \,.
\end{align}

The first term is again an elliptic integral of the form encountered previously. The other two integrals are the most terrible integrals to be encountered in this work, and are hence denoted $T_+$ and $T_-$. They may be put in terms of Carlson's elliptic integral $R_J$\cite{CarlsonThird}:
\begin{align}
T_\pm &= \int_{r_3}^\infty \frac{dr}{(r-r_\pm)\sqrt{r(r-r_1)(r-r_2)(r-r_3)}} \nonumber\\
&= \frac{S - 2I_r}{r_\pm - r_1} - \frac{2 r_1 (r_3 - r_1)(r_2 - r_1)}{3(r_\pm - r_1)^2} R_J(X,Y,Z, W^2) \,.
\end{align}
where:
\begin{align}
P^2 &= (r_3 - r_\pm)^2 \nonumber \,,\\
Q^2 &= \frac{\left(r_\pm-r_3\right) \left(r_1 r_3 - r_\pm \left(r_1-r_2+r_3\right)\right)}{r_\pm-r_1} \,,\nonumber\\
W^2 &= \frac{\left(r_1-r_3\right) \left(r_1 r_3-r_\pm\left(r_1-r_2+r_3\right)\right)}{r_\pm-r_1} \,, \nonumber\\
\text{and} \, S &= \sqrt{\frac{r_\pm - r_1}{r_\pm(r_2 - r_\pm)(r_3-r_\pm)}}\cosh^{-1}\left(\sqrt{\frac{(r_\pm - r_1)(r_3 - r_\pm)}{r_1 r_3 + r_\pm(r_1 - r_2 + r_3)}}\right) \,.
\end{align}
Given all of the necessary integrals, the total deflection is:
\begin{equation}
\phi - \phi_0 = \frac{4h}{\sqrt{\mathcal{E}^2 -1}} I_r + \frac{2 \alpha \mathcal{E}}{\sqrt{\mathcal{E}^2 - 1}} \left(\frac{\alpha^2 - \alpha h/\mathcal{E} + r_{+}^2}{r_{+}-r_{-}} T_+ -\frac{\alpha^2 - \alpha h/\mathcal{E} + r_{-}^2}{r_{+}-r_{-}} T_-\right) \,.
\label{eqdef}
\end{equation}
\begin{figure}
\centering
\includegraphics{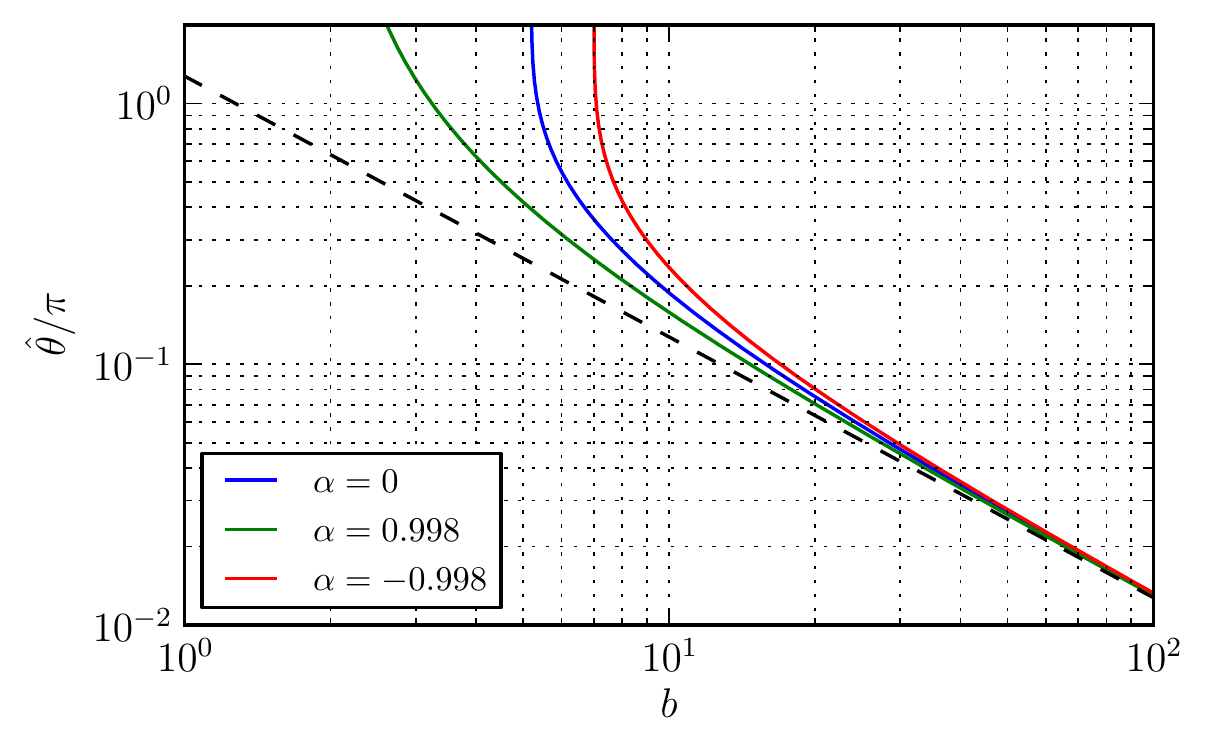}
\caption[Deflection angle for equatorial massless particles]{Deflection angle of a massless particle in the equatorial plane of a Kerr black hole. If the particle travels prograde to the spin, it is deflected less, while if it travels retrograde it is deflected more. As $b\rightarrow \infty$, all deflection angles approach the first order asymptotic expansion (dashed line).}
\end{figure}

\subsection{Arbitrary Kerr Orbits}

The algorithm for calculating scattering angles for arbitrary $\mathcal{E}$, $\alpha$ and $\theta_0$ is as follows:
\begin{enumerate}
\item Determine the roots of $\mathcal{R}(r)$
\item Using these roots, calculate $\int \frac{dr}{\mathcal{R}(r)}$
\item Use equation \ref{ThetaDeflection} to invert the elliptic $\mu$ integral with a Jacobi elliptic function to determine the final coordinate $\mu_f$
\item With the $\mu$ limits of integration determined, calculate the integral over $\mu$ of equation \ref{PhiDeflection}
\item Calculate the integral in $r$ in equation \ref{PhiDeflection} the same way as the equatorial case and add to the $\mu$ integral to obtain $\phi - \phi_0$
\end{enumerate}
\subsubsection{Calculation of $\mu$}
The calculation of the roots of $\mathcal{R}(r)$ proceeds the same way as before, however in the general case where $\mathcal{Q} \neq 0$, $\mathcal{R}(r) = 0$ becomes a quartic equation. Assuming the impact parameters lie outside the capture region, $\mathcal{R}(r)$ will have 4 real roots satisfying $r_1 < r_2 < 1 + \sqrt{1-\alpha^2} < r_3  < r_4$. The periastron coordinate $r_p$ is therefore once again the largest root of $\mathcal{R}(r)$, and the polynomial can once again be factored and reduced to a Carlson integral:
\begin{align}
\int \frac{dr}{\mathcal{R}(r)} &= 2 \int_{r_4}^\infty \frac{dr}{\sqrt{\mathcal{R}(r)}} = \frac{4}{\mathcal{E}^2 - 1} R_F\left(X, Y, Z\right) \,, \\
\text{where} \,\,X &= (r_4 - r_1)(r_4 - r_2)\nonumber \,,\\
Y &= (r_4 - r_1)(r_4 - r_3) \nonumber \,,\\
\text{and}\,\,\, Z &= (r_4 - r_3)(r_4 - r_2) \nonumber \,.
\end{align}

Now it is possible to use equation \ref{ThetaDeflection} to find $\mu_f$. $\mathcal{M}(\mu)$ can be factored into a product of quadratics:
\begin{align}
\mathcal{M}(\mu) &= \alpha^2(\mathcal{E}^2 -1)(\mu^2 - M_1)(M_2 - \mu^2) \,.
\end{align}
For scattering orbits, $M_1$ is always negative while $M_2$ lies on the interval $\left[0, 1\right]$, and the turning points in the $\mu$ motion are hence the real roots of $\mathcal{M}$, $\mu_\pm = \pm \sqrt{M_2}$.

The $\mu$ integral on the left side of equation \ref{ThetaDeflection} is less straightforward because the number of turning points in the $\mu$ coordinate is may take any value. As such, the integral is the sum of parts from an initial approach, an intermediate part during which the body oscillates between turning points, and final escape:
\begin{equation}
\int \frac{d\mu}{\mathcal{M}(\mu)} = \int_{\mu_0}^{\mu_\pm} + N \int_{\mu_-}^{\mu_+} + \int_{\mu_\pm}^{\mu_f} \equiv \frac{1}{|\alpha|\sqrt{\mathcal{E}-1}} \left(I_{\mu_i} + N I_{\mu_c} + I_{\mu_f}\right) \,.
\end{equation}
The limits of the initial and intermediate parts are known, so the integrals may be obtained from the following formula \cite{NIST}:
\begin{align}
\int_{|x|}^{\mu_+} \frac{d\mu}{\mathcal{M}(\mu)} &= \frac{1}{|\alpha|\sqrt{\mathcal{E}^2 - 1}} \int_{|x|}^{\mu_+} \frac{\mu}{\sqrt{(\mu^2 - M_1)(M_2 - \mu^2)}} \nonumber\\
&= \frac{1}{|\alpha|\sqrt{\mathcal{E}^2 - 1}}\dfrac{F\left(\arcsin\left(\dfrac{|x|}{\mu_+}\right), \dfrac{M_2}{M_1}\right)}{\sqrt{-M_1}}\,.
\end{align}
This implies:
\begin{align}
I_{\mu_i} &= \dfrac{K\left(\frac{M_2}{M_1}\right) +  \mathrm{sign}(b_y) F\left(\arcsin\left(\dfrac{|\mu_0|}{\sqrt{M_2}}\right), \dfrac{M_2}{M_1}\right)}{\sqrt{-M_1}} \,\\
\text{and}\,\, I_{\mu_c} &= \dfrac{2K\left(\dfrac{M_2}{M_1}\right)}{\sqrt{-M_1}}\,.
\end{align}
Because $I_{\mu_f} < I_{\mu_c}$, we have:
\begin{align}
N &= \lfloor \frac{I_r - I_{\mu_i}}{I_{\mu_c}} \rfloor \,,\\
I_{\mu_f} &= I_r - N I_{\mu_c} - I_{\mu_i} \,.
\end{align}
Finally, the elliptic integral $I_{\mu_f}$ can be inverted with the Jacobi elliptic function $\mathrm{cn}(x, m)$ to compute $\mu_f$:
\begin{equation}
\mu_f = \sqrt{M_2}\,\mathrm{cn}\left(\sqrt{M_2-M_1}\, I_{\mu_f}\,,\frac{M_2}{M_2 - M_1}\right) \,.
\end{equation}

\subsubsection{Calculation of $\phi$}
It remains to compute the integral in equation \ref{PhiDeflection}. The integral over $r$ is computed exactly the same as the equatorial case, except using the values of $X$, $Y$ and $Z$ which were used to calculate $I_r$. The integral over $\mu$ is similar to the previous one in that it consists of initial, intermediate and final parts (along with $I_r$, which we have already computed):
\begin{align}
\int ^\mu \frac{h/(\mu^2 - 1) - \alpha \mathcal{E}}{\sqrt{\mathcal{M}(\mu)}} d\mu =& \int^\mu \frac{h d\mu}{(\mu^2 - 1)\sqrt{\mathcal{M}(\mu)}} - \alpha \mathcal{E} \int^\mu \frac{d\mu}{\mathcal{M}(\mu)} \nonumber \\
=& \frac{h}{|\alpha|\left(1 - M_2 \right)\sqrt{\left(M_2 - M_1\right) \left(\mathcal{E}^2 - 1 \right)}} \left(\Phi_{\mu_i} + N \Phi_{\mu_c} + \Phi_{\mu_f}\right) \nonumber\\
&- \frac{\alpha \mathcal{E}}{\sqrt{\mathcal{E}^2 - 1}} I_r \,.
\end{align}
The $\Phi$ integrals can be expressed in terms of the Legendre elliptic integral of the third kind $\Pi(\varphi, n, k)$, again with one complete and two incomplete integrals:
\begin{align}
\Phi_{\mu_c} &= 2 \Pi\left(n, k \right) \\
\Phi_{\mu_i} &= \left\{
     \begin{array}{lr}
       \Pi\left(\varphi_i, n, k\right) & \quad\text{for}\quad  \mu_0 \dot{\mu}_0 > 0\\
       \Phi_{\mu_c} - \Pi\left(\varphi_i, n, k\right) & \quad\text{for}\quad  \mu_0 \dot{\mu}_0 < 0 \\
     \end{array}
   \right.\\
\Phi_{\mu_f} &= \left\{
     \begin{array}{lr}
       \Pi\left(\varphi_f, n, k\right) & \quad\text{for}\quad I_{\mu_f} < \frac{1}{2}I_{\mu_c}\\
       \Phi_{\mu_c} - \Pi\left(\varphi_f, n, k\right) & \quad\text{for}\quad  I_{\mu_f} > \frac{1}{2}I_{\mu_c} \\
     \end{array}
   \right. \\
\varphi_i &= \arccos\left(\frac{|\mu_0|}{\mu_+}\right) \\
\varphi_f &= \arccos\left(\frac{|\mu_f|}{\mu_+}\right) \\
n &= \frac{-M_2}{1-M_2} \\
k^2 &= \frac{M_2}{M_2 - M_1}
\end{align}

With these integrals computed, the final answer is:
\begin{align}
\phi - \phi_0 &= \frac{h}{|\alpha|\left(1 - M_2 \right)\sqrt{\left(M_2 - M_1\right) \left(\mathcal{E}^2 - 1 \right)}} \left(\Phi_{\mu_i} + N \Phi_{\mu_c} + \Phi_{\mu_f}\right)\nonumber \\
&+ \frac{2 \alpha \mathcal{E}}{\sqrt{\mathcal{E}^2 - 1}} \left(\frac{\alpha^2 - \alpha h/\mathcal{E} + r_{+}^2}{r_{+}-r_{-}} T_+ -\frac{\alpha^2 - \alpha h/\mathcal{E} + r_{-}^2}{r_{+}-r_{-}} T_-\right)\,.
\label{GenDef}
\end{align}

\subsection{Weak Deflection Limit}

The solutions so far derived, while precise, are not particularly illuminating: one cannot easily discern the individual effects that different parameters may have on the deflection angle simply by looking at equations \ref{schwdef}, \ref{eqdef} or \ref{GenDef}. It is therefore desirable to seek expansions in the limit of large $b$, in the spirit of Einstein's famous weak deflection formula for a light ray in the Schwarzschild metric:
\begin{equation}
\hat{\theta} = \frac{4}{b} + \mathcal{O}\left(\frac{1}{b^2}\right) \,.
\end{equation}
With an analytic solution for timelike geodesics of arbitrary energy, it is possible to work backwards to obtain higher order asymptotic expansions of $\hat{\theta}$ in powers of $\displaystyle\frac{1}{b}$. Computationally, such an expansion has the advantage of requiring only a single polynomial evaluation, making it much less expensive than evaluating one or several elliptic integrals. They are also much simpler to work with analytically. The drawback is of course that the expansion fails entirely to capture the divergence of the deflection angle for orbits near the capture region: each additional order in $\displaystyle\frac{1}{b}$ merely edges the domain of usefulness closer to $b_{min}$ with diminishing returns. Starting first with the Schwarzschild solution, the series expansion of equation \ref{schwdef} is:
\begin{align}
\hat{\theta} = &\left(4+\frac{2}{\epsilon^2}\right) \frac{1}{b} + \frac{3}{4} \pi  \left(\frac{4}{\epsilon ^2}+5\right) \frac{1}{b^2} + \left(-\frac{2}{3 \epsilon^6}+\frac{8}{\epsilon ^4}+\frac{48}{\epsilon ^2}+\frac{128}{3}\right) \frac{1}{b^3}\nonumber\\ &+ \frac{105 \pi  \left(33 \epsilon ^4+48 \epsilon ^2+16\right)}{64 \epsilon ^4} \frac{1}{b^4} + \left(\frac{2}{5 \epsilon ^{10}}-\frac{4}{\epsilon ^8}+\frac{64}{\epsilon ^6}+\frac{640}{\epsilon ^4}+\frac{1280}{\epsilon^2}+\frac{3584}{5}\right) \frac{1}{b^5}\nonumber\\ &+ \frac{1155 \pi  \left(221 \epsilon ^6+468 \epsilon ^4+312 \epsilon ^2+64\right)}{256 \epsilon ^6} \frac{1}{b^6} + \mathcal{O}\left(\frac{1}{b^7}\right) \,.
\label{SchwWeakDef}
\end{align}
Here $\epsilon = \sqrt{\mathcal{E}^2-1} =\displaystyle \frac{v_0^2}{1-v_0^2}$. This can be readily compared with the Newtonian solution:
\begin{align}
\hat{\theta}_{Newtonian} &= \pi - 2 \arctan\left(b v_0^2\right) \nonumber\\
&= \frac{2}{b v_0 ^2}-\frac{2}{3 b^3 v_0 ^6}+\frac{2}{5 b^5 v_0^{10}}+\mathcal{O}\left(\frac{1}{b^7}\right)\,.
\end{align}
The Newtonian solution consists of the terms which dominate in the limit $v_0 \rightarrow 0$, corresponding to slow motion compared to $c$. In the ultrarelativistic limit $v_0 \rightarrow \infty$ the Newtonian terms vanish and only the terms which are constant with respect to $v_0$ remain.

The weak deflection expansion including spin terms for null geodesics was derived by Sereno and de Luca for general orbits\cite{KerrWeakDef}. Here we consider just the equatorial case, which is sufficient to isolate the effect of the spin. The series expansion of \ref{eqdef} is:
\begin{align}
	\hat{\theta} =& \frac{4}{b} + \frac{\frac{15 \pi }{4}-4a}{b^2} + \frac{4 a^2-10 \pi  a+\frac{128}{3}}{b^3} + \frac{\frac{15}{64} \pi  \left(76 a^2+231\right)-4 a\left(a^2+48\right)}{b^4} \\
	&+ \frac{4 \left(a^2+128\right) a^2-\frac{9}{2} \pi\left(6 a^2+77\right)a +\frac{3584}{5}}{b^5} + \mathcal{O}\left(\frac{1}{b^6}\right)\,. \nonumber
\end{align}
The spin dependence enters at next-to-leading order. This is intuitive from the gravitomagnetic analogy: at large distances, the effect of the dipole field due to the spin should fall off an order faster than that of the mass monopole. The effect of spin on the deflection angle is easily recognized: orbits prograde to the spin are deflected less, while those retrograde to the spin are deflected more. There are also terms which increase the deflection of both types of orbits equally, due to the black hole's higher mass multipole moments.

\input{DiffSigma}

%% file: DiffSigma.tex
\section{Differential Cross Section}

\subsection{Schwarzschild and Polar Orbits}
It is easiest to obtain the differential scattering cross section in the cases where the target is symmetric about the axis of approach; this is the case both when there is no spin and when we are firing particles down the axis of symmetry of a spinning black hole. In this case, the Jacobian of the scattering angle function $\phi$ depends only on $b$. The formula for the scattering cross section reduces to:
\begin{equation}
\frac{d\sigma}{d\Omega} = \sum_{n=1}^\infty \left|\frac{b_n(\hat{\theta})}{\sin \hat{\theta}} \frac{db_n}{d\hat{\theta}}\right|\,.
\end{equation}

As $\hat{\theta}(b)$ is a very complicated algebraic function of $b$ involving nested radicals and an elliptic integral, it is generally not possible to invert it to obtain $b(\hat{\theta})$ to calculate $\displaystyle\frac{d\sigma}{d\Omega}$ directly. To obtain the differential cross section as a function of $\hat{\theta}$ therefore requires a numerical approach or an approximation. An algorithm to calculate $\displaystyle\frac{d\sigma}{d\Omega}$ for a given value of $\mathcal{E}$ is as follows:
\begin{enumerate}
\item Choose a discrete range of impact parameters $b_n$ ranging from $b_{min}$ to some $b_{max}$ large enough that the regime $\hat{\theta} \approx 0$ is reached. Also choose a discrete range of deflection angles $\hat{\theta}_m$ ranging from $0$ to $\pi$, with enough points to achieve the desired resolution. The precision of the subsequent numerical differentiation improves with greater $b$-resolution.
\item Calculate $\hat{\theta}(b_n)$.
\item Approximate $\frac{db}{d\hat{\theta}}$ using the centered difference formula:
\begin{equation}
\left(\frac{db}{d\hat{\theta}}\right)_n = \frac{b_{n+1} - b_{n-1}}{\hat{\theta}(b_{n+1}) - \hat{\theta}(b_{n-1})} + \mathcal{O}(\Delta \hat{\theta} ^2)\,.
\end{equation}
\item Calculate $\frac{d\sigma}{d\Omega}(\hat{\theta}(b_n))$.
\item Partition the data into branches of index $k$ based on the value of $\hat{\theta}(b_n)$.
\item Interpolate the values of the individual branches to the $\hat{\theta}_m$ grid to approximate $\frac{d\sigma_k}{d\Omega}(\hat{\theta}_m)$.
\item Sum the values of each branch to approximate $\frac{d\sigma}{d\Omega}(\hat{\theta}_m)$
\end{enumerate}

If the total scattering cross section $\sigma$ is desired, $\displaystyle\frac{d\sigma}{d\Omega}(\hat{\theta}_m)$ can be integrated over the sphere using a numerical quadrature rule.

\begin{figure}[h!]
\centering
\includegraphics{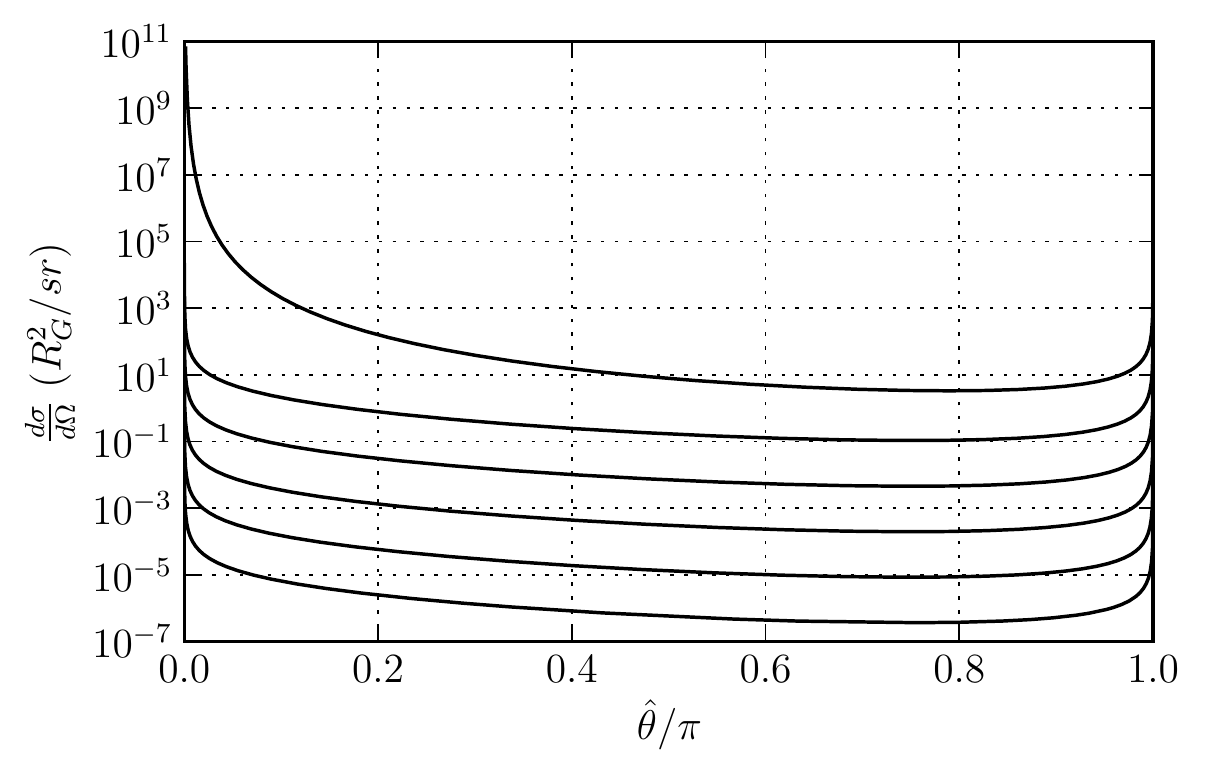}
\caption[Differential cross section for a photon in the Schwarzschild spacetime]{The first 5 branches of $d\sigma/d\Omega$ for a massless or ultrarelativistic particle orbiting a Schwarzschild black hole. The branches decay in magnitude approximately exponentially, so the curve representing the sum of all branches would be visually indistinguishable from that of the first branch on this graph.}
\label{PhotonBranches}
\end{figure}
\begin{figure}
\centering
\includegraphics{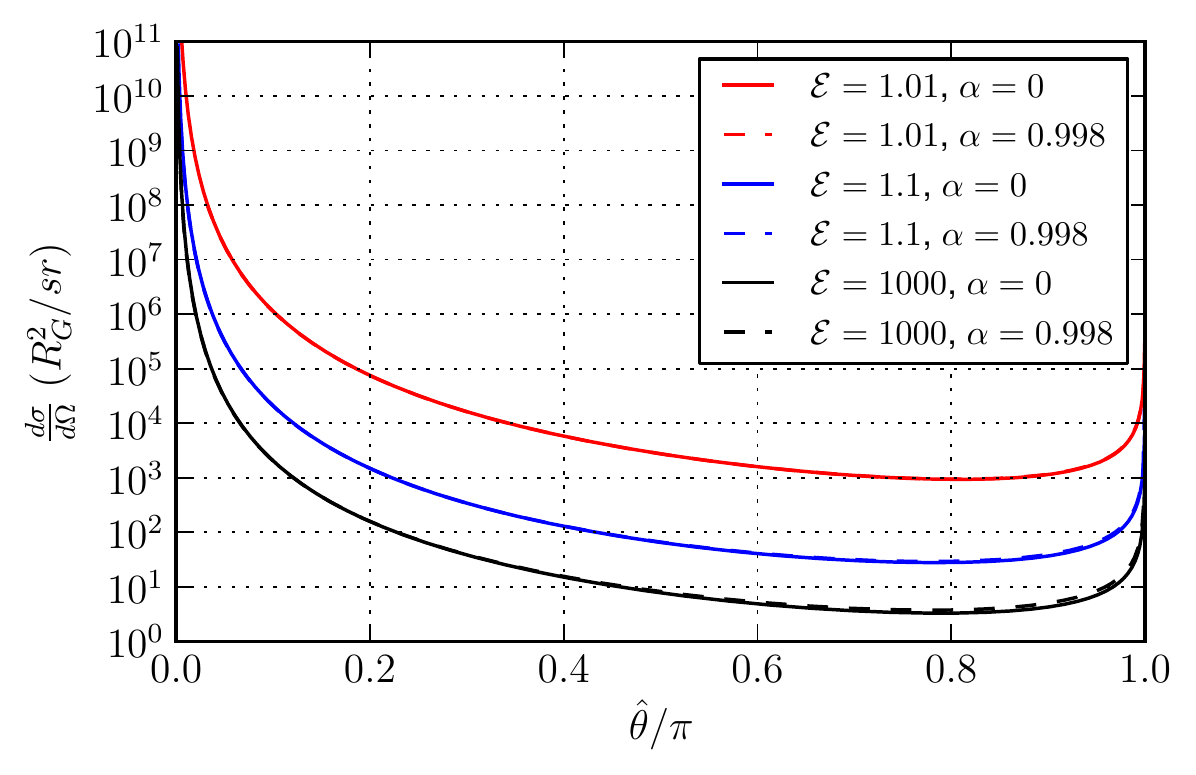}
\caption[Dependence of $\displaystyle\frac{d\sigma}{d\Omega}$ on spin and energy for polar orbits]{For particles fired along the spin axis of the black hole, the differential cross section is weakly dependent on spin, with the relative dependence the most pronounced for massless particles. We also see that as $E \rightarrow mc^2$ the differential cross section diverges in a way much like the capture cross section.}
\label{PolarVsSchw}
\end{figure}
Figures \ref{PhotonBranches} and \ref{PolarVsSchw} demonstrate properties common to the scattering cross section at all energies. Firstly, the contributions of branches of higher order than the first are much smaller than that of the first. This is because only orbits which get within a few gravitational radii orbit around the black hole before escaping, so the cross section to be on such an orbit is relatively small.

The other salient feature in the differential cross section is that it diverges at $\hat{\theta} = \pi$. This scattering phenomenon is known as a ``glory'' \cite{FordWheeler1959}, and in the case of light scattering is a type of optical caustic. It occurs because all trajectories through the circles of radius $b_n(\pi)$ in the $b$-plane are deflected to the same angle, \emph{ie}. directly backwards relative to the approach vector. It is a feature common to any scattering phenomenon for which deflection angles greater than $\pi$ are possible and the cross section is solely a function of $\hat{\theta}$. When this symmetry is broken by spin, we shall see that caustic-like structures remain, but spread out from the single point $\hat{\theta}=\pi$ into a series of 2D curves on the sphere.

\subsection{Weak Deflection Limit}

It is also possible to derive fully analytic approximations to $\displaystyle\frac{d\sigma}{d\Omega}$ from the weak deflection formula \ref{SchwWeakDef} in the previous section. While it would be very challenging to invert $\hat{\theta}(b)$ for the exact answer, to do it for the asymptotic expansion is a trivial application of series reversion. The differential scattering cross section for a Schwarzschild black hole is:
\begin{align}
\displaystyle\frac{d\sigma}{d\Omega} \sin\hat{\theta} =& \displaystyle\frac{4 \left(1-2 \mathcal{E}^2\right)^2}{\left(\mathcal{E}^2-1\right)^2 \hat{\theta} ^3}-\displaystyle\frac{3 \pi  \left(1-5 \mathcal{E}^2\right)}{4 \left(\mathcal{E}^2-1\right) \hat{\theta}^2} \nonumber\\
&-\displaystyle\frac{1}{1024 \left(\mathcal{E}^2-1\right)\left(1-2 \mathcal{E}^2\right)^4}\left\lbrace 27 \pi ^3 \left(\mathcal{E}^2-1\right)^2 \left(5 \mathcal{E}^2-1\right)^3 \right.  \nonumber \\
&\left.-4 \pi  \left(2 \mathcal{E}^2-1\right) \left(3310 \mathcal{E}^8-7073 \mathcal{E}^6+4095 \mathcal{E}^4-515\mathcal{E}^2+55\right)\right\rbrace \nonumber\\
&+ \mathcal{O}\left(\hat{\theta}\right) \,.
\end{align}
or more succinctly in the the ultrarelativistic limit $\mathcal{E} \rightarrow \infty$:
\begin{align}
\displaystyle\frac{d\sigma}{d\Omega} \sin\hat{\theta} =& \displaystyle\frac{16}{\hat{\theta} ^3}+\displaystyle\frac{15 \pi }{4\hat{\theta} ^2}-\displaystyle\frac{5 \pi  \left(675 \pi^2-5296\right)}{16384} \nonumber\\
&+\left(\displaystyle\frac{225 \pi ^2 \left(1125 \pi ^2-10592\right)}{1048576}-\displaystyle\frac{16}{15}\right) \hat{\theta} \nonumber\\
&-\displaystyle\frac{\pi  \left(208999424+16875 \pi ^2 \left(945 \pi ^2-10592\right)\right)}{67108864}\hat{\theta} ^2 + \mathcal{O}(\hat{\theta}^3)\,.
\end{align}
\begin{figure}
\centering
\includegraphics{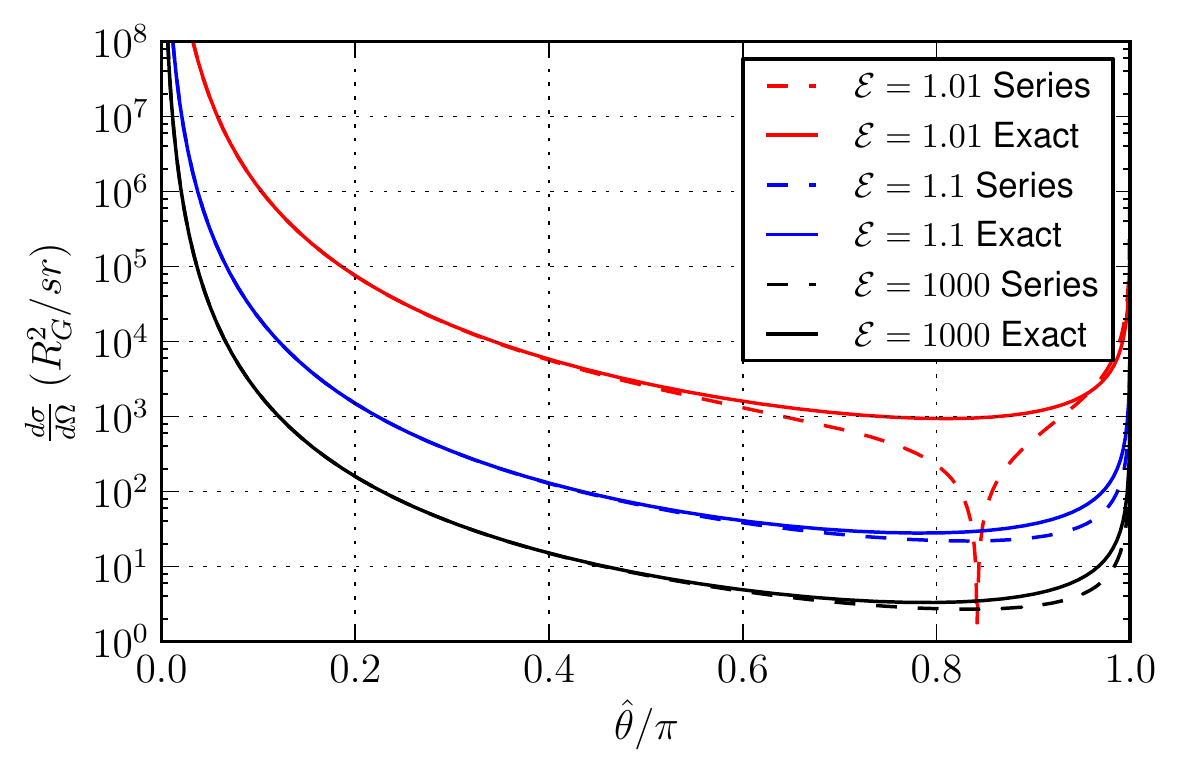}
\caption[Comparison of exact and series approximated differential cross section]{The series expansion about $\hat{\theta} = 0$ is accurate for weak deflection, with the approximation worsening for increasing $\hat{\theta}$ and decreasing $\mathcal{E}$.}
\label{ExactVsSeries}
\end{figure}
As figure \ref{ExactVsSeries} demonstrates, these expressions accurately reproduce the differential cross section in the regime of weak deflection.

\subsection{General Case}

If the black hole is spinning and the orbits do not approach along the axis of rotation, the differential cross section will depend on both $\hat{\theta}$ and $\hat{\phi}$, and requires that the full Jacobian determinant of the scattering function be computed. This can be done again using finite differencing, as was done by Bozza when investigating the optical caustics present in the differential cross section for light rays \cite{KerrCaustics}. A simpler method than 2D finite differencing, which is able to visualize the entire cross section with high resolution, is to run a virtual Monte Carlo scattering experiment. The idea is to ``fire'' particles from random, uniformly distributed locations in a subset of the $b$-plane, most easily a disk. This simulates a scattering experiment in which a circular beam of particles of uniform number flux density is fired at the black hole. If the resulting deflection angles are then binned into a histogram on $\left[0,\pi\right]\times\left[0,2\pi\right]$, the histogram will converge to a distribution proportional to $\displaystyle\frac{d\sigma}{d\Omega}$ as the number of particles approaches infinity. The obvious drawback of this method is that it requires a very large number of deflection angles to be computed to achieve decent resolution, however the speedup from using elliptic integrals instead of an ODE solver makes this feasible.

\begin{figure}
\centering
\includegraphics{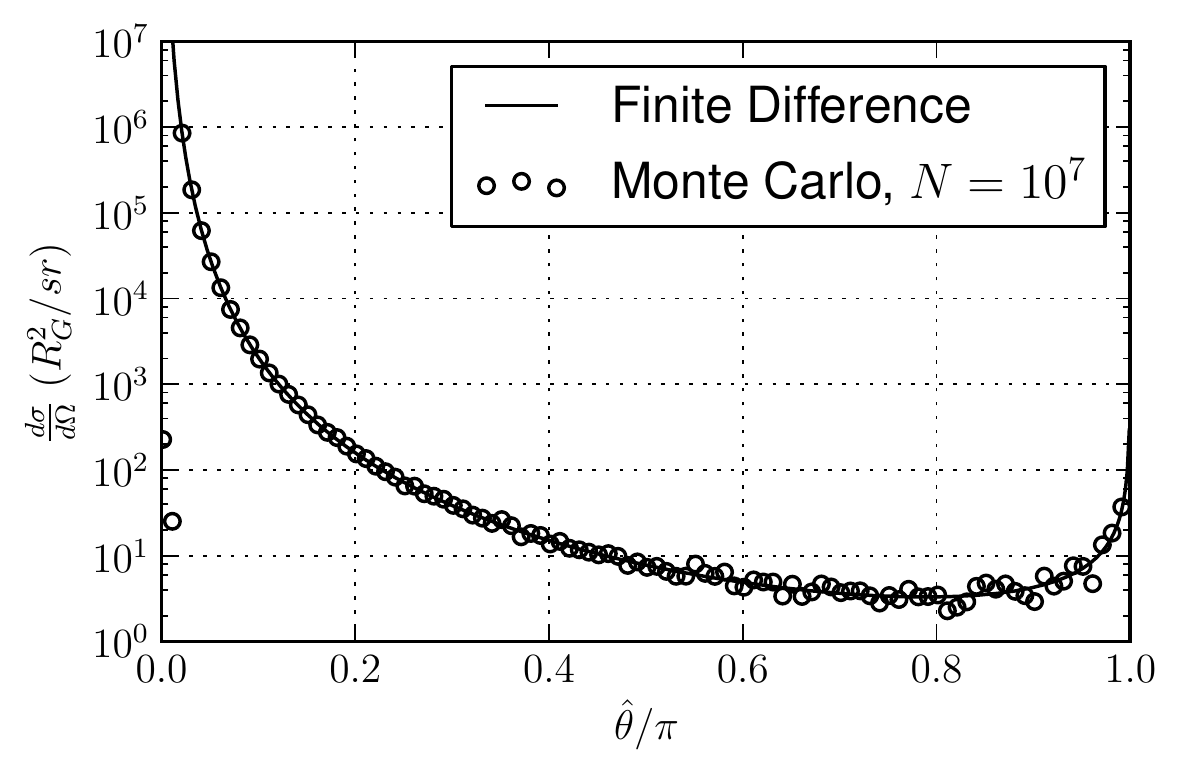}
\caption[Comparison of Monte Carlo and finite difference method for visualizing $\displaystyle\frac{d\sigma}{d\Omega}$]{Comparison between estimates of $\displaystyle\frac{d\sigma}{d\Omega}$ for massless particles using the finite difference and Monte Carlo methods. The Monte Carlo method converges to the exact answer as more and more particles are fired, eventually reaching a point where it is adequate for visualization purposes. It drops near $\hat{\theta}=0$ because the ``beam'' of particles has a finite radius. The accuracy deteriorates here at higher deflection angles because the regime where the cross section is small receives a proportionally small number of particles.}
\end{figure}

We partition the sphere of deflection angles into an $m \times n$ grid with cell boundaries $\hat{\theta}_i$ and $\hat{\phi}_j$ and count the number of particles $n_{ij}$ that land on the coordinate patch $[\hat{\theta}_i,\hat{\theta}_{i+1}] \times [\hat{\phi}_j, \hat{\phi}_{j+1}]$. Given the total cross-sectional area $A$ of the particle beam, and the total number of particles $N$, we can associate with cell $(i,j)$ a cross sectional area $\Delta \sigma_{ij} = \displaystyle\frac{A}{N} n_{ij}$. The solid angle $\Delta\Omega_{ij}$ associated with the cell is approximately $\Delta \hat{\theta} \Delta \hat{\phi} \sin \hat{\theta}$. Hence:
\begin{equation}
\left(\displaystyle\frac{d\sigma}{d\Omega}\right)_{ij} \approx \displaystyle\frac{\Delta \sigma_{ij}}{\Delta \Omega_{ij}} = \displaystyle\frac{A n_{ij}}{N \Delta \hat{\theta}\Delta\hat{\phi} \sin \hat{\theta}_i} = \displaystyle\frac{A  n m}{2 \pi^2 N  \sin \hat{\theta}_i} n_{ij}\,.
\end{equation}

\begin{figure}
\centering
\includegraphics[width=\textwidth]{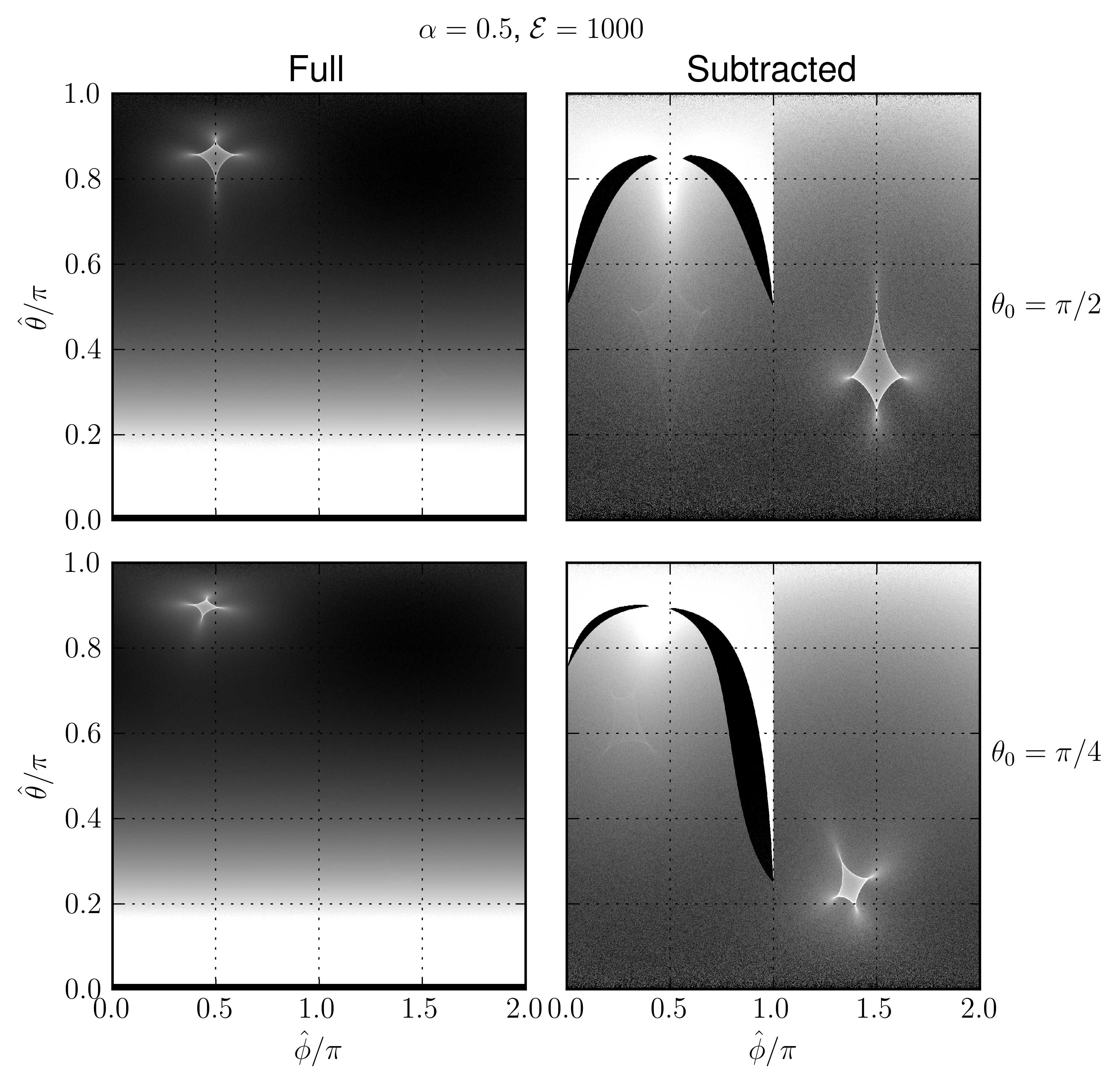}
\caption[Differential scattering cross section of an ultrarelativistic particle around a black hole with moderate spin]{Differential scattering cross section of an ultrarelativistic particle around a black hole with moderate spin.}
\label{first}
\end{figure}

\begin{figure}
\centering
\includegraphics[width=\textwidth]{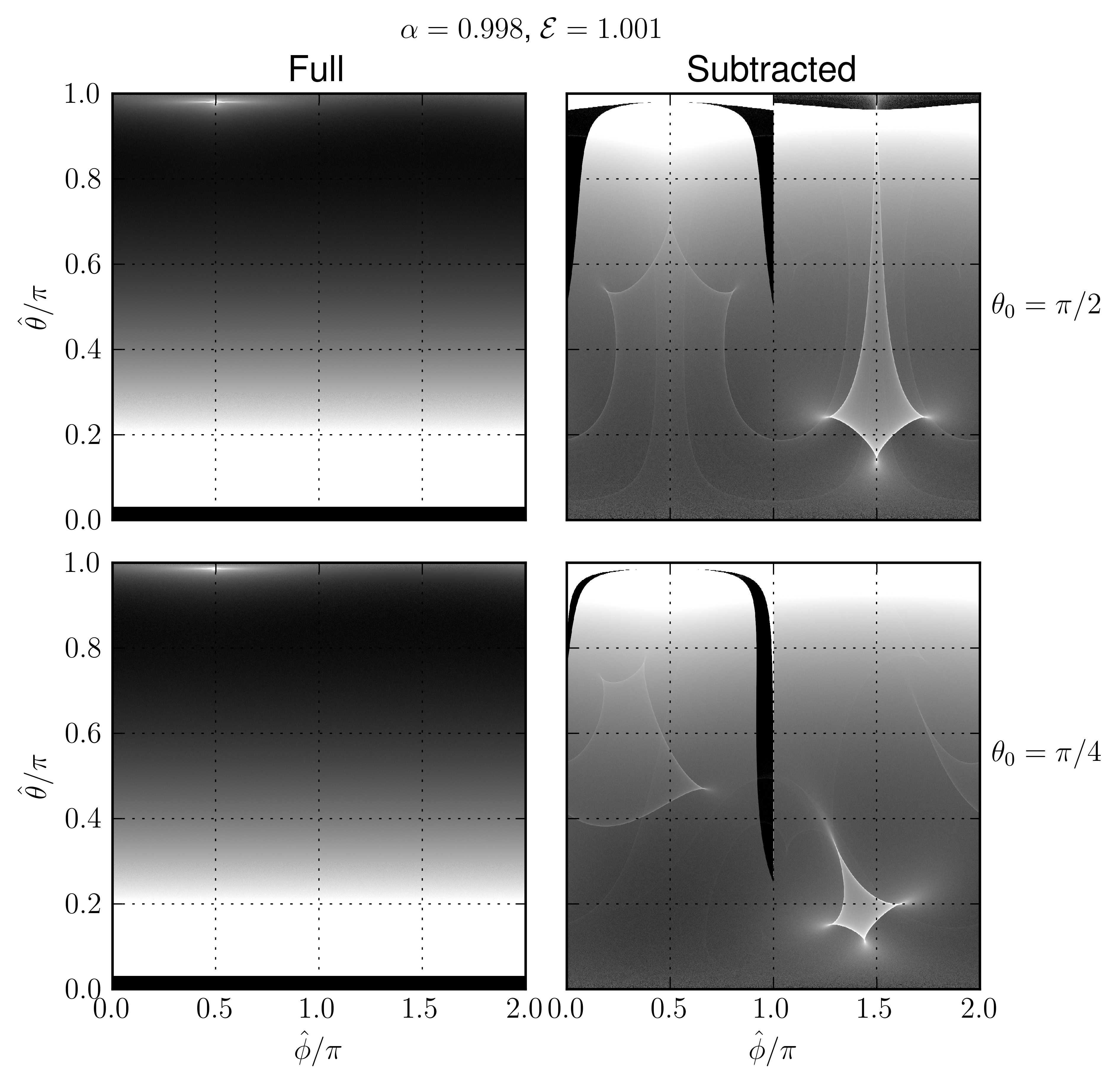}
\caption[Differential scattering cross section of a weakly relativistic particle around a near-extremal black hole]{Differential scattering cross section of a weakly relativistic particle around a near-extremal black hole.}
\label{last}
\end{figure}

In computing the differential cross section, the parameter space was sampled at high ($\alpha = 0.998$) and moderate ($\alpha = 0.5$) spins. For each spin, the differential cross section was computed for orbits approaching side-on to the black hole ($\theta_0=\pi/2$) and at a 45 degree angle ($\theta_0=\pi/4$) at weakly relativistic ($E = 1.001mc^2$), moderately relativistic ($E = 1.1 m c^2$) and ultrarelativistic ($E = 1000 mc^2$) initial velocities. The ultrarelativistic case is effectively indistinguishable from the result for massless particles moving at the speed of light. For each differential cross section, both the full version (the actual value) and a ``subtracted'' version were computed. The ``subtracted'' version has the contribution from the main branch cut subtracted, allowing the structure due to higher order orbits to be visualized. The results of these scattering experiments, computed for a representative sample of the parameter space, are presented in figures \ref{first}-\ref{last}. 

The most striking feature of the cross section with non-zero spin is the presence of 1 dimensional optical caustics. Naturally, in the small-deflection limit $\hat{\theta} \rightarrow 0$ the cross section also diverges as seen in all previous cases. In the axisymmetric cases, it also diverged at $\hat{\theta}=\pi$, but this is no longer true once the axisymmetry has been broken: the 0-dimensional caustic spreads out into a series of curves upon which the value of $\displaystyle\frac{d\sigma}{d\Omega}$ approaches infinity, which become larger in size at higher spin. 

The full structure of these optical caustics in the case of light trajectories is discussed in detail in \cite{KerrCaustics}. Although the cross section for orbits which circle the black hole many times becomes vanishingly small, they are present up to the highest order branches that were resolved with the Monte Carlo scattering data. Each one is result of the point caustic at each $\hat{\theta} = \pi + 2\pi N$ singularity spreading out into a closed curve, so there are in fact infinitely many.

\begin{figure}
\centering
\includegraphics[width=\textwidth]{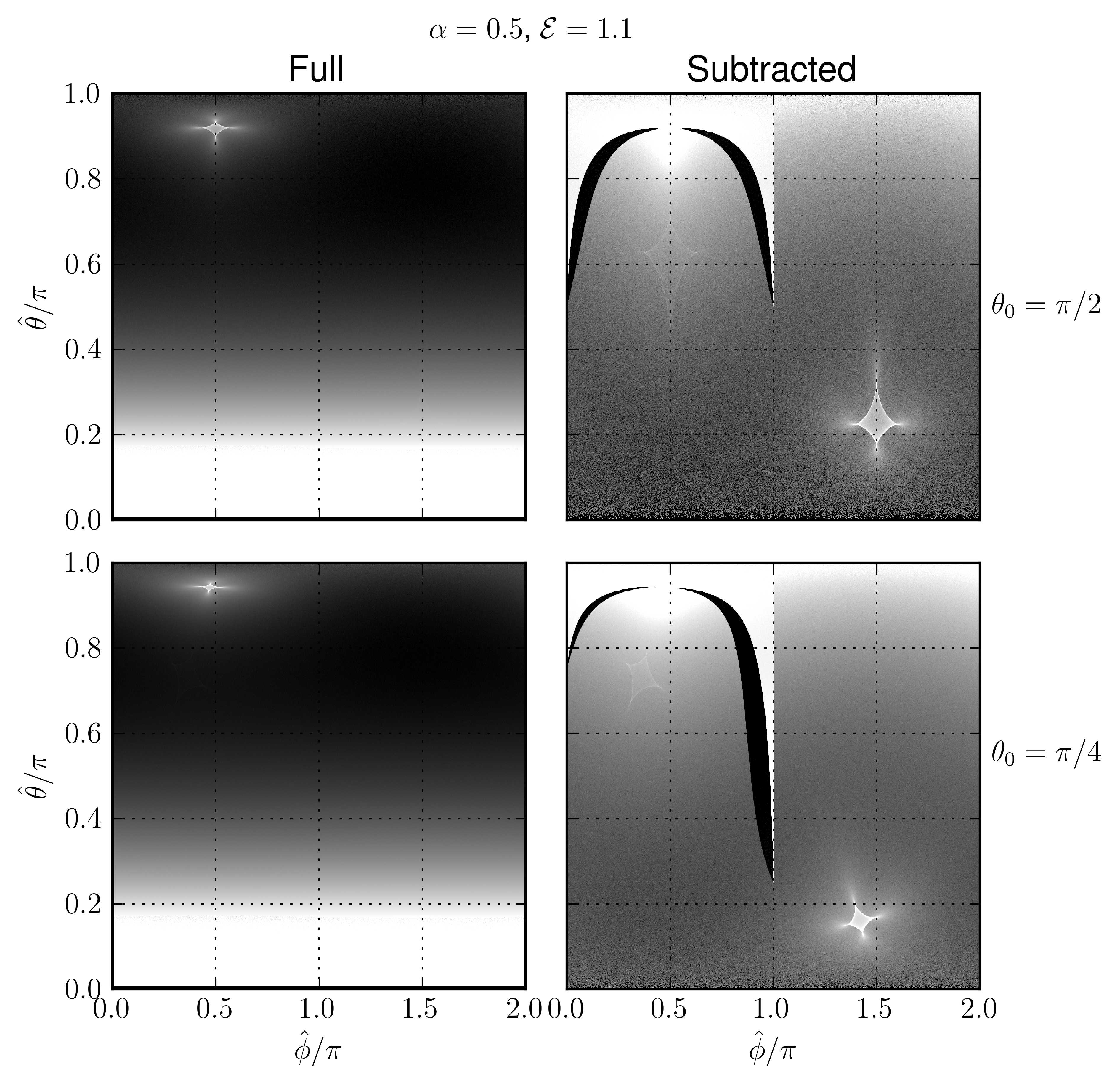}
\caption[Differential scattering cross section of a moderately relatvistic particle around a black hole with moderate spin]{Differential scattering cross section of a moderately relativistic particle around a black hole with moderate spin.}
\end{figure}
\begin{figure}
\centering
\includegraphics[width=\textwidth]{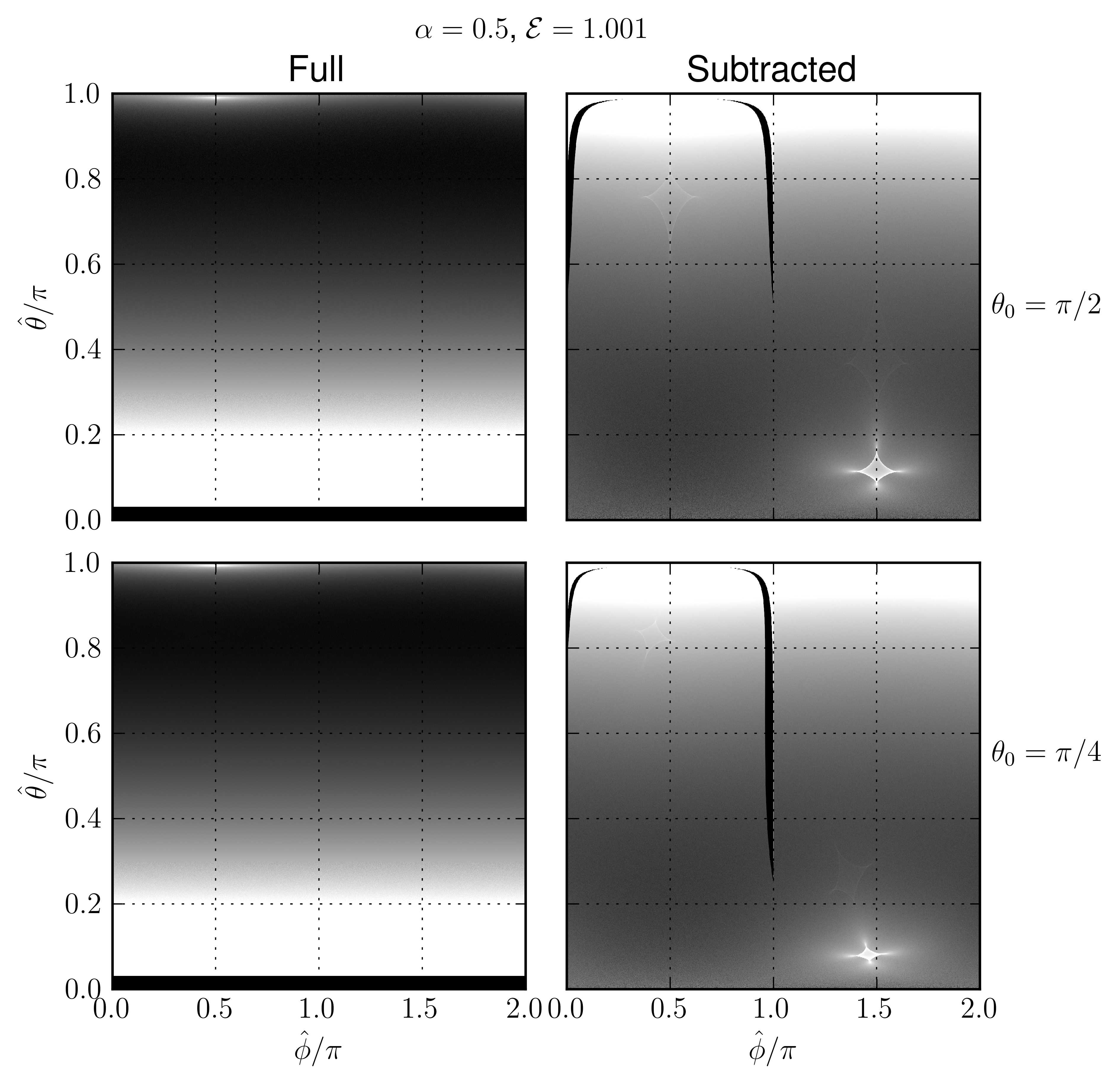}
\caption[Differential scattering cross section of a weakly relativistic particle around a black hole with moderate spin]{Differential scattering cross section of a weakly relativistic particle around a black hole with moderate spin.}
\end{figure}
\begin{figure}
\centering
\includegraphics[width=\textwidth]{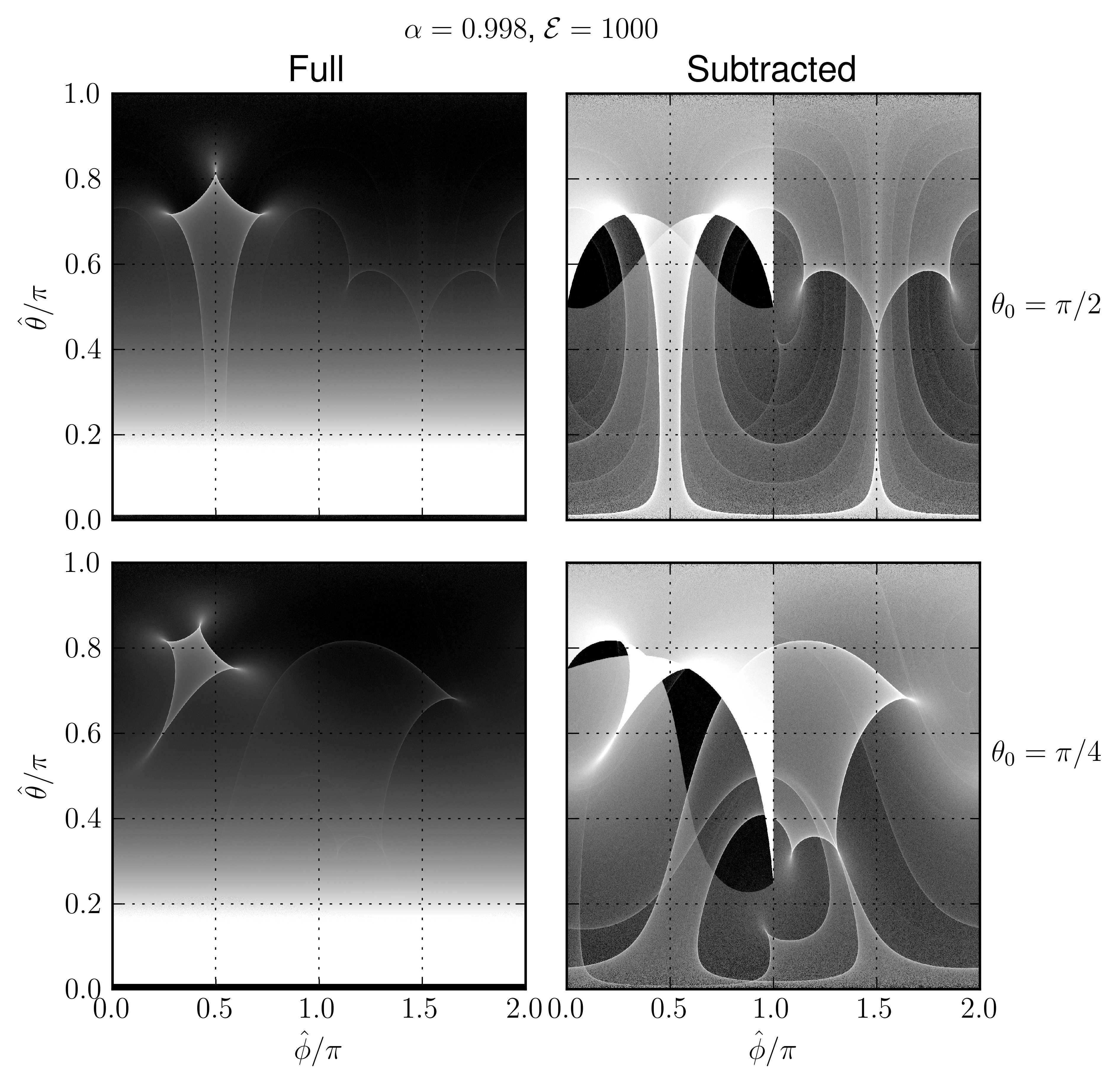}
\caption[Differential scattering cross section of an ultrarelativistic particle around a near-extremal black hole]{Differential scattering cross section of a weakly relativistic particle around a near-extremal black hole.}
\end{figure}
\begin{figure}
\centering
\includegraphics[width=\textwidth]{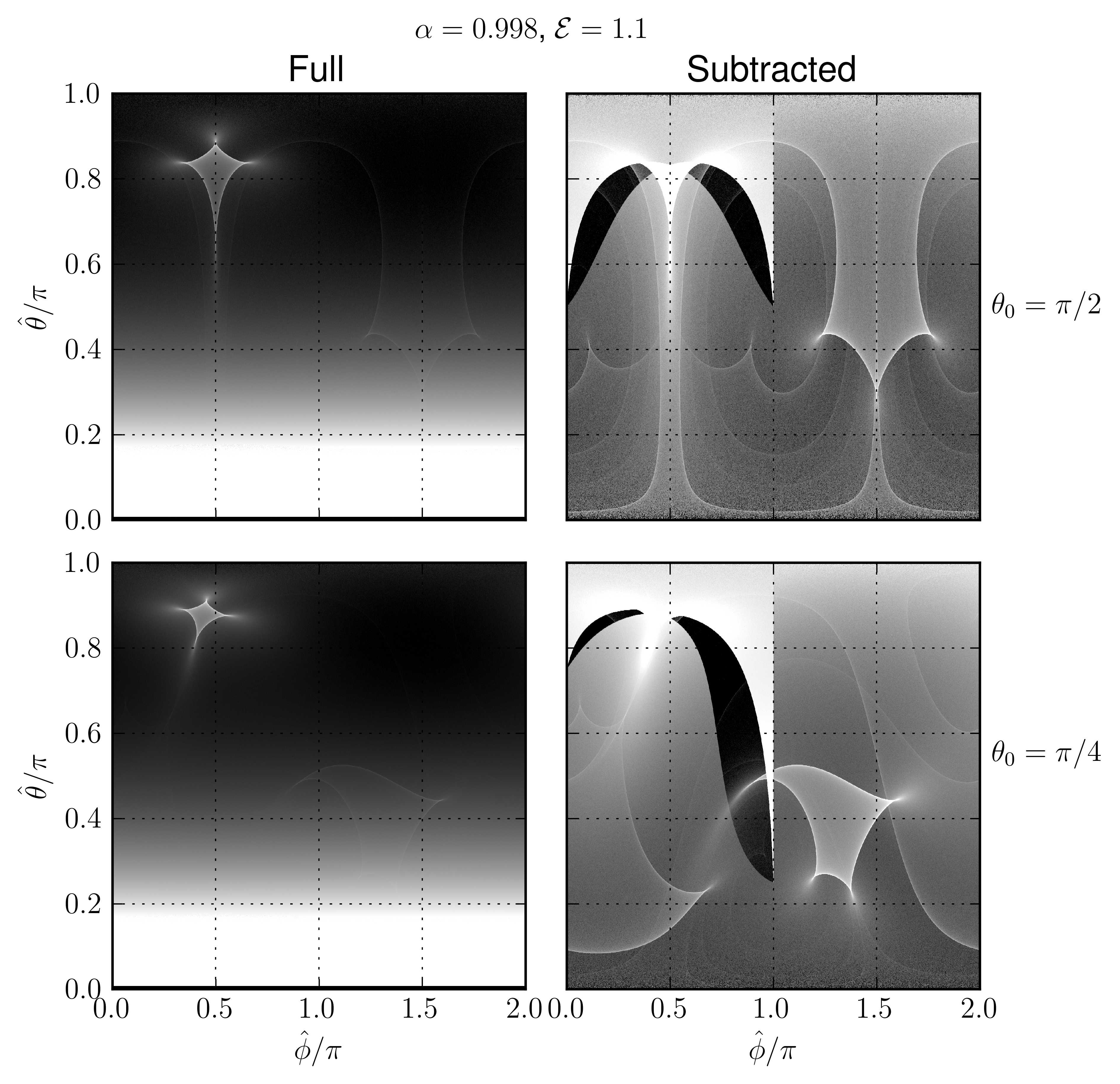}
\caption[Differential scattering cross section of a moderately relatvistic particle around a near-extremal black hole]{Differential scattering cross section of a weakly relativistic particle around a near-extremal black hole.}
\end{figure}

%% file: PNResults.tex
\chapter{Scattering and Capturing of Compact Object Binaries}
\epigraph{``I don't care about what anything was \emph{designed} to do; I care about what it \emph{can} do.''}{Gene Kranz, \it{Apollo 13}}
\section{Numerical Method and Validation}
In practice, the ODE of equation \ref{PNEqn} must be integrated numerically. There exists a method specifically tailored for integrating this flow, which preserves the PN Hamiltonian and has superior global error properties to general integration schemes when integrating bound orbits \cite{SymplecticPN}. However, as we are interested in unbound orbits, which involve a wide range of timescales, a method which allows adaptive timestepping is much more efficient. An implementation of the \texttt{dopri853} method (described in \cite{NR}) was found to be efficient at computing deflection angles to acceptable precision (relative errors of $10^{-12}$ when just the Newtonian term was integrated and compared with the exact solution of equation \ref{NewtonDeflection}). 

The binary is started at some large separation $R_0$ to approximate approaching from infinity. To minimize the error in the deflection angle, $R_0$ should be chosen to be several orders of magnitude larger than the impact parameter, and if adaptive timestepping is used, choosing larger values of $R_0$ does not significantly increase computation time. The integration is performed over an adequate time interval for the binary to reach periastron and escape to a radius of the same order of magnitude as $R_0$. In the Newtonian case, the exact time interval required to approach from $R_0$ and return is:
\begin{equation}
\Delta t = 2 \int_{r_p}^{R_0} \frac{dr}{\dot{r}} = 2 \int_{r_p}^{R_0} \frac{dr}{\sqrt{v_0^2 + \displaystyle\frac{2}{r} - \frac{h^2}{r^2}}}
\end{equation}
For PN orbits, integrating over this time interval multiplied by a safety factor of 2 works well. The deflection angle is taken to be the angle between the final and initial relative 3-velocities.

The post-Newtonian formalism may be unreliable in situations of strong gravitational fields. To estimate the parameter space in which the results can be trusted, we test the post-Newtonian result on a problem for which the exact solution is known, namely the motion of a test particle around a Schwarzschild black hole. By substituting $\eta=0$ into equation \ref{PNEqn} and letting $v_0=1$, we obtain the PN approximation to the equation of motion of a light ray. As conventional wisdom says that the PN error will be large where the gravitational field is strong, one expects the error to be a function of the periastron coordinate $r_p$.
\begin{figure}[ht!]
\centering
\includegraphics[width=\textwidth]{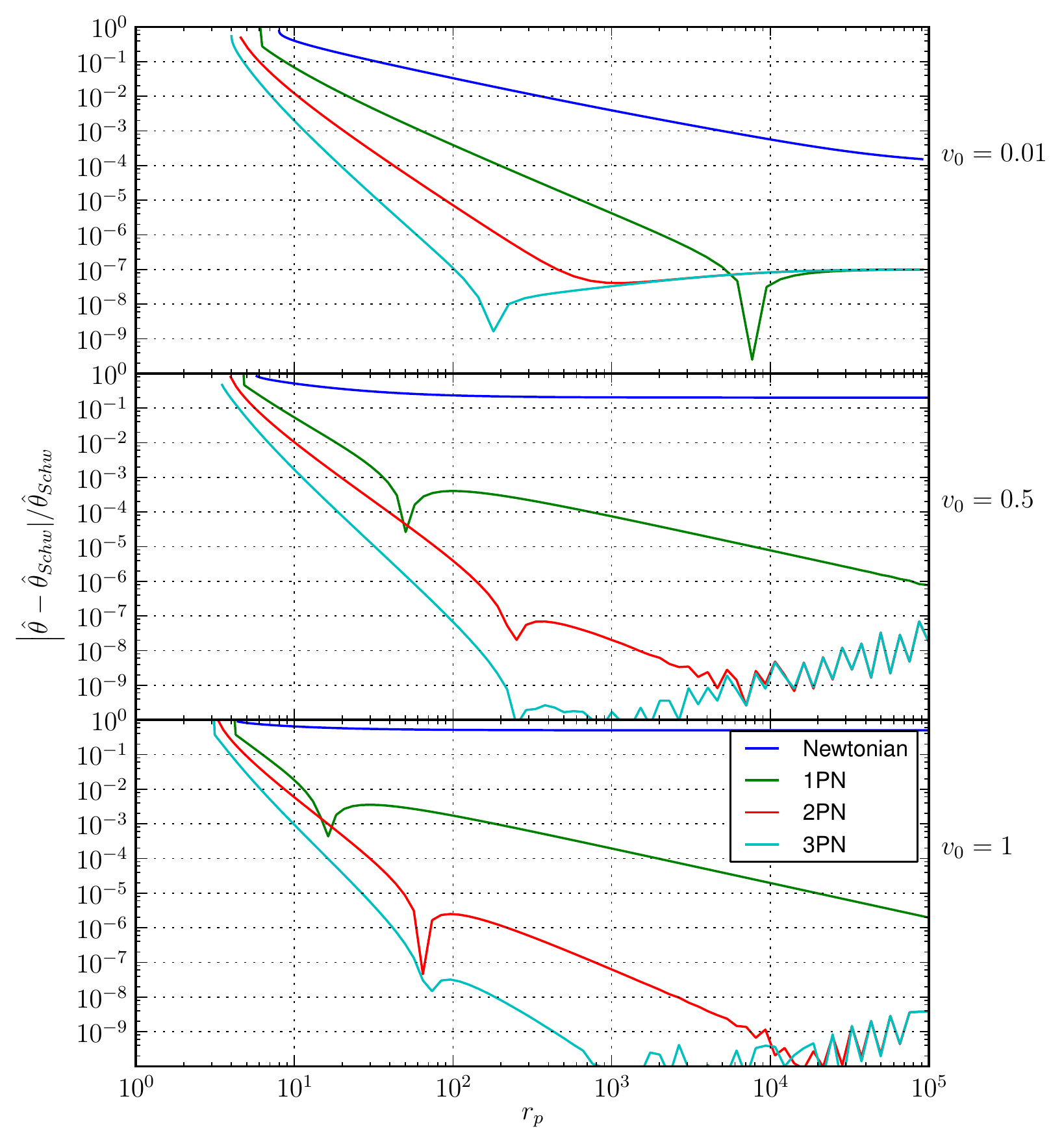}
\caption[Error in post-Newtonian test particle deflection angle]{Relative error in the deflection angle of a test particle compared to the exact Schwarzschild solution. The PN result reproduces weak deflection faithfully at all velocities, but for encounters within $10R_G$ the relative error may be 10 percent or greater. The relative error ceases to improve where the higher PN terms become vanishingly small compared to floating point precision.}
\label{PNvsSchw}
\end{figure}
\begin{figure}
\centering
\includegraphics{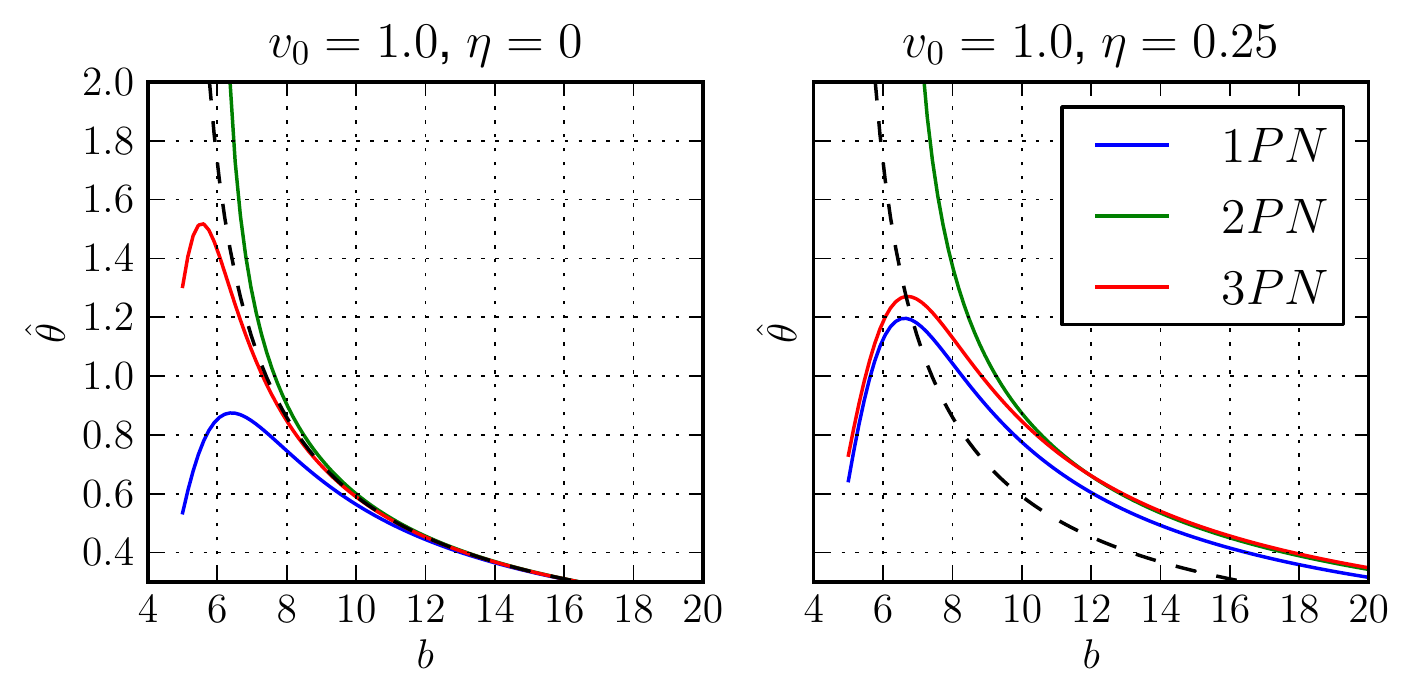}
\caption{Strong-field deflection angle in the test-particle and equal-mass cases at the various PN orders. The Schwarzschild geodesic result (dashed) is included for comparison.}
\label{PN_Strong}
\end{figure}
Comparing the PN deflection angle with the exact Schwarzschild deflection up to 3PN order (figure \ref{PNvsSchw}) reveals that, at all initial speeds, the relative error in the deflection angle becomes large for approximately the same values of $r_p$, confirming that the PN approximation breaks down for orbits that approach to within a few gravitational radii. It is therefore possible to use $r_p$ as a practical barometer of the accuracy of the PN approximation. As figure \ref{PN_Strong} demonstrates, in the strong-field regime the 2PN approximant begins to greatly overestimate the deflection angle compared to the Schwarzschild solution, while the 3PN approximant reaches an upper bound and then begins to \emph{decrease}. Clearly, such results are unphysical. In all results presented in this chapter, the 2PN deflection angle is within 10 percent of the 3PN one; this error bound gives physically reasonable results. Beyond the test particle limit, the corrections to the motion due to the mass ratio are subtle, so this assessment should also hold for any mass ratio. In practice this error condition is satisfied for all impact parameters greater than some value of $b$ which is at least $b_{min}$, the critical impact parameter obtained from equation \ref{Schwhmin}, letting $\mathcal{E}=\left(1-v_0^2\right)^{-\frac{1}{2}}$.

\section{Capture Cross Section}

Though the PN orbits do not generally admit closed-form orbital equations, it is possible to estimate the total gravitational flux emitted along a hyperbolic orbit by simply integrating the lowest-order flux contribution along a Newtonian orbit, as was originally done by Hansen \cite{Hansen1972}. The instantaneous rate of energy loss, to lowest order, is \cite{Kidder1995}:
\begin{equation}
\dot{\mathcal{E}} = -\frac{8\eta^2}{15 r^4}\left(12 v^2 - 11 \dot{r}^2 \right)\,.
\end{equation}
The total energy loss is then:
\begin{equation}
\Delta \mathcal{E} = \int \dot{\mathcal{E}}\,dt = \int \frac{\dot{\mathcal{E}}}{\dot{\phi}}\, d\phi\,.
\end{equation}
Substituting the Newtonian angular momentum relation $\dot{\phi}=\displaystyle\frac{h}{r^2}$ and the Newtonian orbital equation $r=\frac{h^2}{1+e\cos\phi}$ where $e=\sqrt{1+h^2 v_0^2}$ yields the expression:
\begin{equation}
\Delta \mathcal{E} = \frac{54\pi \eta^2}{b^7 v_0^7} + \mathcal{O}\left(\frac{1}{v_0^3}\right)\,.
\end{equation}
Setting this equal to the kinetic energy at infinity gives, to lowest order, the impact parameter required for the binary to radiate enough energy to end up in a bound orbit:
\begin{equation}
b_{capture} = \left(\frac{108\pi\eta}{v_0^9}\right)^{1/7}
\label{PNbcapture}
\end{equation}
Therefore, for small $v_0$, the capture cross section from gravitational radiation is the area of a disk of radius $b_{capture}$, and hence is proportional to $v_0^{-18/7}$. It is also proportional to $\eta^{2/7}$, which approaches 0 as the mass ratio $q$ approaches 0 and is greatest for an equal-mass binary. It therefore recovers the obvious result that the cross section for radiative capture of a test particle is zero. However, somewhere between the equal-mass limit and the test-particle limit the PN approximation is expected to break down, as in the extreme mass ratio limit the smaller mass must travel deep into the larger one's gravity well to radiate enough energy to be captured. Some smooth transition between the $v_0^{-18/7}$ dependence in the equal mass case and the $v_0^{-2}$ dependence (see figure \ref{SigmaVsE}) in the test mass case is expected to occur.

The value of $b_{capture}$ for a given $v_0$ may be computed numerically by finding a lower bound value where the orbital energy of the corresponding orbit eventually passes below zero and an upper bound whose orbital energy remains greater than zero. One then uses the bisection method to determine the value of $b$ for which the final orbital energy is 0. In practice an error is introduced by the necessity of integrating the trajectory from a finite distance to another finite distance instead of integrating it to and from infinity. However, because the gravitational luminosity falls off as $\displaystyle\frac{1}{r^4}$, by far the largest portion of the energy loss occurs near periastron, so the error from neglecting the energy radiated as the binary escapes to infinity is very small, and, in the implementation used, smaller than what can be resolved at machine precision.
\begin{figure}
\centering
\includegraphics{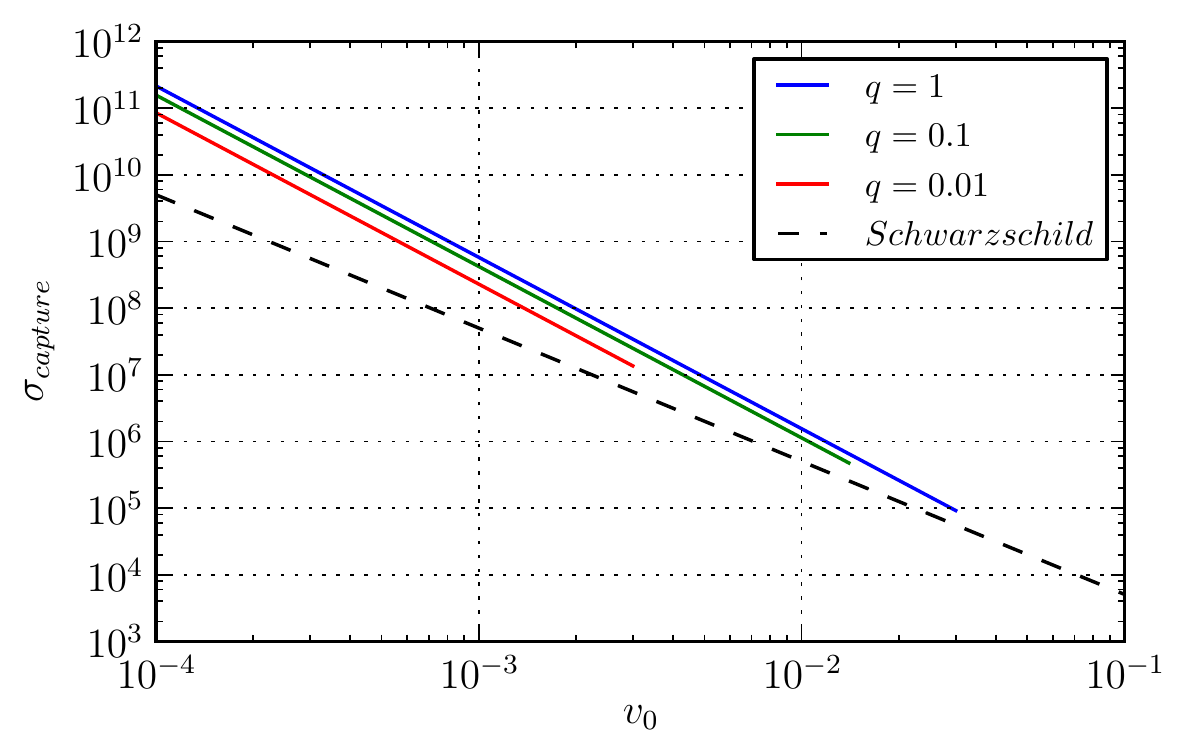}
\caption[Capture cross section due to gravitational radiation]{Capture cross section due to gravitational radiation, computed to 3.5PN order. In the limit of small velocity the agreement with the analytic formula is good, however as the capture cross section becomes of comparable order of magnitude to the Schwarzschild geodesic capture cross section (dotted) the PN approximation begins to diverge.}
\end{figure}
Computing $\sigma_{capture}$ this way gives results which agree quite well with equation \ref{PNbcapture} in the Newtonian regime. In the parameter space where the cross section is close to the Schwarzschild cross section, toward large $v$ and/or small $q$, the PN approximation ceases to converge.

Astrophysically speaking, the largest plausible encounter velocity of two black holes would likely be on the order of $10^3$\,km/s \cite{BBH_Kick}, or $\sim0.01c$, so equation \ref{PNbcapture} can be expected to be accurate in any physically realistic situation where the mass ratio is not too extreme. In terms of quantities relevant to stellar astrophysics, equation \ref{PNbcapture} can be written:
\begin{equation}
\frac{\sigma_{capture}}{\pi R_\odot^2} \approx 1.78 \left(\frac{M}{20M_\odot}\right)^2 \left(4\eta\right)^{1/7}\left(\frac{v_0}{100\mathrm{km/s}}\right)^{-18/7}\,.
\end{equation}
where M is the total mass of the system. Hence, for example, the capture cross section of two $10M_\odot$ black holes moving toward each other at $100km/s$ relative velocity is about 1.78 times the cross section of the Sun. Even assuming a stellar density of $10^3$pc${}^{-3}$, comparable to a dense globular cluster core, this would imply a mean free path two orders of magnitude greater than the Hubble length. This suggests that such a capture event is, at least at small redshift, quite rare.

\section{Scattering Orbits}
\begin{figure}[h!]
\centering
\includegraphics{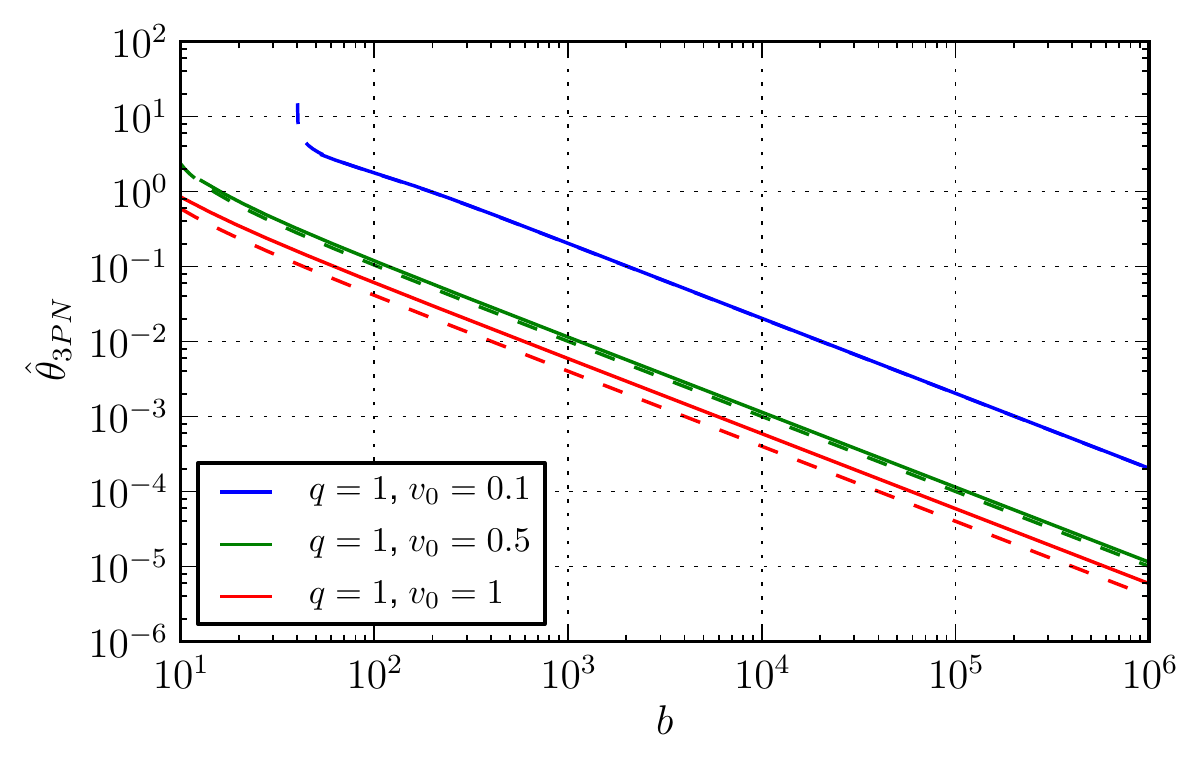}
\caption[3PN-accurate deflection angle for an equal mass binary as a function of impact parameter]{3PN-accurate deflection angle for an equal mass binary as a function of impact parameter, compared with the Schwarzschild deflection angle for the same $v_0$ (dashed).}
\label{3PNDef}
\end{figure}
\begin{figure}
\centering
\includegraphics{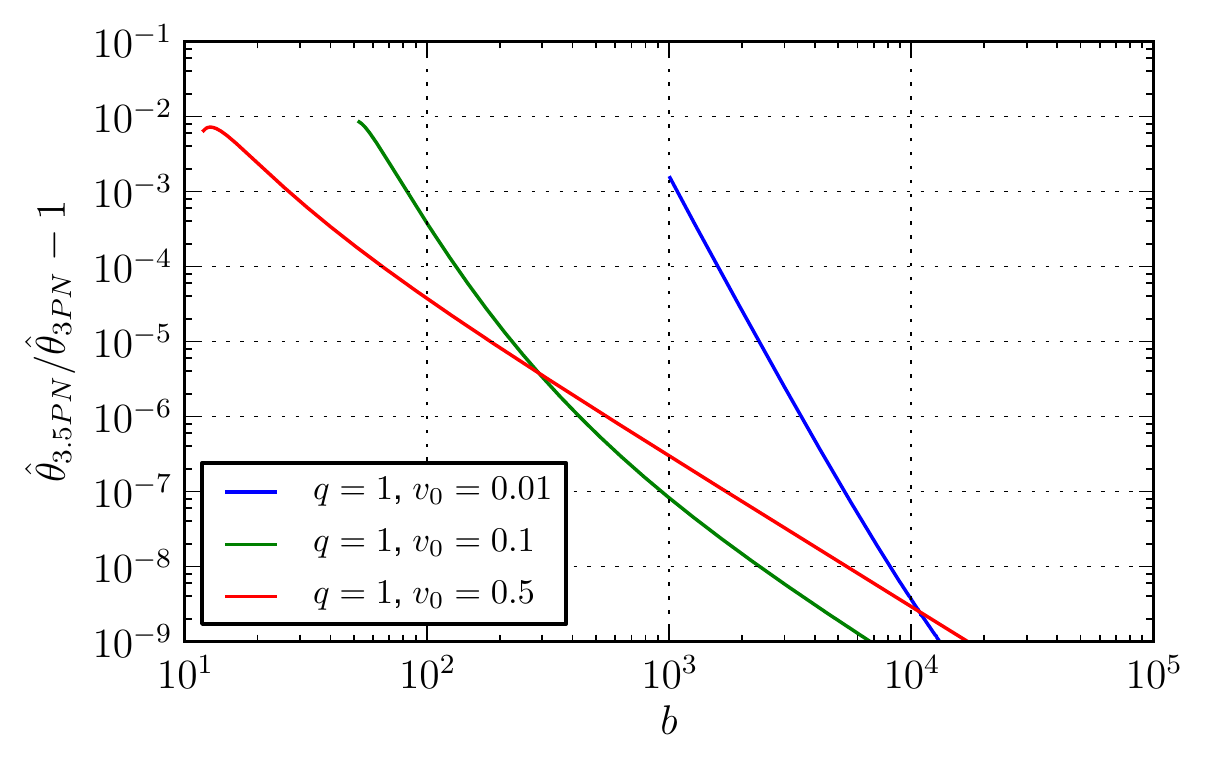}
\caption[Effect of radiation action on deflection angle]{Relative increase in deflection angle computed at 3.5PN order compared to 3PN order with only conservative terms. The upper bounds on the deflection of the $v_0=0.1$ and $v_0=0.5$ cases are due to the PN series starting to diverge. The upper bound on the $v_0=0.01$ case corresponds to the largest possible deflection angle of an orbit of that energy; reducing $b$ any further results in a capture.}
\label{RadReaction}
\end{figure}
At least within the PN regime, radiation reaction always increases the deflection angle. This can be understood intuitively with the patched-conic approximation to orbital mechanics. Because the magnitude of the reaction force is by far the greatest near periastron, to first order the effect is equivalent to a sudden drop in orbital energy and angular momentum exactly at periastron. This decreases the eccentricity and displaces the asymptote of the outgoing hyperbolic orbit by a small amount in the prograde direction.
The effect of radiation reaction is, however, very subtle even at the very edge of the PN regime. As can be seen from figure \ref{RadReaction}, even as the approximation falters or the regime of capturing orbits is approached, the relative increase in deflection does not exceed $10^{-2}$. Therefore, the angle is very well approximated by just the conservative part of the PN dynamics.

\begin{figure}
\centering
\includegraphics{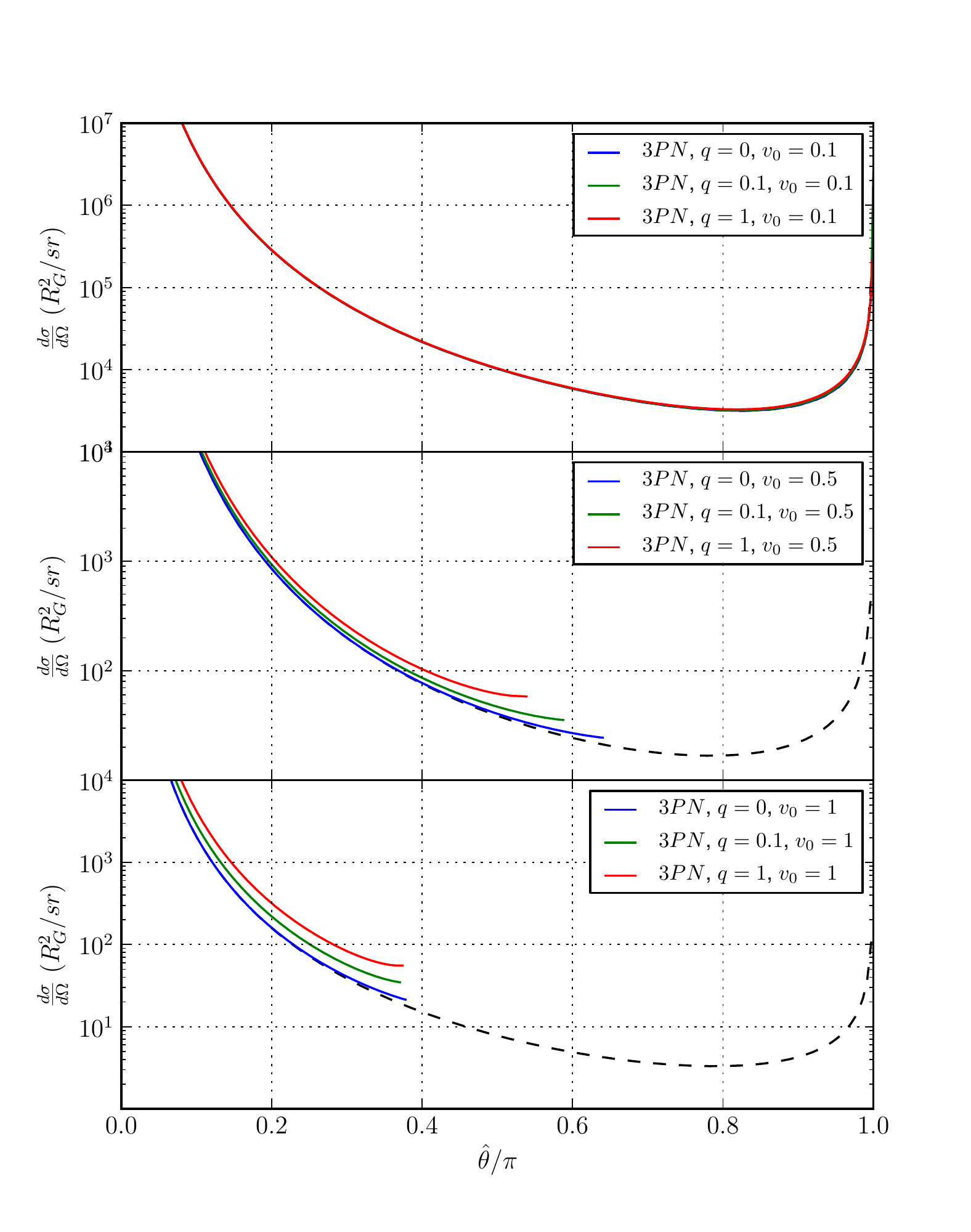}
\caption[3PN differential cross section]{Differential scattering cross section computed to 3PN order, within the domain where the 3PN approximant converges sufficiently well. At larger velocities, the convergence at large deflection angles (or equivalently, close orbits) deteriorates.}
\label{PNdSigmadOmega}
\end{figure}
When comparing the finite mass ratio case with the test particle case, it is generally true that, for a given impact parameter and initial speed, the deflection is greatest for equal mass and smallest for the test particle case. This is in contrast to results found for black holes on bound orbits, for which the relativistic periastron precession is always less than the test particle limit \cite{Abdul2011}. The weak deflection limit still shows the familiar $\displaystyle \frac{1}{b}$ dependence, however the coefficient of the $\displaystyle\frac{1}{b}$ term apparently has some dependence on mass ratio and initial speed. The discrepancy between the equal mass and test mass deflections is most pronounced for orbits which are more highly relativistic, which not surprising because the Newtonian solution predicts no mass ratio dependence. This motivates a fitted weak-deflection approximant of a form analogous to equation \ref{SchwWeakDef}:
\begin{align}
\hat{\theta} &\approx \left(4 + c_1\eta + c_2 \eta^2 + \frac{2}{\mathcal{E}^2 - 1}\right) \frac{1}{b} \,,
\end{align}
\noindent where $c_1=8.23$ and $c_2=-5.39$. As the deflection angle is always greater by a certain constant factor in the weak-deflection regime (Figure \ref{3PNDef}), it can then be expected that $\frac{d\sigma}{d\Omega}$ be correspondingly larger by the square of this factor. Figure \ref{PNdSigmadOmega} demonstrates that this is in fact the case, with the increase with mass ratio again being most pronounced for more highly relativistic velocities.


\section{Discussion}
The PN expansion at 3.5PN order reliably describes the scattering dynamics of binary compact objects on \emph{weakly} deflected hyperbolic orbits. For small (and astrophysically realistic) initial velocities it also can accurately compute larger scattering angles all the way into capturing orbits. The shortcomings of the method become evident in any case where the binary approaches to within about 10$R_G$ in harmonic coordinate distance. This includes strongly deflected or near-capture orbits at relativistic speeds and extreme mass ratios.

There are myriad other approaches to this problem. In the limit of large mass ratio, the character of the radiation reaction force becomes rather different: it can be viewed as the smaller body's subtle reaction to its own gravitational perturbation. Calculations with the so-called gravitational self-force formalism provide a more accurate description of the nearly-geodesic motion of a massive object orbiting a much more massive black hole all the way down to close orbits \cite{GSF}. Therefore, in the limit of extreme mass ratio where radiation reaction is subtle but nonzero it can be used to compute the cross section for gravitational radiative capture, as well as the deflection of unbound orbits. In this limit the capture cross section should be only marginally larger than the geodesic capture cross section, effectively a perturbation of it with the smaller mass as the perturbation parameter. Furthermore, given how subtle an effect radiation reaction has even on equal-mass scattering angles, in the limit of extreme mass ratio the difference between the geodesic and forced trajectories would be extremely subtle. Therefore, no particularly dramatic physics is expected to arise from such calculations.

Numerical relativity can, in theory, be used to compute scattering results in any region of parameter space. In particular, it is likely necessary to solve the full nonlinear Einstein equation to determine the capture cross section of binaries of comparable mass approaching at relativistic speed. Toward more extreme mass ratios such calculations would be more challenging from a practical standpoint, as the wider range of length scales requires greater resolution. With the na\"{i}ve approach, this makes the necessary simulation time proportional to the ratio between the larger and smaller masses. Extending the usefulness of numerical relativity to large mass ratios and problems with multiple length scales is currently an active area of research.

The effects of spin-spin and spin-orbit interactions have been neglected from the scope of this work. The spin effects would likely be analogous to the test particle case, with prograde and retrograde orbits being deflected less and more respectively. When at least one spin is not normal to the orbital plane, the orbital plane will precess noticeably for close orbits as spin angular momentum is exchanged for orbital or vice versa. Therefore, as in the test particle case, it is expected that spin effects would break the axisymmetry of the scattering angle function and the corresponding cross sections. Indeed, caustic-like structures in the differential cross sections of spinning binaries possibly also exist, however the number of caustics would be finite because higher-order orbits would end up as capturing orbits.

%% file: conclusion.tex
\chapter{Conclusion}
\epigraph{``End? No, the journey doesn't end here.''}{Gandalf the White}
The relativistic Kepler problem, while in certain limits agreeing with Kepler's prescription of conic sections, also encompasses far more dramatic physics which were unprecedented before general relativity. This includes the mechanisms of event horizon and radiative capture, as well as periastron precession. The strong-field orbital dynamics near a black hole can only be described by asymptotic expansions like Einstein's lensing formula to a certain extent, with only the exact solution encoding the rich structure of the scattering angle function. Solving the problem in terms of elliptical integrals provides a way to compute large numbers of trajectories for little computational expense, enabling otherwise demanding computations such as virtual scattering experiments to to be performed.

The capture and scattering cross sections of a black hole are part of the toolkit of physical observables by which the theory of gravity could be tested. One could hypothetically perform the same sort of scattering experiment as was performed in Monte Carlo simulations, and this would constitute a method of probing the strong gravitational field near the black hole. An interesting line of research is the inverse problem: what information about the spacetime geometry can actually be recovered from the scattering angle function? The answer to such a problem may be of astrophysical interest when it becomes possible to resolve distant black holes on the scale of the gravitational radius, as is being attempted with the Event Horizon Telescope project. The ability to infer the geometry around of a black hole from the lensed images of other objects (or from the shape and size of the hole's shadow) would provide a way to test the strong-field predictions of GR.

The results of chapter 4 provide an idea of how far the post-Newtonian approximation can be trusted when studying objects on hyperbolic orbits. At astrophysically realistic orbital velocities, it fares quite well all the way down to capturing orbits. At encounter velocities that are a significant fraction of $c$, it is incapable of reliably predicting the gravitational capture process, or even accurately predicting moderate deflection angles. It is possible that numerical relativity is able to overcome PN's deficiencies for binaries of comparable mass, while toward more extreme mass ratios the problem can be approached with black hole perturbation theory.

The physics of compact objects in bound orbits has been a much more active area of research than unbound: compact object binary inspirals and mergers are by far the most promising candidate events for gravitational wave astronomy, whereas hyperbolic orbits at realistic velocities either lack the gravitational luminosity to be detectable with the gravitational wave detectors of the near future or have very small event rates, as is quantified in chapter 4. Nevertheless, hyperbolic orbits constitute a largely neglected area of the parameter space of the relativistic Kepler problem, and are interesting from the perspective of fundamental gravitational physics.

The coming decades will be critical for the understanding of strong gravitational fields. The theory of general relativity will tested through an arsenal of observational techniques including gravitational wave astronomy (using both pulsar timing arrays and laser interferometry) and VLBI, which will provide the first images of compact objects resolved on the scale $R\approx R_G$. Depending on what is found, Einstein's relativity may enjoy a reign as the standard classical theory of gravity at least as long as Newton's, and advances in computing power will make it more useful and tractable for astrophysics than ever. However, at this point there is no telling what may be found, and it is possible that discrepancies will be found which will further refine the understanding of the mechanism of celestial motion. Like Newton's theory before it, finding evidence pointing to GR's successor requires first that GR's predictions can be precisely computed, measured, and understood.

%

%% file: AppendixB.tex
\chapter{Computational Details}
All numerical algorithms used to obtain the results presented in this work were implemented either in Python, making heavy use of the \texttt{numpy} and \texttt{scipy} libraries, or in C++ called from the Python framework via the \texttt{scipy.weave} interface. The source code implementing these algorithms is presented in this section.
\section{Kerr Geodesics}
When performing Monte Carlo scattering simulations, it was important that the scattering angle calculation be as fast as possible to achieve good angular resolution in a reasonable amount of time. A non-negligible portion of the computational expense of the solution is the determination of the roots of the quartic polynomial $\mathcal{R}(r)$. General polynomial solvers often use a linear algebra subroutine to obtain the eigenvalues of the polynomial's companion matrix. This works well, but because the polynomial is only fourth order, a significant savings can be made by solving it analytically. A numerically stable algorithm for doing this is described in \cite{quartic}. All elliptic integrals were computed using functions provided by the \texttt{Boost} library. These functions implement the duplication algorithm described in \cite{CarlsonDuplication} and \cite{NR}.

The problem of computing a large number of scattering angles is ``embarrassingly parallel''. That is, it consists of computing a large number of independent results. Therefore, a simple optimization is to employ all available logical CPU cores in parallel using OpenMP. Overall, the implementation used was able to compute on the order of $10^5$ deflection angles per second on a modern laptop with four logical cores.

\definecolor{mygreen}{rgb}{0,0.6,0}
\definecolor{mygray}{rgb}{0.5,0.5,0.5}
\definecolor{mymauve}{rgb}{0.58,0,0.82}

\lstset{ %
  backgroundcolor=\color{white},   
  basicstyle=\ttfamily\footnotesize,       
  breakatwhitespace=false,         
  breaklines=true,                 
  commentstyle=\color{mygreen},    
  deletekeywords={...},            
  escapeinside={\%*}{*)},          
  extendedchars=true,              
  keepspaces=true,                 
  keywordstyle=\color{blue},       
  language=Python,                 
  morekeywords={*,...},            
  numbers=left,                    
  numbersep=5pt,                   
  numberstyle=\tiny\color{mygray}, 
  rulecolor=\color{black},         
  showspaces=false,                
  showstringspaces=false,          
  showtabs=false,                  
  stepnumber=2,                    
  stringstyle=\color{mymauve},     
  tabsize=4,                       
}
\begin{spacing}{1}
\subsection{KerrDeflection.py}
\begin{lstlisting}
import numpy as np
import math
from scipy import weave, integrate, linalg, optimize
import CarlsonR
from CarlsonR import *

pi = np.pi

def f(r, e, a, t):
    """ Given the periastron radius r, returns (r-1)*by^2. """
    return (r**2*(2*e*(-2 + r)*r*np.sqrt(r + (-1 + e**2)*r**2) + a**2*(1 - r + 2*e**2*r + 2*e*np.sqrt(r + (-1 + e**2)*r**2))) + a**4*(-1 + e**2)*(-1 + r)**2*np.cos(t)**2 - 2*a**4*e*np.sqrt(r + (-1 + e**2)*r**2)/np.tan(t)**2 + 4*a**2*e*r*np.sqrt(r + (-1 + e**2)*r**2)/np.tan(t)**2 - 4*e*r**3*np.sqrt(r + (-1 + e**2)*r**2)/np.tan(t)**2 + 2*e*r**4*np.sqrt(r + (-1 + e**2)*r**2)/np.tan(t)**2 - a**2*(a**2*(e**2 + r + (-1 + e**2)*r**2) + 2*r**2*(-((-2 + r)*(-1 + r)) + e**2*(-1 + (-2 + r)*r)))/np.tan(t)**2 + r**3*(-4 + r*(8 + (-5 + r)*r - e**2*(5 + (-4 + r)*r)))/np.sin(t)**2)/(a**2*(-1 + e**2))

def g(r, e, a, t):
    """ Given the periastron radius r, returns bx. """
    return ((a*e*(-a**2 + r**2) - a*(a**2 + (-2 + r)*r)*np.sqrt(r + (-1 + e**2)*r**2))/np.sin(t))/(a**2*np.sqrt(-1 + e**2)*(-1 + r))

def rLimits(a, E, theta0):
    """ Returns the maximum and minimum periastron radius for orbits on the edge of the capture region. """
    h = lambda r: (2*a**2*r**2*(-2 - E**2*(-1 + r) + r) - a**4*(-r + E**2*(1 + r)) + r**3*(4 + r*(-4 - E**2*(-3 + r) + r)))/((-1 + r)*(a**2 + (-2 + r)*r))     #Quintic polynomial in equation 3.4

    r_mid = optimize.newton(h, 5)     #Solve for real root with 5 as initial guess
    if theta0 == 0:
        return r_mid
    
    #Find the minimum periastron on (1, r_mid] and maximum on [r_mid, 6) by solving by==0 using Brent's method
    a1, a2 = optimize.brentq(f,r_mid, 6, args=(E, a, theta0)), optimize.bisect(f, 1, r_mid, args=(E, a, theta0))
    return a1, a2

def CaptureCrossSection(a, E, theta0):
    """Computes the capture cross section of a black hole with spin a for a particle E approaching from colatitude theta0. """
    if a==0:
        return np.pi*bmin(E)**2
    if theta0==0:
        r = rLimits(a, E, 0.0)
        return ((a**2 + r**2)*(a**2*(-1 + E**2) + r*(2 + (-1 + E**2)*r)))/((-1 + E**2)*(a**2 + (-2 + r)*r)) * np.pi
    rp, rm = rLimits(a, E, theta0)
    eta = np.linspace(0, np.pi, 1000)
    r = 0.5*(rp*(1-np.cos(eta)) + rm*(1+np.cos(eta)))
    by = np.sqrt(f(r, E, a, theta0).real)/(r-1)*np.sign(eta)
    bx = g(r, E, a, theta0)
    notnan = np.invert(np.isnan(bx))*np.invert(np.isnan(by))
    bx, by, r = bx[notnan], by[notnan], r[notnan]
    return 2*np.abs(integrate.simps(by, x=bx))

def EtaBxBy(a, E, theta0, N):
    """ Generates the (bx,by) coordinates of an N-point grid along the capture region boundary """
    eta = np.linspace(-np.pi, np.pi, N)
    if a==0:
        bm = bmin(E)
        return bm*np.cos(eta), bm*np.sin(eta)
    if theta0 == 0:
        r = rLimits(a, E, 0.0)
        bm = np.sqrt(((a**2 + r**2)*(a**2*(-1 + E**2) + r*(2 + (-1 + E**2)*r)))/((-1 + E**2)*(a**2 + (-2 + r)*r)))
        return bm*np.cos(eta), bm*np.sin(eta)

    rp, rm = rLimits(a, E, theta0)
    r = 0.5*(rp*(1-np.cos(eta)) + rm*(1+np.cos(eta)))
    by = np.sqrt(f(r, E, a, theta0).real)/(r-1)*np.sign(eta)
    bx = g(r, E, a, theta0)
    notnan = np.invert(np.isnan(bx))*np.invert(np.isnan(by))
    return bx[notnan], by[notnan]    

def SphericalToCartesian(coords, theta0):
    """ Converts Boyer-Lindquist coordinates to the (x,y,z) frame with the x axis aligned with the approach velocity. """
    theta, phi = coords[0], coords[1]
    x, y, z = np.cos(phi)*np.sin(theta), np.sin(phi)*np.sin(theta), np.cos(theta)
    x, z = np.cos(np.pi-theta0)*x + np.sin(np.pi-theta0)*z, -np.sin(np.pi-theta0)*x + np.cos(np.pi-theta0)*z
    return x, y, z

def CubicRoots(e, b):
    """ Solves R(r)=0 in the equatorial case, where it can be reduced to a cubic. """
    if type(b) != np.ndarray:
        b = np.array([b])
    r3 = np.zeros(b.shape)
    r1 = np.copy(r3)
    r2 = np.copy(r3)
    code = """                                                                                               
    int i;                                                                                                   
    double B, C, Q, R, Q3, theta, SQ;
    double A = 2.0/(e + 1.0)/(e - 1.0);
    double TAU = 2*3.141592653589793116;                                                                     
    for (i = 0; i < Nb[0]; ++i){    
        B = -b[i]*b[i];          
        C = -2*B; 
        Q = (A*A - 3*B)/9;
        R = (2*pow(A, 3) - 9*A*B + 27*C)/54.0;
        Q3 = pow(Q, 3);
        theta = acos(R/sqrt(Q3));      
        SQ = sqrt(Q);
        r1[i] = -2*SQ*cos(theta/3) - A/3;
        r3[i] = -2*SQ*cos((theta + TAU)/3) - A/3;                                 
        r2[i] = -2*SQ*cos((theta - TAU)/3) - A/3;                                                            
    }                                                                                                        
    """
    weave.inline(code, ['e', 'b', 'r3', 'r1', 'r2'])
    return r1, r2, r3

def bmin(e):
    """ Returns critical impact parameter for Schwarzschild orbits at a given energy """
    return math.sqrt((8 - 36*e**2 + 27*e**4 + e*(9*e**2 - 8)**(3.0/2.0))/2)/(e**2 - 1)

def SchwarzDeflection(E, b):
    """ Computes the deflection angle in the orbital plane in the Schwarzschild case """
    bm = bmin(E)
    result = np.zeros(b.shape)
    fall_in= (b<=bm)
    result[fall_in] = np.NaN
    r1, r2, r3 = CubicRoots(E,b[np.invert(fall_in)])
    x = (r3 - r1)*(r3 - r2)
    y = r3*(r3 - r2)
    z = r3*(r3 - r1)
    ellipf = CarlsonR.BoostRF(x, y, z)
    result[np.invert(fall_in)] = 4*b[np.invert(fall_in)]*ellipf
    return result

def WeakDeflection(E, b):
    """ Computes the asymptotic approximation to SchwarzDeflection (6th order) """
    e2 = E**2 - 1.0
    wd_coeffs = np.array([pi*(255255/256. + 1155/(4.*e2**3) + 45045/(32.*e2**2) + 135135/(64.*e2)),
                        716.8 + 2/(5.*e2**5) - 4/e2**4 + 64/e2**3 + 640/e2**2 + 1280/e2,
                        pi*(3465/64. + 105/(4.*e2**2) + 315/(4.*e2)),
                        42.666666666666664 - 2/(3.*e2**3) + 8/e2**2 + 48/e2,
                        pi*(15/4. + 3/e2),
                        4.0 + 2.0/e2**2,
                        pi
                        ])

    return np.polyval(wd_coeffs, 1.0/b)

def EquatorialDeflection(a, E, b, roots):
    """ Computes deflection angle in the equatorial plane of a Kerr black hole """
    C1 = math.sqrt(1-a**2)
    L = b*math.sqrt(E**2-1)
    rplus, rminus = 1 + C1, 1 - C1
    r1, r2, r3, r4 = roots
    int1 = InvSqrtQuartic(r1, r2, r3, r4, r4)
    int2 = TerribleIntegral(r1, r2, r3, r4, rplus, r4)
    int3 = TerribleIntegral(r1, r2, r3, r4, rminus, r4)
    part1 = (L - a*E)/math.sqrt(E**2 - 1) * int1
    part3 = a*E/math.sqrt(E**2 - 1) * (int1 + (a**2 -a*L/E + rplus**2)/(rplus - rminus)*int2  - (a**2 -a*L/E + rminus**2)/(rplus - rminus)*int3)
    phi_result = 2*(part1 + part3)
    return phi_result

def SchwTiltCoords(deflection, theta, bx, by):
    """ Given the deflection angle in the orbital plane, converts the scattering angles to the Boyer-Lindquist frame for a Schwarzzschild orbit. """
    t1 = np.arctan2(by,bx)
    y = np.cos(t1)*np.sin(deflection)
    x = np.cos(deflection)*math.sin(theta) - np.sin(deflection)*math.cos(theta)*np.sin(t1)
    phi_result = np.arctan2(y, x)
    theta_result = np.arccos(np.cos(deflection)*math.cos(theta)+math.sin(theta)*np.sin(deflection)*np.sin(t1))
    return phi_result, theta_result

def KerrDeflectionC(a, theta, E, bx, by, force_compile=False):
    """ Wrapper for C++ implementation of the deflection angle calculation. """
    
    if type(bx) != np.ndarray:
        bx = np.array([bx,])
    if type(by) != np.ndarray:
        by = np.array([by,])
    phi_result = np.empty(bx.shape)
    theta_result = np.empty(bx.shape)

    #Schwarzschild case
    if a==0.0:
        sch_def = SchwarzDeflection(E, np.sqrt(bx**2 + by**2))
        phi_result, theta_result = SchwTiltCoords(sch_def, theta, bx, by)
        return phi_result%(2*pi), theta_result%(pi)

    code = """
    int nn = Nbx[0];
    int i;
    #pragma omp parallel for
    for (i = 0; i < nn; i++)
    {
        KerrDeflection(a, E, theta, bx[i], by[i], theta_result[i], phi_result[i]);
    }
    """

    #C++ subroutine for full Kerr case
    weave.inline(code,
                 ['a','E','theta','bx','by','phi_result','theta_result'],
                 headers=["<algorithm>",
                          "<cmath>",
                          "<boost/math/special_functions/ellint_rf.hpp>",
                          "<boost/math/special_functions/jacobi_elliptic.hpp>",
                          "<boost/math/special_functions/ellint_3.hpp>",
                          "<boost/math/special_functions/ellint_rj.hpp>",
                          "<boost/math/special_functions/ellint_rc.hpp>",
                          "</usr/include/quintic_C.c>",
                          "</usr/include/KerrDeflection.cpp>",
                          "<omp.h>"],
                extra_compile_args =['-O3 -fopenmp -mtune=native -march=native -ffast-math -msse3 -fomit-frame-pointer -malign-double -fstrict-aliasing'],
                extra_link_args=['-lgomp'],
                force=force_compile
                )
    return theta_result, phi_result

def KerrTrajectory(a, theta, E, bx, by, N):
    """ Computes N points along scattering trajectory which are more or less evenly spaced out in the phi coordinate """
    mu0 = math.cos(theta)
    C1 = E**2 - 1
    C2 = math.sqrt(1-a**2)
    rplus, rminus = 1 + C2, 1 - C2
    L = bx*math.sqrt(C1)*math.sin(theta)
    Q = (E**2 - 1)*((bx**2 - a**2)*mu0**2 + by**2)
    zeros = np.zeros(N)
    ones = np.ones(N)

    if -1 < mu0 < 1:
        if by==0:
            if mu0 > 0:
                s_mu = -1
            else:
                s_mu = 1
        elif by > 0:
            s_mu = 1
        else:
            s_mu = -1
    else:
        raise Exception( "Polar orbits not implemented.")

    r_coeffs = np.array([C1, 2, a**2*C1 - L**2 - Q, 2*((-(a*E) + L)**2 + Q), -a**2*Q])    
    r1, r2, r3, r4 = r_roots = np.sort(np.roots(r_coeffs))
    
    if np.sum(r_roots.imag) > 0.0 or np.max(r_roots.real) < 1+C2:
        raise Exception( "Capture orbits not implemented.")

    discriminant = np.sqrt((bx**2 + by**2)**2 + 2*a**2*(bx - by)*(bx + by)*(-1 + mu0**2) + a**4*(-1 + mu0**2)**2)
    A = -a**2
    B = a**2 * (1 + mu0**2) - bx**2 - by**2
    C = by**2 + (bx**2 - a**2)*mu0**2
    q = -0.5*(B + np.sign(B)*np.sqrt(B**2 - 4*A*C))
    M1, M2 = np.sort((q/A, C/q),axis = 0)
    aSqrM2 = (-bx**2 - by**2 + discriminant + a**2*(1+mu0**2))/2
    aSqrM1 = (-bx**2 - by**2 - discriminant + a**2*(1+mu0**2))/2
    mu_max = np.sqrt(M2)
    mu_min = -np.sqrt(M2)
    kSqr = M2/(M2 - M1)
    n = M2/(1-M2)

    #r-coordinates to calculate
    r_full = 2*InvSqrtQuartic(r1, r2, r3, r4, r4)
    r_integral = np.linspace(0,r_full,N)
    
    f = lambda u, n: InvSqrtQuartic(r1, r2, r3, r4, 1/u) - r_integral[n]
    r = np.empty(N)
    r[0] = np.inf
    r[1:N/2] = 1/np.array([optimize.brentq(f, 1e-16, 1/r4, args = (m,)) for m in xrange(1,N/2)])
    r[N/2:] = r[:N/2][::-1]
    
    mu_complete_integral = 2*CarlsonR.BoostRF(0.0, (bx**2+by**2+discriminant - a**2*(1+mu0**2))/2.0, discriminant)

    mu_initial_integral = CarlsonR.BoostRF(mu0**2, M2*(mu0**2 - M1)/(M2-M1), M2)*np.sqrt((np.abs(M2 - mu0**2))/(M2-M1))/a

    if by*mu0 < 0:
        mu_initial_integral = mu_complete_integral - mu_initial_integral

    case1 = r_integral < mu_initial_integral
    case2 = np.invert(case1)
        
    nTurns = np.empty(N)
    nTurns[case1] = 0
    nTurns[case2] = np.floor((r_integral[case2] - mu_initial_integral)/mu_complete_integral)
    
    integral_remainder = np.abs(r_integral - nTurns*mu_complete_integral - mu_initial_integral)
    
    alpha = s_mu*(-1)**nTurns

    J = np.sqrt(M2-M1)*integral_remainder*a

    mu_final = mu_max*CarlsonR.JacobiCN(J, ones*np.sqrt(kSqr))*alpha

# Do mu-integrals for phi deflection
    xSqr_init = np.abs(1 - mu0**2/M2)
    xSqr_final = np.abs(1 - mu_final**2/M2)
    P = 1/np.sqrt(M2 - M1)/(1-M2)

    pi_complete = P*2*CarlsonR.LegendrePiComplete(-n, kSqr)
    pi_init = P*CarlsonR.LegendrePi(-n, xSqr_init, kSqr)
    pi_final = P*CarlsonR.LegendrePi(-n*ones, xSqr_final*ones, kSqr*ones)

    if mu0*s_mu < 0:
        pi_init = pi_complete - pi_init
    
    A = integral_remainder > mu_complete_integral/2
    pi_final[A] = pi_complete - pi_final[A]
    
    mu_phi_integral = np.empty(N)
    mu_phi_integral[case1] = np.abs(pi_init - pi_final[case1])*L/a
    mu_phi_integral[case2] = np.abs(pi_init + pi_final[case2] + nTurns[case2]*pi_complete)*L/a
    mu_phi_integral = (mu_phi_integral - a*E*r_integral)/math.sqrt(C1)

    r_phi_int_full = PhiTerribleIntegral(r1, r2, r3, r4, a, E, L)

    r_phi_int2 = TerribleIntegral(r1, r2, r3, r4, rplus, r[1:N/2])
    r_phi_int3 = TerribleIntegral(r1, r2, r3, r4, rminus, r[1:N/2])

    r_phi_integral = np.empty(N)
    r_phi_integral[0] = 0.0
    r_phi_integral[1:N/2] = a*E/math.sqrt(E**2-1)*(r_integral[1:N/2] + (a**2 -a*L/E + rplus**2)/(rplus - rminus)*r_phi_int2  - (a**2 -a*L/E + rminus**2)/(rplus - rminus)*r_phi_int3)
    r_phi_integral[N/2:] = r_phi_int_full - r_phi_integral[:N/2][::-1]

    phi = mu_phi_integral + r_phi_integral

    return r, phi, np.arccos(mu_final)
\end{lstlisting}
\lstset{
	language=C++
}
\pagebreak
\subsection{KerrDeflection.cpp}
\begin{lstlisting}
const double PI = 3.141592653589793;

// TerribleIntegral
// Computes the integral in equation 3.16:
// 
// \int_{r_4}^\infty \frac{dr}{(r-r5)\sqrt{(r-r1)(r-r2)(r-r3)(r-r4)}}

inline double TerribleIntegral(double r1, double r2, double r3, double r4, double r5){
  double U12sqr = (r4 - r1)*(r4 - r2),
    U13sqr = U12sqr - (r4 - r1)*(r3-r2),
    U14sqr = U12sqr - (r3-r1)*(r4-r2);

  double Wsqr = U12sqr - (r3-r1)*(r4-r1)*(r5-r2)/(r5-r1),
    Qsqr = (r4 - r5)/(r4-r1)*Wsqr,
    Psqr = Qsqr + (r5 - r2)*(r5 - r3)*(r5-r4)/(r5-r1),
    rc = acosh(sqrt(((r1 - r5)*(r5 - r4))/(r2*r3 - r1*r4 + (r1 - r2 - r3 + r4)*r5)))/sqrt(-((-r2 + r5)*(-r3 + r5)*(-r4 + r5))/(r1 - r5));
    return (2*(r2 - r1)*(r3 - r1)*(r4 - r1)/3.0/(r5-r1) * boost::math::ellint_rj(U12sqr, U13sqr, U14sqr, Wsqr) + 2*rc)/(r5 - r1);
}

// KerrDeflection
// Computes the deflection angles of an arbitrary Kerr orbit, returning NaN if it is a capture orbit
// 

inline void KerrDeflection(double a, double E, double theta, double bx, double by, double &theta_result, double &phi_result){
  //First, some useful quantities
  const double C1 = E*E - 1,
    C2 = sqrt(1-a*a),
    aSqr = a*a,
    rplus = 1 + C2,
    rminus = 1 - C2;

  double mu0 = cos(theta);
  if (mu0 < -1.0) mu0 = -1.0; //insurance against roundoff error
  if (mu0 > 1.0) mu0 = 1.0;
  const double mu0Sqr = mu0*mu0;

  const double bxSqr = bx*bx,
    bySqr = by*by,
    L = bx*sqrt(C1)*sin(theta),
    Q = C1*((bxSqr - aSqr)*mu0Sqr + bySqr);

  // s_mu is the initial sign of \dot{\mu}
  double s_mu;
  if (-1.0 < mu0 && mu0 < 1.0){
    if (by == 0.0) s_mu = -copysign(1.0, mu0);
    else s_mu = copysign(1.0, by);
  } else {
    s_mu = -copysign(1.0, mu0);
  }

  // Solve for the roots of R(r)
  double rootsr[4], rootsi[4];
  double coeffs[] = {-aSqr*Q, 2*(pow(L - a*E, 2) + Q), aSqr*C1 - L*L - Q, 2.0, C1};

  int info;
  quartic(coeffs, rootsr, rootsi, &info);
    
  std::sort(rootsr, rootsr + 4);

  // Check whether capture trajectory, if so return NaN. We get a capture trajectory if there are complex roots,
  // so we add the imaginary parts and check if it is 0
  double root_sum = fabs(rootsi[0])+fabs(rootsi[1])+fabs(rootsi[2])+fabs(rootsi[3]);
  if (root_sum != 0.0 || rootsr[3] < rplus){
    phi_result = NAN;
    theta_result = NAN;
    return;
  }
  
  double r1 = rootsr[0],
    r2 = rootsr[1],
    r3 = rootsr[2],
    r4 = rootsr[3];

  //Solve biquadratic polynomial equation M(mu) == 0
  double disc = sqrt(pow(bxSqr + bySqr, 2) + 2*aSqr*(bxSqr - bySqr)*(mu0Sqr - 1.0) + aSqr*aSqr*pow(mu0Sqr - 1.0, 2)),
    A = -aSqr,
    B = aSqr*(mu0Sqr + 1.0) - bxSqr - bySqr,
    C = Q/C1;
  double q = -0.5*(B + copysign(1.0, B)*sqrt(B*B - 4*A*C));
  double M1 = std::min(q/A, C/q),
    M2 = std::max(q/A, C/q);
  //protect against roundoff error by clipping M2 to 1 if it is greater
  if (M2 > 1.0) M2 = 1.0;

  double k = sqrt(M2/(M2-M1)),
    n = M2/(1-M2);

  double U12sqr = (r4 - r1)*(r4 - r2),
    U13sqr = U12sqr - (r4 - r1)*(r3-r2),
    U14sqr = U12sqr - (r3-r1)*(r4-r2);

  // Compute I_r (Equation 3.19)
  double r_integral = 4*boost::math::ellint_rf(U12sqr, U13sqr, U14sqr);

  // Compute I_{\mu_c} (Equation 3.24)
  double mu_complete_integral = 2*boost::math::ellint_rf(0.0, (disc - B)/2.0, disc);
  double mu_initial_integral;

  // Compute I{\mu_i} (Equation 3.23)
  if(fabs(M2 - mu0Sqr)/M2 > 1e-15){
    mu_initial_integral = boost::math::ellint_rf(mu0Sqr, M2*(mu0Sqr - M1)/(M2-M1), M2)*sqrt(fabs((M2-mu0Sqr)/(M2-M1)))/fabs(a);
  } else {
    mu_initial_integral = mu_complete_integral;
  }

  if (mu0*by < 0.0) mu_initial_integral = mu_complete_integral - mu_initial_integral;

  int N = int((r_integral - mu_initial_integral)/mu_complete_integral);
  double integral_remainder = r_integral - N*mu_complete_integral - mu_initial_integral;
  double alpha = s_mu*pow(-1.0, N);

  // Get final mu coordinate with Jacobi elliptic function (Equation 3.27)
  double cn, dn;
  boost::math::jacobi_elliptic(k, sqrt(M2-M1)*integral_remainder*fabs(a), &cn, &dn);
  double mu_final = sqrt(M2)*cn*alpha;
  theta_result = acos(mu_final);

  // Now we use the limits of integration in mu to do the integrals for phi
  // First some useful quantities
  double xSqr_init = 1 - mu0Sqr/M2,
    xSqr_final = 1- mu_final*mu_final/M2,
    P = 1/sqrt(M2 - M1)/(1-M2);

  //Insurance against roundoff error: both of these guys should always be non-negative
  xSqr_init = std::max(0.0, xSqr_init);
  xSqr_final = std::max(0.0, xSqr_final);
  
  // Compute \Phi_{\mu_i}, \Phi_{\mu_c}, \Phi_{\mu_f} (Equation 3.29-3.31)
  double Phi_complete, Phi_init, Phi_final;
  if (fabs(bx/by) > 1e-8){
    // Non-polar orbits
    Phi_complete = 2*boost::math::ellint_3(k, -n);
    Phi_init = boost::math::ellint_3(k, -n, asin(sqrt(xSqr_init)));
    Phi_final = boost::math::ellint_3(k, -n, asin(sqrt(xSqr_final)));
  } else {
    //Zero angular momentum orbits which pass over the poles
    Phi_complete = PI;
    Phi_init = 0;
    Phi_final = 0;
  }

  if (mu0*s_mu < 0.0) Phi_init = Phi_complete - Phi_init;  
  if (integral_remainder > mu_complete_integral/2) Phi_final = Phi_complete - Phi_final;

  double mu_phi_integral;
  if (fabs(bx/by) > 1e-8){
    mu_phi_integral = P*(Phi_init + Phi_final + N*Phi_complete)*L/fabs(a)/sqrt(C1); //Again, non-polar orbits
  } else {
    mu_phi_integral = Phi_init + Phi_final + N*Phi_complete; // polar orbits
  }

  //Evaluate the Terrible Integrals T_+ and T_- 
  double Tminus = TerribleIntegral(r1, r2, r3, r4, rminus) - r_integral/2/(rminus-r1),
    Tplus = TerribleIntegral(r1, r2, r3, r4, rplus)- r_integral/2/(rplus-r1);

  // second line of Equation 3.36
  double r_phi_integral = 2*a*E/sqrt(C1) * ((aSqr -a*L/E + rplus*rplus)/(rplus - rminus)*Tplus  - (aSqr -a*L/E + rminus*rminus)/(rplus - rminus)*Tminus);
  
  phi_result = mu_phi_integral + r_phi_integral;
}
\end{lstlisting}
\pagebreak

\section{Post-Newtonian Computations}
The post-Newtonian results from chapter 4 were obtained by numerically integrating the PN equation. An adaptive integration scheme, the implementation of the Dormand-Prince 853 method in \texttt{scipy}'s \texttt{integrate.ode} class, was used to perform the integration. 
\lstset{
	language=Python
}
\subsection{PostNewtonian.py}
\begin{lstlisting}
def PN_Deriv(X, t, order, eta, radiation=0):
    """ Calls the C++ subroutine and returns the PN derivative """
    
    derivative = np.empty(6)
    code = """
    derivative[0] = X[3];
    derivative[1] = X[4];
    derivative[2] = X[5];
    PNDeriv(X[0], X[1], X[2], X[3], X[4], X[5], derivative[3], derivative[4], derivative[5], eta, order, radiation);
"""
    weave.inline(code, ['X', 'derivative','order', 'eta','radiation'],headers=["<cmath>","<PostNewtonian.cpp>"], extra_compile_args = ['-O3 -mtune=native -march=native -ffast-math -msse3 -fomit-frame-pointer -malign-double -fstrict-aliasing'])
    return derivative

def PNDeflection(v0, b, eta, order, R0=None, radiation=0):
    """ Computes the PN deflection angle """
    
    if R0==None:
        R0 = 1e6*b

    # Newtonian time to get to periastron and back
    T = ((2*v0*np.sqrt(2*R0 + (-b**2 + R0**2)*v0**2) + np.log(1 + b**2*v0**4) - 2*np.log(1 + v0*(R0*v0 + np.sqrt(2*R0 + (-b**2 + R0**2)*v0**2))))/v0**3)

    # Initial conditions
    x, y, z = R0, b, 0.0
    vx, vy, vz = -v0, 0.0, 0.0
    X0 = np.array([x, y, z, vx, vy, vz])
    
    t = np.linspace(0, T, 2)
    X = IntegratePN(X0, t, eta, order, radiation)

    if type(X) == int: return np.nan
    elif np.sum(X[:3]**2) < R0/10.0 or Energy(X,eta) < 0.0: return np.nan
    else:
        result = (np.arctan2(X[4],X[3]))%(2*np.pi)
        return result

def NewtonianDeflection(v0, b):
    e = np.sqrt(1 + v0**4 * b**2)
    return np.sign(b)*(2*np.pi - 2*np.arccos(1/e))

def IntegratePN(X, t, eta, order, radiation=0):
    #Y = np.empty((len(t), 6))
    f = lambda t, x, order, eta, radiation: PN_Deriv(x, t, order, eta, radiation)
    r = integrate.ode(f).set_integrator('dop853',rtol=1e-12, nsteps=1000)#'dop853',rtol=1e-12,nsteps=10000)
    r.set_initial_value(X, t[0]).set_f_params(order,eta, radiation)
    r.integrate(t[-1])
    if r.successful(): return r.y
    else: return -1
    
def PNTrajectory(v0, b, eta, order, R0, N, radiation=0):
    T = ((2*v0*np.sqrt(2*R0 + (-b**2 + R0**2)*v0**2) + np.log(1 + b**2*v0**4) - 2*np.log(1 + v0*(R0*v0 + np.sqrt(2*R0 + (-b**2 + R0**2)*v0**2))))/v0**3)
    x, y, z = R0, b, 0.0
    vx, vy, vz = -v0, 0.0, 0.0
    X0 = np.array([x, y, z, vx, vy, vz])
    
    t = np.linspace(0, T, N)
    X = IntegratePNTraj(X0, t, eta, order, radiation)
    return X

def IntegratePNTraj(X, t, eta, order, radiation=0):
    Y = np.zeros((len(t), 6))
    f = lambda t, x, order, eta, radiation: PN_Deriv(x, t, order, eta, radiation)
    r = integrate.ode(f).set_integrator('dop853',rtol=1e-8, nsteps=1000)
    r.set_initial_value(X, t[0]).set_f_params(order,eta, radiation)
    i = 1
    Y[0] = X
    while r.successful() and i<len(t):
        r.integrate(t[i])
        Y[i] = r.y
        i += 1
    return Y

def FinalEnergy(v0, b, eta, order, radiation=0):
    R0 = 1e9
    T = ((2*v0*np.sqrt(2*R0 + (-b**2 + R0**2)*v0**2) + np.log(1 + b**2*v0**4) - 2*np.log(1 + v0*(R0*v0 + np.sqrt(2*R0 + (-b**2 + R0**2)*v0**2))))/v0**3)
    x, y, z = R0, b, 0.0
    vx, vy, vz = -v0, 0.0, 0.0
    X0 = np.array([x, y, z, vx, vy, vz],dtype=np.float64)
    t = np.linspace(0, T, 2)
    X = IntegratePN(X0, t, eta, order, radiation)
    if type(X) == int or Energy(X, eta) > Energy(X0, eta): return -1.0
    else: return  Energy(X,eta)

def FindBmin(v0, eta, order, radiation=0, b0=None):
    if order == 3 and radiation==0:
        return PN3_Bmax(v0, eta, radiation=radiation)
    R0 = 1e9
    bmax = KerrDeflection.bmin((1-min(v0,0.999)**2)**-0.5)**(9.0/7.0)
    if b0==None:
            b0 = 0.5*KerrDeflection.bmin((1-min(v0,0.999)**2)**(-1./2))
    f = lambda b: FinalEnergy(v0, b, eta, order, radiation)
    return optimize.brentq(f, b0, bmax)
    
def MonotonicAngles(angles):
    for i in range(len(angles)-1)[::-1]:
        while angles[i]-angles[i+1] < 0.0:
            angles[i] += 2*np.pi
            
def Energy(state, eta):
    vSqr = np.sum(state[3:]**2)
    rSqr = np.sum(state[:3]**2)
    r = np.sqrt(rSqr)
    rdot = np.sum(state[:3]*state[3:])/r
    
    E_N = 0.5*vSqr - 1.0/r
    E_PN = 3.0/8.0*(1-3*eta)*vSqr*vSqr + 0.5*(3 + eta)*vSqr/r + 0.5*eta/r*rdot**2 + 0.5/rSqr
    E_2PN = 5.0/16.0*(1 - 7*eta + 13*eta**2)*vSqr**3 - 3.0/8.0*eta*(1-3*eta)/r*rdot**4 + 1.0/8.0*(21-23*eta-27*eta**2)*vSqr**2/r+ 1.0/8.0*(14 - 55*eta + 4*eta**2)/rSqr*vSqr + 0.25*eta*(1-15*eta)/r*vSqr*rdot**2- 0.25*(2+15*eta)/r**3 + 1.0/8.0*(4 + 69*eta + 12*eta**2)/rSqr*rdot**2
    return E_N + E_PN+E_2PN

def EnergyFlux(v0, b, eta, order=2):
    R0 = 1e9
    T = ((2*v0*np.sqrt(2*R0 + (-b**2 + R0**2)*v0**2) + np.log(1 + b**2*v0**4) - 2*np.log(1 + v0*(R0*v0 + np.sqrt(2*R0 + (-b**2 + R0**2)*v0**2))))/v0**3)
    x, y, z = R0, b, 0.0
    vx, vy, vz = -v0, 0.0, 0.0
    X0 = np.array([x, y, z, vx, vy, vz],dtype=np.float64)
    t = np.linspace(0, T, 2)
    X = IntegratePN(X0, t, eta, order, radiation=1)
    if type(X) == int: return np.nan
    else: return  Energy(X0,eta) - Energy(X, eta)
            
def CrossSection(b, theta):
    db_dtheta = np.gradient(b)/np.gradient(theta)
    return np.abs(b/np.sin(theta)*db_dtheta)

def PN3_Bmax(v0, eta, rad=0):
    if v0==1.0:
        sbm = 3*np.sqrt(3)
    else:
        sbm = KerrDeflection.bmin((1-v0**2)**-0.5)
    
    f = lambda b: PNDeflection(v0, b, eta, 3, radiation=rad)- PNDeflection(v0, b - 1e-3, eta, 3, radiation=rad)
    return optimize.newton(f, sbm)

def PNError(v0, eta, b, rad=0):
    pn3 = PNDeflection(v0, b, eta, 3, radiation = rad)-np.pi
    pn2 = PNDeflection(v0, b, eta, 2, radiation = rad)-np.pi
    #print b, pn2, pn3
    return min(np.abs(pn3-pn2)/np.abs(pn3),np.abs(pn3-pn2-2*np.pi)/np.abs(pn3))
    
def PNValidRegion(v0, eta, rad=0):
    f = lambda b: np.abs(PNError(v0, eta, b, rad=rad) - 0.1)
    E = (1-min(v0,0.999)**2)**-0.5
    b0 = KerrDeflection.bmin(E)
    bm2 = FindBmin(v0, eta, 2)
    return optimize.minimize_scalar(f, method='bounded', bounds=(bm2*(1+1e-3), 2*bm2)).x
\end{lstlisting}
\lstset{
	language=C++
}
\pagebreak
\subsection{PostNewtonian.cpp}
\begin{lstlisting}
const double PI = 3.141592653589793;

inline void PNDeriv(double x, double y, double z, double vx, double vy, double vz, double &ax, double &ay, double &az, double eta, int order, int radiation){
  const double rSqr = x*x + y*y + z*z,
    r = sqrt(rSqr),
    vSqr = vx*vx + vy*vy + vz*vz,
    v4 = vSqr*vSqr,
    v6 = v4*vSqr,
    rdot = (x*vx + y*vy + z*vz)/r,
    rdotSqr = rdot*rdot,
    rdot4 = rdotSqr*rdotSqr,
    rdot6 = pow(rdotSqr,3);
 
  const double A1 = -3.0/2.0*rdotSqr*eta + (1+3*eta)*vSqr -(4+2*eta)/r,
    A2 = 0.75*(12 + 29*eta)/rSqr + eta*(3-4*eta)*vSqr*vSqr + 15./8*eta*(1-3*eta)*rdotSqr*rdotSqr 
    - 1.5*eta*(3-4*eta)*vSqr*rdotSqr - 0.5*eta*(13 - 4*eta)*vSqr/r - (2 + 25*eta + 2*eta*eta)/r*rdotSqr,
    A25 = rdot*eta*(-24.0/5.0*vSqr/r - 136.0/15.0/rSqr),
    A3 = 1.0/16.0*eta*rdot6*(-35 + 175*(eta - eta*eta)) + rdot4*vSqr*eta*(15.0/2.0 - 135.0/4.0*eta + 255.0/8.0*eta*eta) + rdotSqr*v4*eta*(-15.0/2.0 + 237.0/8.0*eta - 45.0/2.0*eta*eta) +eta*v6*(11.0/4.0 -49.0/4.0*eta + 13.0*eta*eta) + 1.0/r*(rdot4*eta*(79 - 69.0*eta/2.0 - 30*eta*eta) + rdotSqr*vSqr*eta*(-121.0 + 16.0*eta + 20*eta*eta) + v4*eta*(75.0/4.0 + 8*eta - 10*eta*eta)) + 1.0/rSqr * (rdotSqr*(1 + 22717.0/168.0*eta + 11.0/8.0*eta*eta - 7.0*eta*eta*eta + 615.0/64.0*eta*PI*PI) + vSqr*(-20827.0/840.0*eta + eta*eta*eta - 123.0/64.0*eta*PI*PI)) + 1.0/rSqr/r*(-16 - 1399.0/12.0*eta - 71.0/2.0*eta*eta + 41.0/16.0*PI*PI*eta),
    A35 = rdot*eta/r*((v4*(366.0/35.0 + 12*eta) + vSqr*rdotSqr*(-114 - 12*eta) + 112*rdot4) + 1.0/r*(vSqr*(692.0/35.0 - 724.0/15.0*eta) + rdotSqr*(294.0/5.0 + 376.0/5.0*eta)) + 1.0/rSqr*(3956.0/35.0 + 184.0/5.0*eta));

  const double B1 = -2*(2-eta)*rdot,
    B2 = -0.5*rdot*(eta*(15.0 + 4.0*eta)*vSqr - (4.0 + 41.0*eta + 8.0*eta*eta)/r - 3.0*eta*(3.0+2.0*eta)*rdotSqr),
    B25 = eta*(8.0/5.0*vSqr/r + 24.0/5.0/rSqr),
    B3 = rdot4*rdot*eta*(-45.0/8.0 + 15*eta +15.0/4.0*eta*eta) + rdotSqr*rdot*vSqr*eta*(12.0 - 111.0/4.0*eta -12*eta*eta) + rdot*v4*eta*(-65.0/8.0 + 19*eta + 6*eta*eta) + 1.0/r*(eta*rdot*rdotSqr*(329.0/6.0 + 59.0/2.0*eta + 18*eta*eta) + rdot*vSqr*eta*(-15 -27*eta - 10*eta*eta)) + 1.0/rSqr*rdot*(-4.0 - 5849.0/840.0*eta + 25.0*eta*eta + 8*eta*eta*eta - 123.0/32.0*eta*PI*PI),
    B35 = eta/r*(v4*(-626.0/35.0 - 12.0/5.0*eta) + vSqr*rdotSqr*(678.0/5.0 + 12.0/5.0*eta) - 120*rdot4 + 1.0/r*(vSqr*(164.0/21.0 + 148.0/5.0*eta) + rdotSqr*(-82.0/3.0 - 848.0/15.0*eta)) + 1.0/rSqr*(-1060.0/21.0 - 104.0/5.0*eta));
  
  double A = 0, B = 0;
  if (order > 2) {
    if (radiation){
      A += A35;
      B += B35;
      }
    A += A3;
    B += B3;
  }
  if (order > 1) {
    if (radiation) {
      A += A25;
      B += B25;
    }
    A += A2;
    B += B2;
  }
  if (order > 0) {
    A += A1;
    B += B1;
  }

  ax = -((1+A)*x/r + B*vx)/rSqr;
  ay = -((1+A)*y/r + B*vy)/rSqr;
  az = -((1+A)*z/r + B*vz)/rSqr;
}
\end{lstlisting}
\end{spacing}

%% file: bibliography.tex
\bibliography{mybib}{}
\bibliographystyle{unsrt}
\addcontentsline{toc}{chapter}{Bibliography}


